\documentclass{jfm}
\usepackage{amsmath}
\usepackage{amsfonts}
\usepackage{bm}
\usepackage{enumerate}
\usepackage{graphicx}
\usepackage{mathrsfs}
\usepackage{subfigure}
\usepackage{url}
\usepackage{verbatim}
\usepackage{amssymb}
\usepackage[pdftex,bookmarksnumbered]{hyperref}
\usepackage{color}

\usepackage{epstopdf, epsfig}

\newtheorem{proposition}{Proposition}

\shorttitle{The approach to the long time asymptotic of passive scalar}

\title{  Long-Time Asymptotics of Passive Scalar Transport in Periodically Modulated Channels}

\author{Lingyun Ding\aff{1}  \corresp{\email{dingly@g.ucla.edu}}}

  \affiliation{\aff{1}Department of Mathematics, University of California, Los Angeles, Los Angeles, CA, 90095, United States}

\begin{document}

\maketitle
\begin{abstract}
This work investigates the long-time asymptotic behavior of a diffusing passive scalar advected by fluid flow in a straight channel with a periodically varying cross-section. The goal is to derive an asymptotic expansion for the scalar field and estimate the timescale over which this expansion remains valid, thereby generalizing Taylor dispersion theory to periodically modulated channels. By reformulating the eigenvalue problem for the advection-diffusion operator on a unit cell using a Floquet-Bloch-type eigenfunction expansion, we extend the classical Fourier integral of the flat-channel problem to a periodic setting, yielding an integral representation of the scalar field. This representation reveals a slow manifold that governs the algebraically decaying dynamics, while the difference between the scalar field and the slow manifold decays exponentially in time. Building on this, we derive a long-time asymptotic expansion of the scalar field. We show that the validity timescale of the expansion is determined by the real part of the eigenvalues of a modified advection-diffusion operator, which depends solely on the flow and geometry within a single unit cell. This framework offers a rigorous and systematic method for estimating mixing timescales in channels with complex geometries. We show that non-flat channel boundaries tend to increase the timescale, while transverse velocity components tend to decrease it. The approach developed here is broadly applicable and can be extended to derive long-time asymptotics for other systems with periodic coefficients or periodic microstructures.
 \end{abstract}

\begin{keywords}
  Passive scalar, Multiscale analysis, Taylor dispersion, Advection–diffusion equation, Long-time asymptotics, Slow manifold
\end{keywords}

\section{Introduction}
Taylor dispersion plays a crucial role in the transport and mixing of passive scalars (such as temperature or concentration) in fluid flows, with applications spanning various scientific and engineering disciplines. Originally studied by \cite{taylor1953dispersion}, it describes how a solute in shear flow undergoes enhanced mixing due to the interplay between advection and diffusion. Classical Taylor dispersion theory has been widely utilized, as it enables dimensional reduction by approximating the governing equations with fewer independent variables (\cite{young1991shear,ding2023shear,teng2023diffusioosmotic,guan2024streamwise,david2024magneto}). However, since Taylor dispersion is an asymptotic approximation valid at long times, a fundamental question arises: at what time scale does this approximation become valid?

To address this question, it is useful to examine the evolution of the scalar field in a straight channel with uniform cross-section, from its initial introduction into the flow to its eventual homogenization. This process can be characterized by three distinct time scales: dispersion, longitudinal normality, and transverse uniformity, which we outline below.

In the early stage after the solute is injected into the channel, it is primarily advected along streamlines and  undergoes stretching. During this stage, advection dominates while diffusion begins to play a role, leading to a complex evolution of the scalar field \cite{taghizadeh2020preasymptotic,camassa2010analysis}.  The first significant time scale arises when the variance of the solute concentration begins to grow linearly in time, marking the onset of Taylor dispersion \cite{aris1956dispersion, camassa2010exact}. This signals that diffusion has started to act in concert with the background flow to redistribute the solute along the channel. At this stage, the rate of variance growth exceeds that of pure molecular diffusion due to the coupling between shear-induced advection and transverse diffusion. One might expect that the scalar field could already be described by a diffusion equation, since the solution of a diffusion equation is a Gaussian function with linearly increasing variance. However, at this stage, transverse variations in the scalar field persist, and the cross-sectionally averaged scalar field may still deviate from a Gaussian profile, exhibiting skewness that depends on the channel geometry and shear flow characteristics \cite{aminian2016boundaries}.

As molecular diffusion continues, the cross-sectionally averaged scalar field approaches a Gaussian distribution, marking the second time scale. Once this stage is reached, the evolution of the cross-sectionally averaged scalar field can be effectively described by a one dimensional diffusion equation with an enhanced diffusivity \cite{chatwin1970approach}. We refer to this governing equation as the effective equation and to the enhanced diffusion coefficient as the effective diffusivity. However, even at this stage, the solute concentration may still exhibit nonuniformity across the channel cross-section.

The third and final time scale corresponds to the attainment of transverse uniformity across the channel cross-section. At this point, diffusion has sufficiently spread the solute so that the concentration becomes nearly uniform across the cross-section. \cite{wu2014approach} demonstrated that this time scale can be up to ten times larger than that for longitudinal normality, highlighting the extended duration required for complete transverse homogenization. At this stage, the whole scalar field can be well approximated by a one-dimensional effective equation.

How do these time scales depend on physical parameters? The enhanced mixing in Taylor dispersion arises from the diffusion of concentration variations across the channel’s cross-section. The time scale for this enhanced mixing is related to the rate at which these variations decay into a uniform concentration through diffusion. Dimensional analysis shows that the diffusion time scale is $L^2/\kappa$, where $L$ is the characteristic length scale of the channel cross-section, and $\kappa$ is the molecular diffusivity. Interestingly, many studies of Taylor dispersion in a straight channel with flat boundaries show that the variance starts to increase linearly at an earlier time, approximately $0.1L^2/\kappa$, for steady shear flows in various geometries, such as parallel plate channels \cite{camassa2010exact}, circular pipes \cite{barton1983method}, and even time-dependent flows \cite{vedel2012transient, vedel2014time}. The variance evolution can be computed exactly using the method of moments, first proposed by \cite{aris1956dispersion}. These calculations refine the dispersion time scale to $L^2/(\kappa \lambda)$, where $\lambda$ is the smallest nonzero eigenvalue of the Laplacian on the cross-section domain. For a parallel plate channel, $L$ is the gap width, $\lambda = \pi^2$, and $1/\lambda \approx 0.1$. For a circular pipe with radial symmetry, $L$ is the radius, $\lambda = 14.682$ (the square of the roots of $J_1(x)$, the Bessel function of the first kind of order one), yielding $1/\lambda \approx 0.0681$.

In addition to studies focusing on the variance, others investigate the approximation of the entire scalar field.  \cite{mercer1990centre, watt1995accurate} demonstrated that the scalar field converges exponentially to the center manifold of the system, and the rate of this convergence is governed by the Laplacian eigenvalue. This implies that the time scale for achieving longitudinal normality is also closely related to $L^2/(\kappa \lambda)$. \cite{stokes1990concentration,phillips1996uniformly} applied a Laplace transform in time and a Fourier transform in the longitudinal direction to derive an integral representation of the concentration field, enabling accurate asymptotic expansions at both short and long times. Using a representation that combines the Fourier transform with eigenfunction expansion, \cite{ding2022ergodicity} derived an effective equation and effective diffusivity for a family of stochastic shear flows, demonstrating that the time scale is governed by the Laplacian eigenvalues associated with the random scalar field.

While the theory for channels with flat walls is well-developed, real-world systems often involve more complex geometries, such as channels with periodically varying cross-sections \cite{roggeveen2023transport}. These geometries are crucial in applications like passive microfluidic mixers and flow through porous media. Micro passive mixers enable rapid homogenization, essential for chemical and biological processes at the microscale. Specially designed channel boundaries can enhance mixing by inducing chaotic advection (\cite{liu2000passive,stroock2002chaotic,stone2004engineering,oevreeide2020curved}), leading to faster homogenization compared with shear-driven mixing in straight channels. In addition, a sinusoidally varying channel radius serves as an idealized model for flow through porous media, where constrictions and expansions mimic pore throats and pore bodies (\cite{richmond2013flow,liu2024scaling}). In both cases, generalized Taylor dispersion theory provides a framework for estimating the longitudinal dispersion coefficient and approximating the scalar field (\cite{brenner1980dispersion,hoagland1985taylor,amaral1997dispersion,degroot2011closure,haugerud2022solute}). Thus, determining the characteristic time for the validity of Taylor dispersion theory is of fundamental importance.

Several numerical studies have attempted to estimate the characteristic time scale. For instance, \cite{bouquain2012impact} investigated sinusoidally varying channels and estimated the characteristic time by fitting the numerically computed time evolution of the normalized dispersion coefficient to an exponential relaxation function. \cite{haugerud2022solute} investigated channels with periodic square boundary roughness.  Non-flat boundaries can induce circulation regions that trap particles and affect the dispersion process. \cite{haugerud2022solute} showed that particle residence times in these circulation regions are directly related to the characteristic mixing time. However, these approaches rely on simulating the scalar field over the entire domain, which is computationally expensive, particularly in geometries with complex boundaries. Despite valuable insights, a rigorous and systematic theoretical framework for estimating mixing times in such domains remains an open challenge.

As mentioned earlier, in the case of scalar transport in a channel with flat boundaries, the approach combining the Fourier transform in the longitudinal direction with eigenfunction expansion has been successful in analyzing characteristic times. However, two main challenges arise when generalizing this method to a channel with periodically varying cross-sections.  First, the Fourier transform in the axial direction is no longer applicable due to the complex geometry. Second, while the flow may be periodic and solvable within a unit cell, the scalar field itself is not periodic. As a result, the eigenfunctions of the advection-diffusion operator must be defined over the entire domain rather than a single cell. Since exact solutions to this eigenvalue problem are typically unavailable, numerical methods are required. However, solving on an infinite domain poses computational challenges, making this approach impractical.

To overcome the challenges of solving the advection-diffusion equation in a periodic channel, we introduce a Floquet–Bloch–type eigenfunction representation that separates the non-periodic and periodic components of the scalar field. This reduces the eigenvalue problem from the full domain to a simpler one defined on a single unit cell. These eigenfunctions form a complete basis for square-integrable functions, enabling an integral representation of the scalar field. This representation highlights the mode governing the system's slow dynamics, leading to a reduced model whose difference from the full system decays exponentially in time. This reduced model allows us to rigorously determine the timescale over which Taylor dispersion theory remains valid. By applying the method of steepest descent to the integral representation of the slow manifold, we derive the long-time asymptotic expansion of the scalar field. This approach reveals the structure of the advection-diffusion solution in a periodic channel and clarifies the relationships among key dispersion timescales: longitudinal normality, transverse uniformity, and effective dispersion. The dominant timescale depends solely on the flow and domain geometries and can be computed from the solution within a single unit cell. Together, these results offer a rigorous and systematic framework for estimating mixing timescales in complex geometries.

This paper is organized as follows: In Section \ref{eq:Governing Equation and asymptotic analysis}, we first formulate the governing equations for the scalar field and fluid flow and outline the nondimensionalization process. We then define the eigenvalue problem and present the eigenfunction expansion of the scalar field. Based on this, we identify the system's slow manifold and derive the long-time asymptotic expansion of the scalar field. Next, we discuss the relationship between different timescales in the Taylor dispersion problem. Section \ref{sec:Characteristic time for converging to the slow manifold} examines the characteristic time for the scalar field to converge to the slow manifold through several examples.  In Section~\ref{sec:Extension to Porous Media}, we briefly discuss extending the proposed method to porous media. In Section \ref{sec:conclusion}, we summarize our findings and explore potential directions for future research. Appendix \ref{sec:Eigenvalue} provides the perturbation calculations for the small-wavenumber expansion of the eigenfunctions, while Appendix \ref{sec:Numerical method} details the numerical methods used in this study.

\section{Governing Equation and asymptotic analysis}
\label{eq:Governing Equation and asymptotic analysis}
We consider a straight channel domain of length $L$ with a periodically varying cross-section, defined as
\begin{equation}\label{eq:domain}
\begin{aligned}
\Omega (L) = \left\{ (x, \mathbf{y}) \,|\, -\frac{L}{2} \leq x \leq \frac{L}{2}, \, \mathbf{y} \in \Omega_{c}(x) \right\},
\end{aligned}
\end{equation}where the $x$-direction is the longitudinal axis of the channel and $\Omega_{c}(x) \subset \mathbb{R}^{d}$ denotes the cross-section of the channel, which depends on $x$. We assume that $\Omega_{c}(x)$ is periodic in $x$ with a period $L_{p}$.  We often focus on the case of an infinitely long channel, denoting it as $\Omega = \Omega(\infty)$ for simplicity. The volume of the channel is represented as $|\Omega (L)|$, while the average cross-sectional area is given by $|\Omega_{c}| = \frac{|\Omega (L_{p})|}{L_{p}}$. The simplest case is a domain bounded by two wavy walls: $\Omega = \left\{ (x, y) \,|\, x \in \mathbb{R}, \, h_{1}(x) \leq y \leq h_{2}(x) \right\}$, where $h_{1}(x)$ and $h_{2}(x)$ share the same period $L_{p}$. In this scenario, the average cross-sectional area is calculated as $|\Omega_{c}| = \frac{1}{L_{p}} \int_{0}^{L_{p}} (h_{2}(x) - h_{1}(x)) \, \mathrm{d} x$. Beyond this type of domain, practical examples include axisymmetric channels as studied in \cite{chang2023taylor} and curved microfluidic channels \cite{stroock2002chaotic, liu2000passive}.

\begin{figure}
  \centering
    \includegraphics[width=1\linewidth]{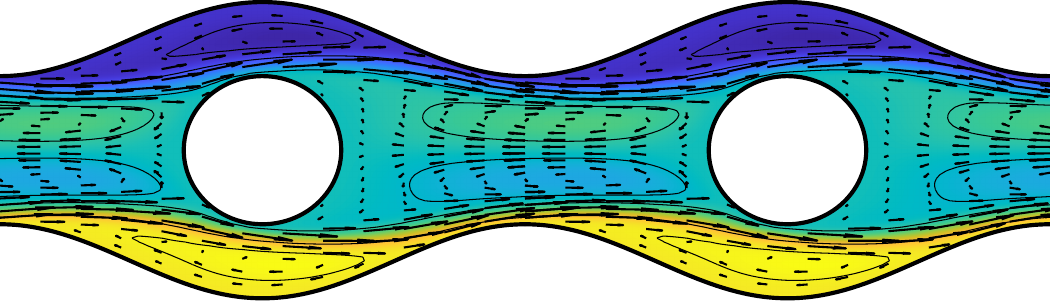}
  \hfill
  \caption[]
  { A pressure-driven flow passes from left to right through a channel with periodically varying cross-sections. Two unit cells of the periodic pattern are displayed.   The velocity magnitude in the recirculating region is about one-tenth of that in the main flow stream; therefore, the arrow lengths have been scaled to enhance the visibility of the recirculation.  The flow field was obtained using FreeFem++ with the algorithm described in Appendix \ref{sec:Numerical method}.  }
  \label{fig:CurveChannelSchematic}
\end{figure}

The passive scalar is governed by the advection-diffusion equation with no-flux boundary conditions, which takes the form
\begin{equation}\label{eq:AdvectionDiffusionEquation}
\begin{aligned}
  &\partial_{t} c + \mathbf{u} \cdot \nabla c = \nabla \cdot (\kappa \nabla c) , \quad c (x, \mathbf{y}, 0) = c_{I} (x, \mathbf{y}), \quad \left. \mathbf{n} \cdot \nabla c \right|_{\mathrm{wall}} = 0,
\end{aligned}
\end{equation}
where $\kappa$ is the diffusivity, and $c_{I}$ is the initial condition, which is assumed to be nonnegative. The vector $\mathbf{n} = (n_{x}, n_{1}, \ldots, n_{d})$ denotes the outward normal to the boundary, and $\mathbf{u}$ is the velocity field. The theory developed in this paper can be readily generalized to the case where $\kappa$ is a spatially periodic function, i.e., $\kappa = \kappa(x, \mathbf{y})$, but for simplicity, we assume it to be constant throughout this work. To ensure mass conservation, we assume that the velocity field is incompressible and satisfies the no-flux boundary condition. Additionally, we consider flows that are periodic in the longitudinal direction of the channel, with a fundamental period $L_{p}$. While the theoretical results developed here apply to any velocity field satisfying these assumptions, the simulations presented use velocity fields governed by the Navier-Stokes equations. For a homogeneous fluid, the velocity field $\mathbf{u} = (u, \mathbf{v})$ satisfies the Navier-Stokes equations:
\begin{equation}\label{eq:NS dimensional}
\begin{aligned}
  &\partial_{t} u + u \partial_{x} u + \mathbf{v} \cdot \nabla_{\mathbf{y}} u = -\frac{1}{\rho_{0}} \partial_{x} p + \nu \left( \partial_{x}^{2} u + \Delta_{\mathbf{y}} u \right)+f_{x}, \\
  &\partial_{t} v_{i} + u \partial_{x} v_{i} + \mathbf{v} \cdot \nabla_{\mathbf{y}} v_{i} = -\frac{1}{\rho_{0}} \partial_{y_{i}} p + \nu \left( \partial_{x}^{2} v_{i} + \Delta_{\mathbf{y}} v_{i} \right)+f_{y,i}, \quad i = 1, \ldots, d, \\ 
  &\partial_{x} u + \nabla_{\mathbf{y}} \cdot \mathbf{v} = 0, \quad \left. \mathbf{u} \right|_{\mathrm{wall}} = 0,
\end{aligned}
\end{equation}
where $u$ is the fluid velocity in the longitudinal $x$-direction, $v_i$ is the fluid velocity aligned with the $y_i$-directions, $\mathbf{v} = (v_{1}, \ldots, v_{d})$, $p$ is the pressure, $\rho_{0}$ is the fluid density, and $\nu$ is the fluid kinematic viscosity. The subscript $\mathbf{y}$ denotes quantities related to the coordinates in the cross-section, where $\mathbf{n}_{\mathbf{y}} = (n_{1}, \ldots, n_{d})$, $\nabla_{\mathbf{y}} = (\partial_{y_{1}}, \ldots, \partial_{y_{d}})$, and $\Delta_{\mathbf{y}} = \sum_{i=1}^{d} \partial_{y_{i}}^{2}$. The forcing terms $f_{x}$ and $f_{y_{i}}$ represent external forces. In pressure-driven flow, $f_{x}$ is a constant while $f_{y_{i}} = 0$. If the domain boundary is periodic, it induces periodicity in the flow, as illustrated in figure \ref{fig:CurveChannelSchematic}, satisfying our assumption about the velocity field structure. While a transient region exists near the inlet, the flow converges exponentially to the periodic state \cite{feppon2025asymptotic}. Therefore, our analysis focuses on the fully developed periodic flow.

\subsection{Nondimensionalization}
We denote the characteristic length scale of the channel cross-section as $L_{y}$.  We then proceed with the following change of variables for nondimensionalization:
\begin{equation}
\begin{aligned}
  &L_{y} \tilde{x}=x, \quad L_{y} \tilde{\mathbf{y}}=\mathbf{y},\quad U \tilde{u}  =u, \quad  U \tilde{\mathbf{v}}=\mathbf{v}, \quad \frac{\rho_{0}\nu U}{L_{y}} \tilde{p}=p,  \quad \frac{L_{y}^{2}}{\kappa}\tilde{t}=t,\quad c_{c}\tilde{c}=c, 
\end{aligned}
\end{equation}
equation \eqref{eq:AdvectionDiffusionEquation} becomes
\begin{equation}
\begin{aligned}
  &\partial_{\tilde{t}} \tilde{c}+\frac{U}{L_{y}}\tilde{\mathbf{u}}\cdot \nabla \tilde{c} =    \frac{\kappa}{L_{y}^{2}} \Delta \tilde{c},
\quad \left. \frac{1}{L_{y}} \mathbf{n} \cdot \nabla \tilde{c} \right|_{\mathrm{wall}}=0,
\end{aligned}
\end{equation}
and equation \eqref{eq:NS dimensional}  becomes
\begin{equation}
\begin{aligned}
& \frac{U \kappa }{L_{y}^{2}} \partial_{\tilde{t}}\tilde{u}+ \frac{U^{2}}{L_{y}} \tilde{u} \partial_{\tilde{x}}\tilde{u} + \frac{U^{2}}{L_{y}} \tilde{\mathbf{v}} \cdot \partial_{\tilde{\mathbf{y}}}\tilde{u}= -\frac{\nu U}{L_{y}^{2}}  \partial_{\tilde{x}}\tilde{p} +  \nu U \left( \frac{1}{L_{y}^{2}} \partial_{\tilde{x}}^{2}\tilde{u} + \frac{1}{L_{y}^{2}} \Delta_{\tilde{\mathbf{y}}}^{2}\tilde{u}  \right),\\
& \frac{U \kappa }{L_{y}^{2}} \partial_{\tilde{t}}\tilde{v}_{i}+ \frac{U^{2}}{L_{y}} \tilde{u} \partial_{\tilde{x}}\tilde{v}_{i} + \frac{U^{2}}{L_{y}} \tilde{\mathbf{v}}\cdot \partial_{\tilde{\mathbf{y}}}\tilde{v}_{i}= -\frac{\nu U}{L_{y}^{2}} \partial_{\tilde{y}_{i}}\tilde{p} +  \nu U \left( \frac{1}{L_{y}^{2}} \partial_{\tilde{x}}^{2}\tilde{v}_{i} + \frac{1}{L_{y}^{2}} \Delta_{\tilde{\mathbf{y}}}^{2}\tilde{v}_{i}  \right),\\
&\frac{U}{L_{y}}\partial_{\tilde{x}}\tilde{u}+\frac{U}{L_{y}}\nabla_{\tilde{\mathbf{y}}} \cdot \tilde{\mathbf{v}}=0.
\end{aligned}
\end{equation}

Now the velocity field $\tilde{\mathbf{u}}$ has the dimensionless fundamental period $\tilde{L}_{p} = L_{p} / L_{y}$.  After dropping the tilde, we obtain  the equation for the scalar field
\begin{equation}\label{eq:AdvectionDiffusionEquationNon1}
  \begin{aligned}
  &\partial_{t} c+ \mathrm{Pe} \mathbf{u} \cdot \nabla c =  \Delta c ,\quad c (x,\mathbf{y},0) =  c_{I} \left( x,\mathbf{y}\right),\quad  \left.  \mathbf{n}\cdot \nabla c \right|_{\mathrm{wall}}=0,
\end{aligned}
\end{equation}
and the equation for the velocity field
\begin{equation}\label{eq:NSEquationNon1}
\begin{aligned}
 &\frac{1}{\mathrm{Sc}}\partial_{t}u+  \mathrm{Re}\left(    u \partial_{x}u +  \mathbf{v} \cdot \nabla_{\mathbf{y}}u \right)= -\partial_{x}p +  \left( \partial_{x}^{2}u + \Delta_{\mathbf{y}}u  \right),\\
&\frac{1}{\mathrm{Sc}}\partial_{t}v_{i}+  \mathrm{Re} \left(  u \partial_{x}v_{i} +  \mathbf{v}\cdot \nabla_{\mathbf{y}}v_{i} \right)= - \partial_{y_{i}}p +   \left(  \partial_{x}^{2}v_{i} +  \Delta_{\mathbf{y}}v_{i}  \right),\\
&\partial_{x}u+\nabla_{\mathbf{y}} \cdot \mathbf{v}=0,\quad \left. \mathbf{u} \right|_{\mathrm{wall}}=0,
\end{aligned}
\end{equation}
where we have defined the  P\'elect number  $\mathrm{Pe}=\frac{L_{y}U}{\kappa}$,  the Reynolds number $\mathrm{Re}=\frac{ L_{y}U}{\nu}$ and the Schmidt number $\mathrm{Sc}=\frac{\nu}{ \kappa}$. Unless otherwise noted, all quantities in the subsequent sections, including $\tilde{L}_{p}$, are taken to be dimensionless, and for simplicity we drop the tilde.

\subsection{Variance and effective longitudinal diffusivity}
Since the solution of the advection–diffusion equation in a channel domain converges to a Gaussian distribution at long times, the variance serves as a key quantity for characterizing the solution. In this section, we present its definition.

As the following discussion involves complex-valued functions, we define the inner product as  
\begin{equation}
\left\langle f, g \right\rangle = \frac{1}{|\Omega(L_p)|} \int\limits_{\Omega} f(x, \mathbf{y}) g^{*}(x, \mathbf{y}) \,\mathrm{d}x \,\mathrm{d}\mathbf{y},
\end{equation}
where the asterisk denotes the complex conjugate. For simplicity, we use the notation $\left\langle u \right\rangle$ as a shorthand for $\left\langle u, 1 \right\rangle$.

Since the solution approaches a Gaussian function at long times, its variance is given by
\begin{equation}
\begin{aligned}
\text{Var}_{c}(t) &=  \frac{\left\langle (x-\left\langle xc \right\rangle)^{2}c \right\rangle}{\left\langle c \right\rangle}=\frac{\left\langle x^{2}c \right\rangle- \left\langle xc \right\rangle^{2}}{\left\langle c \right\rangle}.
\end{aligned}
\end{equation}
Then the effective longitudinal diffusivity is defined as 
\begin{equation}\label{eq:variance limit}
\kappa_{\mathrm{eff}} = \lim\limits_{t\rightarrow \infty} \frac{\text{Var}_{c}(t)}{2t} = \lim\limits_{t\rightarrow \infty} \frac{1}{2} \partial_{t} \text{Var}_{c}(t).
\end{equation}
In some studies on time-dependent flows, the term $\frac{1}{2} \partial_{t} \text{Var}(t)$ is directly defined as the effective diffusivity, as seen in \cite{vedel2012transient,vedel2014time,ding2023dispersion}.

\subsection{Eigenfunction expansion}
Since the advection-diffusion equation \eqref{eq:AdvectionDiffusionEquation} is linear, the eigenfunction expansion of the scalar field is a convenient tool for analyzing scalar dynamics. The eigenvalue $\lambda$ and eigenfunction $\phi$ satisfy the following equation: 
\begin{equation}\label{eq:eigenvalue problem}
\begin{aligned}
&-\lambda \phi = \mathcal{L} \phi = -\mathrm{Pe} \left( u \partial_{x} \phi + \mathbf{v} \cdot \nabla_{\mathbf{y}} \phi \right) + \partial_{x}^{2} \phi + \Delta_{\mathbf{y}} \phi, \quad \left. n_{x} \partial_{x} \phi + \partial_{\mathbf{n}_{y}} \phi \right|_{\mathrm{wall}} = 0.
\end{aligned}
\end{equation}
It forms a basis for square-integrable functions defined on the infinite channel domain that satisfy the no-flux boundary condition $\left. n_{x} \partial_{x} \phi + \partial_{\mathbf{n}_{y}} \phi \right|_{\text{wall}} = 0$. 

There are three properties of eigenvalues and eigenfunctions that will be used later. First, given that the velocity field is real, $\lambda^{*}$ is also an eigenvalue, and $\phi^{*}$ is the associated eigenfunction. Second, the real part of the eigenvalue is non-negative (see the proof in appendix \ref{sec:Nonnegative real part}). Third,  $\lambda = 0$ is the smallest eigenvalue and $\phi = 1$ is the associated eigenfunction.

Since the advection operator is a skew-adjoint operator, $\mathcal{L}$ is not a self-adjoint operator when the velocity field is nonzero. Hence, we introduce $\lambda^{*}$ and $\varphi(x, \mathbf{y})$, which are the eigenvalues and eigenfunctions of the adjoint eigenvalue problem given by
\begin{equation}\label{eq:adjoint eigenvalue problem}
\begin{aligned}
&-\lambda^{*} \varphi = \mathcal{L}^{*} \varphi = \mathrm{Pe} \left( u \partial_{x} \varphi + \mathbf{v} \cdot \nabla_{\mathbf{y}} \varphi \right) + \partial_{x}^{2} \varphi + \Delta_{\mathbf{y}} \varphi, \quad \left. n_{x} \partial_{x} \varphi + \partial_{\mathbf{n}_{y}} \varphi \right|_{\mathrm{wall}} = 0.
\end{aligned}
\end{equation}
In the adjoint eigenvalue problem, $\mathrm{Pe}$ is replaced with $-\mathrm{Pe}$, effectively reversing the flow direction. 

Next, we choose $\phi_{n}$ and $\varphi_{n}$ such that the sets $\left\{ \phi_{n} \right\}_{n=0}^{\infty}$ and $\left\{ \varphi_{n} \right\}_{n=0}^{\infty}$ form a bi-orthogonal system, i.e., $\left\langle \phi_{n}, \varphi_{m} \right\rangle = \delta_{n,m}$, where $\delta_{n,m}$ is the Kronecker delta. The solution of the advection-diffusion equation $\partial_{t} c = \mathcal{L} c$ can be formally represented as 
\begin{equation}
\begin{aligned}
c(x, \mathbf{y}, t) = \sum_{n} \left\langle c_{I}(x, \mathbf{y}), \varphi_{n}(x, \mathbf{y}) \right\rangle \phi_{n}(x, \mathbf{y}) e^{-\lambda_{n} t}.
\end{aligned}
\end{equation}

The exact solution to this eigenvalue problem is typically unavailable, necessitating the use of numerical methods. However, the infinite domain presents challenges for numerical computations. One significant simplification of the problem is to consider the eigenfunctions in the form
\begin{equation}\label{eq:eigenfunction wavenumber}
\begin{aligned}
&\phi(x, \mathbf{y}, k) =\frac{1}{\sqrt{2\pi}} e^{\mathrm{i} k x} \hat{\phi}(x, \mathbf{y}, k), \quad \varphi(x, \mathbf{y}, k) = \frac{1}{\sqrt{2\pi}} e^{\mathrm{i} k x} \hat{\varphi}(x, \mathbf{y}, k),
\end{aligned}
\end{equation}
where $\hat{\phi}(x, \mathbf{y}, k)$ solves the equation 
\begin{equation}\label{eq:eigenvalue problem wavenumber}
\begin{aligned}
&-\lambda \hat{\phi} = \hat{\mathcal{L}}(k) \hat{\phi} = -\mathrm{Pe} \left( \mathrm{i} k u \hat{\phi} + u \partial_{x} \hat{\phi} + \mathbf{v} \cdot \nabla_{\mathbf{y}} \hat{\phi} \right)  + \left( \partial_{x}^{2} \hat{\phi} + 2\mathrm{i} k \partial_{x} \hat{\phi} + (\mathrm{i} k)^{2} \hat{\phi} + \Delta_{\mathbf{y}} \hat{\phi} \right),\\
& \left. n_{x} \mathrm{i} k \hat{\phi} + n_{x} \partial_{x} \hat{\phi} + \partial_{\mathbf{n}_{y}} \hat{\phi} \right|_{\mathrm{wall}} = 0, \quad \hat{\phi}(x, \mathbf{y}, k)=\hat{\phi}(x+L_{p}, \mathbf{y}, k),
\end{aligned}
\end{equation}
and $\hat{\varphi}(x, \mathbf{y}, k)$ solves the adjoint equation
\begin{equation}\label{eq:adjoint eigenvalue problem wavenumber}
\begin{aligned}
&-\lambda^{*} \hat{\varphi} = \hat{\mathcal{L}}^{*}(k) \hat{\varphi} = \mathrm{Pe} \left( \mathrm{i} k u \hat{\varphi} + u \partial_{x} \hat{\varphi} + \mathbf{v} \cdot \nabla_{\mathbf{y}} \hat{\varphi} \right) + \left( \partial_{x}^{2} \hat{\varphi} + 2\mathrm{i} k \partial_{x} \hat{\varphi} + (\mathrm{i} k)^{2} \hat{\varphi} + \Delta_{\mathbf{y}} \hat{\varphi} \right),\\
& \left. n_{x} \mathrm{i} k \hat{\varphi} + n_{x} \partial_{x} \hat{\varphi} + \partial_{\mathbf{n}_{y}} \hat{\varphi} \right|_{\mathrm{wall}} = 0, \quad \hat{\varphi}(x, \mathbf{y}, k)=\hat{\varphi}(x+L_{p}, \mathbf{y}, k),
\end{aligned}
\end{equation}
A key observation is that $\hat{\phi}$ and $\hat{\varphi}$ have a fundamental period of $L_p$, whereas the functions $\phi$ and $\varphi$ may have infinite fundamental periods. The periodic component of the solution is captured by $\hat{\phi}$, while the non-periodic part is represented by the plane wave $e^{\mathrm{i}kx}$. This decomposition reduces the eigenvalue problem from an infinite domain to a finite domain of length $L_p$.

Two additional properties further simplify the analysis.
First, it is straightforward to verify that $\lambda(-k)=\lambda(k)^{*}$, $\phi(x, \mathbf{y}, -k)=\phi(x, \mathbf{y}, k)^{*}$, and $\varphi(x, \mathbf{y}, -k)=\varphi(x, \mathbf{y}, k)^{*}$ are eigenvalues and eigenfunctions. Therefore, it suffices to consider positive real values of $k$.
Second, $\hat{\phi}(x,\mathbf{y},k)$ is periodic in $k$ with period $2\pi/L_p$. To see this, note that
\begin{equation}
\begin{aligned}
\hat{\mathcal{L}} (k) (e^{\mathrm{i} Gx}\hat{\phi}(x, \mathbf{y},k))=e^{\mathrm{i} Gx}\hat{\mathcal{L}} (k+G)\hat{\phi}(x, \mathbf{y},k).
\end{aligned}
\end{equation}
Consequently, multiplying equation \eqref{eq:eigenvalue problem wavenumber} by $e^{-\mathrm{i}x \frac{2\pi}{L_{p}}}$ leads to 
\begin{equation}
  \begin{aligned}
&\lambda (k) e^{-\mathrm{i}x \frac{2\pi}{L_{p}}} \hat{\phi} (x, \mathbf{y},k)=e^{-\mathrm{i}x \frac{2\pi}{L_{p}}} \hat{\mathcal{L}} (k)\hat{\phi} (x, \mathbf{y},k)= \hat{\mathcal{L}} \left( k+\frac{2\pi}{L_{p}} \right) \left( e^{\mathrm{i}x \frac{2\pi}{L_{p}}}\hat{\phi} (x, \mathbf{y},k) \right). \\
\end{aligned}
\end{equation}
Since $e^{\mathrm{i}x \frac{2\pi}{L_{p}}}\hat{\phi} (x, \mathbf{y},k)$ is periodic in $x$ with period $L_{p}$, it follows that
\begin{equation}
\begin{aligned}
\lambda \left( k+\frac{2\pi}{L_{p}} \right)=\lambda (k), \quad \hat{\phi}\left( x, \mathbf{y},k+\frac{2\pi}{L_{p}} \right)= e^{-\mathrm{i}x \frac{2\pi}{L_{p}}}\hat{\phi} (x, \mathbf{y},k).
\end{aligned}
\end{equation}
There are two main reasons for considering the eigenfunctions in the form \eqref{eq:eigenfunction wavenumber}. First, the special form of the eigenfunctions defined in \eqref{eq:eigenvalue problem wavenumber} form a complete bi-orthogonal system for all square-integrable functions defined on the infinite channel domain. A rigorous justification for the completeness of this bi-orthogonal system is presented in appendix \ref{sec:complete orthonormal set}. The second reason is that an alternative approach yields the same form of the eigenfunction. It is intuitive to first consider the eigenvalue problem on a channel domain of finite length $L$, and then investigate the behavior of the eigenfunctions as $L$ approaches infinity. The multiscale asymptotic analysis presented in appendix \ref{sec:Asymptotic expansion} demonstrates that the eigenfunctions take the form \eqref{eq:eigenfunction wavenumber} as $L$ approaches infinity, thereby connecting to the multiscale analysis discussed in \cite{rosencrans1997taylor}.

As a side remark, the eigenfunction representation \eqref{eq:eigenfunction wavenumber} is mathematically equivalent to the Floquet–Bloch form used in solving the Schr\"odinger equation with periodic potentials (\cite{bloch1929quantenmechanik}). This representation was employed by \cite{zwanzig1983effective} to study pure diffusion in periodically modulated channels. \cite{allaire2016comparison} compared the two-scale asymptotic expansion and Bloch-wave approaches for the periodic homogenization of the Poisson equation, showing that the two methods yield identical second-order effective tensors but differ in their fourth-order  corrections. To the best of our knowledge, the application of this form to the Taylor dispersion is new.

With the eigenfunctions expressed in the form \eqref{eq:eigenfunction wavenumber}, the solution to the advection-diffusion equation can be represented as follows:
\begin{equation}\label{eq:eigenfunction expansion}
\begin{aligned}
c(x, \mathbf{y}, t) &= \int\limits_{-\frac{\pi}{L_{p}}}^{\frac{\pi}{L_{p}}} \sum\limits_{n=0}^{\infty} \left\langle c_{I}(x, \mathbf{y}), \varphi_{n}(x, \mathbf{y}, k) \right\rangle \phi_{n}(x, \mathbf{y}, k) e^{-\lambda_{n}(k) t} \mathrm{d}k \\
&= \frac{1}{2\pi}\sum\limits_{n=0}^{\infty} \int\limits_{-\frac{\pi}{L_{p}}}^{\frac{\pi}{L_{p}}} \left\langle c_{I}(x, \mathbf{y}), e^{\mathrm{i} k x} \hat{\varphi}_{n}(x, \mathbf{y}, k) \right\rangle \hat{\phi}_{n}(x, \mathbf{y}, k) e^{\mathrm{i} k x - \lambda_{n}(k) t} \mathrm{d}k.
\end{aligned}
\end{equation}
The ordering of the eigenvalues is determined by their real parts at $k=0$, specifically $0 = \lambda_{0}(0) < \Re \lambda_{1}(0) \leq \ldots \leq \Re \lambda_{n}(0) \ldots$. If the real parts are equal when $k=0$, the ordering is then based on their real parts in the vicinity of zero wavenumber. For nonzero wavenumbers, we order the eigenvalues to ensure that $\lambda_{n}(k)$ remains continuous with respect to $k$. In this context, it is possible to have $\Re \lambda_{n}(k) < \Re \lambda_{m}(k)$ for some $k$ with $n > m$. Figure \ref{fig:FlatChannelShearCos} shows such an example, which will be discussed in the section \ref{sec:Characteristic time for converging to the slow manifold}.  For the flat-walled channel ($L_p = 0$), equation \eqref{eq:eigenfunction expansion} reduces to the Fourier-transform representation used in \cite{stokes1990concentration,phillips1996uniformly,ding2022ergodicity}.

\subsection{Long time asymptotic expansion}
In this section, we use the representation \eqref{eq:eigenfunction expansion} to derive the long-time asymptotic expansion of the scalar field. The key observation is that \(0 = \min_{k} \Re \lambda_{0}(k) < \min_{k} \Re \lambda_{1}(k) \leq \ldots\) implying that all higher-order terms in \eqref{eq:eigenfunction expansion} decay exponentially in time. Hence, the leading term associated with $\lambda_{0} (k)$ dominates the long-time behavior:
\begin{equation}\label{eq:eigenfunctionExpansion0th}
\begin{aligned}
c(x, \mathbf{y}, t) &= \frac{1}{2\pi|\Omega(L_p)|} \int\limits_{-\frac{\pi}{L_{p}}}^{\frac{\pi}{L_{p}}} \phi_{0}(x, \mathbf{y}, k) e^{-\lambda_{0}(k) t} \left( \int\limits_{\Omega} c_{I}(x, \mathbf{y}) \varphi_{0}^{*}(x, \mathbf{y}, k) \mathrm{d}x \mathrm{d}\mathbf{y} \right) \mathrm{d}k + \mathcal{O}\left(e^{-\min\limits_{k} \Re \lambda_{1}(k)t}\right).
\end{aligned}
\end{equation}
% $|\Omega(L_p)|$ is defined after equation \eqref{eq:domain}.
We denote the integral above as $c_{s}$ and refer to it as the slow manifold of the system for several reasons. First, the integral itself is a solution to the advection-diffusion equation, although with a different initial condition. Second, it captures the algebraically decaying dynamics at long times, while other modes decay exponentially. This integral represents a reduced description of the system's long-time behavior, which aligns with the typical characteristics of a slow manifold. The idea behind the approximation  \eqref{eq:eigenfunctionExpansion0th} is utilized in \cite{bronski1997scalar, camassa2021persisting, ding2022ergodicity} to investigate the long-time asymptotic properties in the scalar intermittency problem. Within the center manifold framework adopted in \cite{mercer1990centre, roberts2014model, roberts2015macroscale,ding2022determinism}, the cross-sectional average of the integral in \eqref{eq:eigenfunctionExpansion0th} represents the center manifold of the system. Finally, we note that $\lambda_{0}(k)$ could exceed $\lambda_{1}(k)$ for values of $k$ away from zero. If the initial condition is localized around such modes, the approximation \eqref{eq:eigenfunctionExpansion0th} may no longer be valid. In this work, however, we focus on slowly varying initial conditions, for which the $k=0$ mode is nonzero and the approximation \eqref{eq:eigenfunctionExpansion0th} provides the dominant contribution to the long-time behavior.

Further asymptotic analysis can be conducted by noting that the integral in Equation \eqref{eq:eigenfunctionExpansion0th} is  a Laplace-type integral with respect to the wavenumber \(k\). The asymptotic expansion for large \(t\) can be derived using the steepest descent method, as outlined in \cite{bender1999advanced, ablowitz2003complex}. Intuitively, \(\Re \lambda_{0} = 0\) only when the wavenumber is zero, while it is positive for nonzero wavenumbers. For sufficiently large \(t\), \(e^{-\lambda_{0}(k) t}\) becomes exponentially small for nonzero wavenumbers. Therefore, the integrand near the zero wavenumber contributes dominantly when \(t\) is large. To compute the long-time asymptotic expansion, we expand the eigenvalue and eigenfunction around $k=0$ using the perturbation procedure detailed in Appendix \ref{sec:Eigenfunction expansion for small wavenumbers}:
\begin{equation}\label{eq:eigenvalue eigenvector small k expansion}
\begin{aligned}
\lambda_{0} &= \mathrm{i} \mathrm{Pe} \left\langle u \right\rangle k + \left\langle 1 - \mathrm{Pe} \left( u - \left\langle u \right\rangle \right) \theta_{0}^{(1,1)} + \partial_{x} \theta_{0}^{(1,1)} \right\rangle k^{2} \\
& \quad - \mathrm{i} \left\langle  \mathrm{Pe} \left( u - \left\langle u \right\rangle\right) \theta_{0}^{(2,2)}-\partial_{x} \theta_{0}^{(2,2)}  \right\rangle k^{3} + \mathcal{O}(k^{4}), \\
\phi_{0} &= e^{\mathrm{i} k x} \left( 1 + \mathrm{i} k \theta_{0}^{(1,1)} - k^{2} \left( \theta_{0}^{(2,2)} + \frac{1}{2}\left\langle \theta_{0}^{(1,1)}, q_{0}^{(1,1)} \right\rangle \right) + \mathcal{O}(k^{3}) \right), \\
\varphi_{0} &= e^{\mathrm{i} k x} \left( 1 + \mathrm{i} k q_{0}^{(1,1)} - k^{2} \left( q_{0}^{(2,2)} + \frac{1}{2}\left\langle \theta_{0}^{(1,1)}, q_{0}^{(1,1)} \right\rangle \right) + \mathcal{O}(k^{3}) \right).
\end{aligned}
\end{equation}

The functions $\theta_{0}^{(1,1)}, \theta_{0}^{(2,2)}, q_{0}^{(1,1)}$ are periodic in $x$ with a fundamental period $L_{p}$ and have zero mean over the unit cell domain, satisfying $\left\langle \theta_{0}^{(1,1)}, 1 \right\rangle = \left\langle \theta_{0}^{(2,2)}, 1 \right\rangle = \left\langle 1, q_{0}^{(1,1)} \right\rangle = 0$. The function $\theta_{0}^{(1,1)}$ satisfies:
\begin{equation}
\begin{aligned}
\hat{\mathcal{L}}_{0}(0) \theta_{0}^{(1,1)} &=  \mathrm{Pe} \left( u - \left\langle u \right\rangle \right), \quad \left. n_{x} + n_{x} \partial_{x} \theta_{0}^{(1,1)} + \mathbf{n}_{\mathbf{y}} \cdot \nabla_{\mathbf{y}} \theta_{0}^{(1,1)} \right|_{\mathrm{wall}} = 0.
\end{aligned}
\end{equation}
The function $q_{0}^{(1,1)}$ satisfies:
\begin{equation}
\begin{aligned}
&\hat{\mathcal{L}}_{0}^{*}(0) q_{0}^{(1,1)} = -\mathrm{Pe} \left( u - \left\langle u \right\rangle \right), \; \left. n_{x} + n_{x} \partial_{x} q_{0}^{(1,1)} + \mathbf{n}_{\mathbf{y}} \cdot \nabla_{\mathbf{y}} q_{0}^{(1,1)} \right|_{\mathrm{wall}} = 0.
\end{aligned}
\end{equation}
The function $\theta_{0}^{(2,2)}$ satisfies:
\begin{equation}
\begin{aligned}
&\hat{\mathcal{L}}_{0}(0) \theta_{0}^{(2,2)} = -\mathrm{Pe} \left\langle u \theta_{0}^{(1,1)} \right\rangle + \mathrm{Pe} \left( u - \left\langle u \right\rangle \right) \theta_{0}^{(1,1)} + \left\langle \partial_{x} \theta_{0}^{(1,1)} \right\rangle - 2 \partial_{x} \theta_{0}^{(1,1)}, \\
&\left. n_{x} \theta_{0}^{(1,1)} + n_{x} \partial_{x} \theta_{0}^{(2,2)} + \mathbf{n}_{\mathbf{y}} \cdot \nabla_{\mathbf{y}} \theta_{0}^{(2,2)} \right|_{\mathrm{wall}} = 0.
\end{aligned}
\end{equation}
In addition to the eigenfunction expansion, we need the expansion of the integral related to the initial condition for small wavenumbers:
\begin{equation}\label{eq:Fourier transform initial condition}
\begin{aligned}
&\frac{1}{|\Omega(L_p)|}\int\limits_{\Omega} c_{I}(x, \mathbf{y}) \varphi_{0}^{*}(x, \mathbf{y}, k) \mathrm{d}x \mathrm{d}\mathbf{y} \\
&= \frac{1}{|\Omega(L_p)|}\int\limits_{\Omega} c_{I}(x, \mathbf{y}) e^{-\mathrm{i} k x} \left( 1 - \mathrm{i} k q_{0}^{(1,1)} - k^{2} \left( q_{0}^{(2,2)} + \frac{1}{2}\left\langle \theta_{0}^{(1,1)}, q_{0}^{(1,1)} \right\rangle \right) + \mathcal{O}(k^{3}) \right) \mathrm{d}x \mathrm{d}\mathbf{y} \\
&= \frac{1}{|\Omega(L_p)|}\int\limits_{\Omega} c_{I}(x, \mathbf{y}) \left( 1 - \mathrm{i} k x + \frac{(-\mathrm{i} k x)^{2}}{2} + \mathcal{O}(k^{3}) \right) \\
& \quad \times \left( 1 - \mathrm{i} k q_{0}^{(1,1)} - k^{2} \left( q_{0}^{(2,2)} + \frac{1}{2}\left\langle \theta_{0}^{(1,1)}, q_{0}^{(1,1)} \right\rangle \right) + \mathcal{O}(k^{3}) \right) \mathrm{d}x \mathrm{d}\mathbf{y} \\
&= c_{I}^{(0)} - \mathrm{i} k c_{I}^{(1)} - k^{2} c_{I}^{(2)} + \mathcal{O}(k^{3}),
\end{aligned}
\end{equation}
where 
\begin{equation}
\begin{aligned}
&c_{I}^{(0)}=\frac{1}{|\Omega(L_p)|}\int\limits_{\Omega}c_{I}(x,\mathbf{y}) \mathrm{d}x \mathrm{d}\mathbf{y}, \quad c_{I}^{(1)}=\frac{1}{|\Omega(L_p)|} \int\limits_{\Omega}c_{I}(x,\mathbf{y}) \left( x+q_{0}^{(1,1)} \right)\mathrm{d}x \mathrm{d}\mathbf{y},\\
&c_{I}^{(2)}=\frac{1}{|\Omega(L_p)|} \int\limits_{\Omega}c_{I}(x,\mathbf{y})  \left( \frac{x^{2}}{2}+  q_{0}^{(2,2)} + \frac{1}{2}\left\langle \theta_{0}^{(1,1)}, q_{0}^{(1,1)} \right\rangle  +q_{0}^{(1,1)} x \right)\mathrm{d}x \mathrm{d}\mathbf{y}.  
\end{aligned}
\end{equation}

The steepest descent method requires expanding $\lambda_{0}(k)$ around the saddle point in the complex plane. Notice that the expansion \eqref{eq:eigenvalue eigenvector small k expansion} implies that the saddle point of $\lambda_{0}(k)$ is not located at $k=0$ if $\left\langle u \right\rangle \neq 0$. In fact, this expansion around this saddle point provides us with an asymptotic expansion of equation \eqref{eq:eigenfunctionExpansion0th} as $t \rightarrow \infty$ while keeping other variables fixed. Notably, this type of expansion is not uniform with respect to the spatial variable. To achieve a uniform asymptotic expansion, we should notice that the whole scalar is transported by the flow with  the nonzero mean flow speed. Therefore, we analyze the problem in the moving frame of reference $x_{m} = x - \mathrm{Pe} \left\langle u \right\rangle t$. In this frame, $k=0$ is the saddle point of $\lambda_{0}(k)$ and the steepest decent method yields a uniform asymptotic expansion. 

Substituting the expansions \eqref{eq:eigenvalue eigenvector small k expansion} and \eqref{eq:Fourier transform initial condition} into equation \eqref{eq:eigenfunctionExpansion0th} yields the asymptotic expansion of the scalar field at long times:
\begin{equation}\label{eq:ground state long time expansion}
\begin{aligned}
 &c (x,\mathbf{y},t) =\frac{  e^{-\frac{x_{m}^2}{2 t \lambda ''(0)}}}{|\Omega(L_p)|\sqrt{2 \pi t \lambda ''(0)}}  \int\limits_{\Omega}c_{I}(x,\mathbf{y}) \mathrm{d}x \mathrm{d}\mathbf{y} +\frac{ x_{m}  e^{-\frac{x_{m}^2}{2 t \lambda ''(0)}}}{\sqrt{2 \pi }\left(t \lambda ''(0)\right)^{3/2}} \left( c_{I}^{(1)}-\theta_{0}^{(1,1)} c_{I}^{(0)} \right)\\
  &+\frac{  e^{-\frac{x_{m}^2}{2 t \lambda ''(0)}} \left(x_{m}^2-t \lambda ''(0)\right) \left( \left( \theta_{0}^{(2,2)} + \frac{1}{2}\left\langle \theta_{0}^{(1,1)}, q_{0}^{(1,1)} \right\rangle \right) c_{I}^{(0)}- \theta_{0}^{(1,1)} c_{I}^{(1)}+c_{I}^{(2)} \right)}{\sqrt{2 \pi }\left(t \lambda ''(0)\right)^{5/2}}+\mathcal{O} \left( t^{-\frac{3}{2}} \right). \\
 \end{aligned}
\end{equation}
Note that not all terms of order $\mathcal{O}(t^{-3/2})$ are included above. Capturing the full expansion to this order requires extending the eigenvalue expansion to $\mathcal{O}(k^{5})$ and the eigenfunction expansion to $\mathcal{O}(k^{3})$, resulting in a lengthy expression that is omitted here for brevity.

In Fourier space, multiplication by $k$ corresponds to differentiation in physical space. Since derivatives of the Gaussian function yield Hermite polynomials multiplied by the Gaussian, each higher-order term in \eqref{eq:ground state long time expansion} takes the form of a Hermite polynomial times a Gaussian, with coefficients determined by the eigenfunction expansion.

A more accurate approximation can be obtained by incorporating information about the initial condition. For instance, if the initial condition is a Gaussian distribution $\mathcal{N}(x; x_0,\sigma)$ centered at $x=x_{0}$ with standard deviation $\sigma$,
\begin{equation}\label{eq:Guassian distribution}
  \begin{aligned}
& c_{I} \left( x,\mathbf{y}\right)= \mathcal{N}(x; x_0,\sigma)\equiv \frac{1}{\sqrt{2 \pi \sigma^2}} \exp\left(-\frac{(x - x_0)^2}{2\sigma^2}\right).\end{aligned}
\end{equation}
In this case, instead of approximating $e^{-\mathrm{i} k x}$ using its small $k$ expansion as in \eqref{eq:Fourier transform initial condition}, we can directly evaluate the integral involving the initial condition. This approach yields an expansion that is asymptotically equivalent to Equation \eqref{eq:ground state long time expansion} at long times while significantly improving accuracy at short times. 

The leading-order term in the approximation \eqref{eq:ground state long time expansion} is a Gaussian function, which corresponds to the solution of a one dimensional advection–diffusion equation with an enhanced diffusion coefficient:
\begin{equation}\label{eq:effective equation}
\partial_{t}c + \mathrm{Pe} \left\langle u \right\rangle \partial_{x}c = \kappa_{\mathrm{eff}} \partial_{x}^{2}c, 
\quad \kappa_{\mathrm{eff}} = \frac{\lambda''(0)}{2} = \left\langle 1 - \mathrm{Pe} \left( u - \left\langle u \right\rangle \right) \theta_{0}^{(1,1)} + \partial_{x} \theta_{0}^{(1,1)} \right\rangle.
\end{equation}
This equation serves as the effective long-time approximation of \eqref{eq:AdvectionDiffusionEquationNon1}. The coefficient $\kappa_{\mathrm{eff}}$ represents the effective diffusivity, which is consistent with the definition given in \eqref{eq:variance limit}.  To show that the effective diffusivity is real and nonnegative, we rewrite it as  
\begin{equation}\label{eq:effective diffusivity}
\kappa_{\mathrm{eff}} = \left\langle \nabla_{\mathbf{y}} \theta_{0}^{(1,1)} \cdot \nabla_{\mathbf{y}} \theta_{0}^{(1,1)} + (1 + \partial_{x} \theta_{0}^{(1,1)})^{2} \right\rangle \geq 0,
\end{equation}
where the inequality follows from the fact that $\theta_{0}^{(1,1)}$ is real.  This expression for the effective diffusivity aligns with previously reported results in \cite{rosencrans1997taylor,brenner1980dispersion}. \footnote{The function $\theta_0^{(1,1)}$ satisfies the same cell problem as $\chi_i$ in \cite{rosencrans1997taylor}, p. 1225, and thus equation~\eqref{eq:effective diffusivity} corresponds directly to the expression for the enhanced diffusion coefficient given on p. 1226 of the same reference.}    Furthermore, the validity of this formula has been confirmed through benchmarking against particle tracking methods, as demonstrated in \cite{richmond2013flow,haugerud2022solute}.

Finally, there is an intriguing question that is difficult to address in other frameworks but becomes more tractable in the proposed approach: How does the effective diffusivity change when the flow direction is reversed, i.e., when $\mathrm{Pe}$ is replaced by $-\mathrm{Pe}$?
 This question arose from examining the formula for the effective diffusivity \eqref{eq:effective equation}. At first glance, the forcing term suggests that $\partial_{x} \theta_{0}^{(1,1)}$ scales as $\mathrm{Pe}$, while $\mathrm{Pe}\left(u - \langle u \rangle\right)\theta_{0}^{(1,1)}$ scales as $\mathrm{Pe}^{2}$. This scaling would imply the possible presence of a $\mathrm{Pe}$-order term in the effective diffusivity, potentially making it a non-even function of $\mathrm{Pe}$.

Replacing $\mathrm{Pe}$ by $-\mathrm{Pe}$ essentially switches the roles of the eigenvalue problem \eqref{eq:eigenvalue problem} and its adjoint counterpart \eqref{eq:adjoint eigenvalue problem}. Consequently, the effective diffusivity changes from $\kappa_{\mathrm{eff}} = \frac{\lambda''(0)}{2}$ to $\kappa_{\mathrm{eff}} = \frac{(\lambda''(0))^{*}}{2}$. Since $\lambda''(0)$ is real, the effective diffusivity remains unchanged. One alternative approach to demonstrate this symmetry is to compute a power series expansion of the effective diffusivity in terms of $\mathrm{Pe}$ and show that all odd-order terms vanish, implying that the effective diffusivity is an even function of $\mathrm{Pe}$. However, this calculation is relatively lengthy, whereas the symmetry arises more directly within the framework proposed in this work.

\subsection{Connection to the time scale for dispersion, longitudinal normality, and transverse uniformity}
From the derivation in the previous section, we observe that the effective equation \eqref{eq:effective equation} is obtained through two successive approximations. The first approximation assumes that the scalar field can be described by its slow manifold with exponentially decaying correction terms. This approximation is valid on the time scale estimated by 
\begin{equation}
    t_s = \frac{1}{\min_k \Re\lambda_1(k)},
\end{equation}
which depends solely on the domain geometry and velocity field. A larger \(\min_k \Re\lambda_1(k)\) leads to faster convergence, whereas a smaller \(\min_k \Re\lambda_1(k)\) results in slower convergence.  The second approximation expresses the slow manifold as a Gaussian distribution plus algebraically decaying correction terms. Therefore, validity of the Gaussian distribution approximation depends on the decay of these correction terms, as given in \eqref{eq:ground state long time expansion}, and the initial condition.

To better understand the relationship between the time scales for dispersion, longitudinal normality, and transverse uniformity, we consider an explicit example: a flat channel domain 
\begin{equation}
    \Omega = \{ (x, y) \mid x \in \mathbb{R}, -\frac{1}{2} \leq y \leq \frac{1}{2} \},
\end{equation}
subject to shear flow \( \mathbf{u} = (\frac{3}{2}(1 - 4y^2),0) \) and an initial condition given by \eqref{eq:Guassian distribution}. To characterize the timescale for transverse uniformity, we analyze the long-time asymptotic expansion of the scalar field. For longitudinal normality, we evaluate the asymptotic behavior of the cross-sectionally averaged scalar. To determine the dispersion timescale, we examine the long-time expansion of the scalar variance.

Using the method outlined in the previous section, we obtain the long-time asymptotic expansion of the scalar field:
 \begin{equation}\label{eq:flat channel pressure driven flow scalar expansion}
 \scriptsize
\begin{aligned}
c=&\frac{e^{-\frac{x^2}{2  v}}}{\sqrt{2 \pi v }}\left( 1+\frac{a_{1}\left(240 y^4-120 y^2+7\right)}{480 } +\frac{a_{2} \left(240 y^2 \left(720 y^6-784 y^4+238 y^2-17\right)-413\right)}{3225600}\right.\\
& -\frac{a_{3}\left(1064960 t-520 \left(-15686 y^2+32 \left(10800 y^6-18568 y^4+10725 y^2-2079\right) y^4+8447\right) y^2+240261\right)}{73801728000 } \\
&  \frac{a_{4}}{16862218813440000} \left( 763467599-38993920 t \left(1560 \left(2 y^2-1\right) y^2+283\right)+  2720 y^2  \right.  \\
&\hspace{2.3cm}\times \left(1127049 -26712502 y^2 +101146864 y^4-78215280 y^6  \right.  \\
&\hspace{2.5cm}      \left. \left.  \left.  -259440896 y^8  +779717120 y^{10} -897085440 y^{12} + 377395200 y^{14}\right)  \right) \right)+\mathcal{O} (t^{-\frac{3}{2}}),
\end{aligned}
\end{equation}
where \( v = \sigma^2 + 2 \kappa_{\mathrm{eff}} t \) and \( x_m = x - x_0 - \mathrm{Pe} t \). The coefficients \( a_n \) are given by:
\begin{equation}
    a_n = \frac{\mathrm{Pe}^n}{(2 v)^{n/2}} H_n\left(\frac{x_m}{\sqrt{2 v}}\right),
\end{equation}
where \( H_n \) is the Hermite polynomial of degree \( n \), $H_{1} (x)=2x, H_{2} (x)=4 x^2-2, H_{3} (x)=8 x^3-12 x, H_{4} (x)=16 x^4-48 x^2+12$. Taking the cross-sectional average of \eqref{eq:flat channel pressure driven flow scalar expansion} yields:
\begin{equation}\label{eq:flat channel pressure driven flow averaged scalar expansion}
\begin{aligned}
 \bar{c}=\int\limits_{-\frac{1}{2}}^{\frac{1}{2}} c\mathrm{d} y=\frac{e^{-\frac{x^2}{2  v}}}{\sqrt{2 \pi v }} \left( 1-\frac{a_{2}}{8400}-\frac{a_{3}t}{69300 v^3} -\frac{(685440 t-54991)a_{4}}{1543782240000}\right)+\mathcal{O} (t^{-\frac{3}{2}}) .
\end{aligned}
\end{equation}
Next, we derive the long-time asymptotic expansion of the variance, which requires computing \( \langle x_m^2 c \rangle \). The higher-order terms omitted from \eqref{eq:flat channel pressure driven flow scalar expansion} are Hermite polynomials of degree greater than four. Due to the orthogonality of Hermite polynomials, these terms do not contribute to the variance. Thus, we obtain the exact variance of the slow manifold \( c_s \) and the asymptotic expansion of the scalar field variance with exponential correction terms, which is consistent with equation 3.22 in \cite{camassa2010exact} \footnote{Due to a different choice of characteristic velocity, comparison with the formula in \cite{camassa2010exact} requires replacing $\mathrm{Pe}$ in the present formula with $\tfrac{2}{3}\mathrm{Pe}$.  }:   
\begin{equation}\label{eq:flat channel pressure driven flow scalar variance expansion}
    \begin{aligned}
        & \text{Var}_{c_s}(t) = \left\langle x_m^2 c_s \right\rangle = -\frac{\mathrm{Pe}^2}{4200} + \sigma^2 + 2 \left( \frac{\mathrm{Pe}^2}{210} + 1 \right)t, \\
        & \text{Var}_{c}(t) = \left\langle x_m^2 c \right\rangle = -\frac{\mathrm{Pe}^2}{4200} + \sigma^2 + 2 \left( \frac{\mathrm{Pe}^2}{210} + 1 \right)t + \mathcal{O}\left(e^{-\min_k \Re \lambda_1(k)t}\right),
    \end{aligned}
\end{equation}
where \( \min_k \Re(\lambda_1(k) )= \pi^2 \).   The variance of the slow manifold at \( t=0 \) is given by \( \text{Var}_{c_{s}}(0)=-\frac{\text{Pe}^2}{4200}+ \sigma^{2} \), which differs from the variance of the initial condition of the scalar field, \( \text{Var}_{c}(0)=\sigma^{2} \). This suggests that the slow manifold corresponds to a solution with a different initial condition.  

Now, we can analyze the three different time scales mentioned at the beginning. From equation \eqref{eq:flat channel pressure driven flow scalar variance expansion}, we observe that once the exponential terms have decayed, the variance increases linearly in time. Therefore, the characteristic time scale for dispersion is given by  $t_{s} = 1/\min_{k} \Re \lambda_{1}(k)$.

Equation \eqref{eq:flat channel pressure driven flow averaged scalar expansion} becomes valid after this time scale is reached. The first term represents a Gaussian function, while the remaining terms describe deviations from Gaussianity. Since these correction terms have small coefficients, they become negligible unless \( \mathrm{Pe} \) is large or the initial variance is small. In such cases, after time \( t_s \), the cross-sectionally averaged scalar field is well approximated by a Gaussian function.

Similarly, equation \eqref{eq:flat channel pressure driven flow scalar expansion} is valid after time \( t_s \). The first term represents the Gaussian function, while the second term describes transverse variations. Since the cross-sectional average of this term is zero, it does not appear in equation \eqref{eq:flat channel pressure driven flow averaged scalar expansion}. As time increases, when the second term becomes negligible compared with the first, the transverse uniformity time scale is reached.

\subsection{Validation by the numerical simulation}
After analyzing the example in a flat channel, we now use numerical methods to demonstrate that the proposed theory also holds in a curved channel. This is done by simulating the scalar transport equation \eqref{eq:AdvectionDiffusionEquation} in the following sinusoidal channel:
\begin{equation}\label{eq:sinusoidal channel}
\begin{aligned}
\Omega=\left\{  (x, y) |  x \in \mathbb{R},    0\leq y\leq  h(x) \right\},\quad  h(x) = 1 + A \cos\left(\frac{2\pi x}{L_{p}}\right),
\end{aligned}
\end{equation}
where $A<1$  and $L_p$ is the fundamental period.

%Since directly constructing the slow manifold is numerically challenging, we instead validate the theory by comparing the time derivative of the variance of the numerical solution to \eqref{eq:AdvectionDiffusionEquation} with the theoretical effective diffusivity.

\begin{figure}
  \centering
    \includegraphics[width=0.46\linewidth]{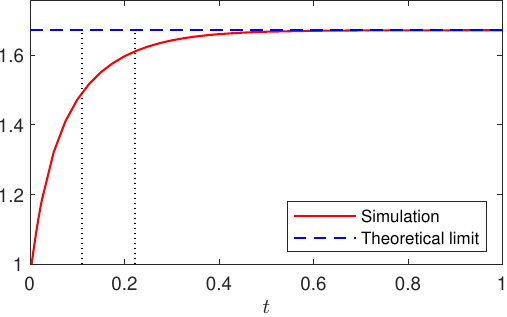}
  \hfill
  \caption[]
  { The quantity $\frac{1}{2} \partial_{t} \text{Var}(t)$ as a function of time is plotted as the red solid curve. The theoretical long-time limit predicted by equation~\eqref{eq:effective equation} is indicated by the blue dashed curve. The vertical dotted lines indicate $t_{s}$ and $2t_{s}$, where $t_{s}$ represents the estimated time scale for the scalar field to converge to the slow manifold. }
  \label{fig:FullVariance}
\end{figure}

In the simulation, we consider the sinusoidal channel \eqref{eq:sinusoidal channel} with $A = 0.2, L_{p}=1$. The computational domain has a length of 42 with periodic boundary conditions at the left and right ends. To confirm that this domain provides an adequate approximation of an unbounded channel, we verify that the scalar concentration at both endpoints remains close to zero throughout the simulation. A steady pressure-driven flow with $\mathrm{Re} = 200$ and a pressure gradient of $f_x = -20.18$ is computed using the method described in Appendix \ref{sec:Numerical method}. The  spatial mean of the resulting flow $\left\langle u \right\rangle=1$. After obtaining the flow, we use it to advect the scalar by solving the advection-diffusion equation \eqref{eq:AdvectionDiffusionEquation} with $\mathrm{Pe} = 6$. The initial condition follows the Gaussian distribution given in equation \eqref{eq:Guassian distribution}, with $\sigma = 0.5$ and $x_{0} = 6$. We find that $\max_k \Re \lambda_1(k) = \lambda_1(0) = 9.0294$, and the estimated time scale for the scalar field to converge to the slow manifold is $t_s = 1/\min_k \Re\lambda_1(k)  =  0.1107$. The effective diffusivity is $\kappa_{\mathrm{eff}} = 1.6721$.

We plot $\frac{1}{2} \partial_{t} \text{Var}(t)$ as a function of time in Figure \ref{fig:FullVariance}, showing its convergence to the theoretical effective diffusivity. Once $\frac{1}{2} \partial_{t} \text{Var}(t)$ stabilizes to a constant value, it indicates that the variance grows linearly in time, marking the onset of Taylor dispersion. The relative differences between $\frac{1}{2} \partial_{t} \text{Var}(t)$ and $\kappa_{\mathrm{eff}}$ are 0.1069, 0.0368, and 0.0132 at $t = t_s$, $2t_s$, and $3t_s$, respectively. The ratio of the adjacent relative differences is $2.7 \sim 2.8$, indicating that $t_s$ is the e-folding time scale \footnote{The e-folding time is the time interval in which an exponentially growing or decaying quantity increases or decreases by a factor of $e$.} for the system, thereby supporting the proposed theory.
\begin{figure}
  \centering
  \subfigure[]{
    \includegraphics[width=0.46\linewidth]{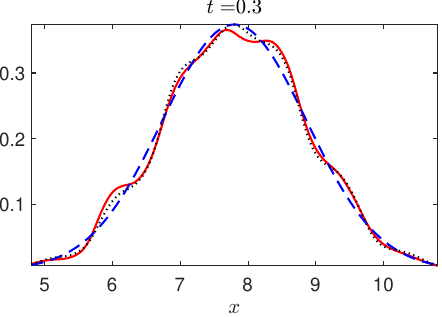}
  }
  \subfigure[]{
    \includegraphics[width=0.46\linewidth]{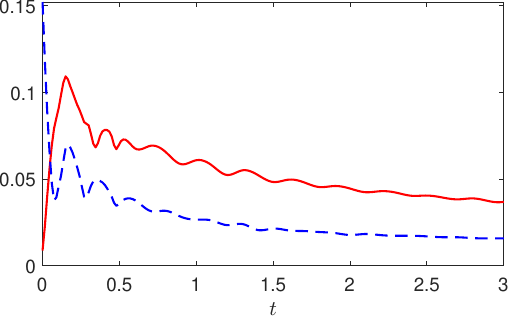}
  }
  \hfill
  \caption[]
 {(a) Red solid curve: $\bar{c}$; blue dashed curve: Gaussian profile with the same mean and variance as $\bar{c}$; black dotted curve: first-order correction term.
(b) Red solid curve: $E_{n,1}(t)$; blue dashed curve: $E_{n,2}(t)$. }
  \label{fig:CrossSectionalAverage}
\end{figure}

Next, we investigate longitudinal normality.
We measure the deviation of the cross-sectional average of the scalar field, $\bar{c} (x) = \frac{1}{h (x)}\int\limits_{0}^{h (x)}c (x,y)\mathrm{d}y$ to the Gaussian function is measured by
from a Gaussian profile by the metric
\begin{equation}
E_{n,1} (t) =\frac{\lVert \bar{c} (x,t) - f (x,t)\rVert_{\infty} }{ \lVert \bar{c}\rVert_{\infty} }, \quad f(x,t)=\mathcal{N}\left( x; \left\langle x c \right\rangle, \sqrt{ \mathrm{Var} (t) } \right),
\end{equation}
where $\lVert \cdot \rVert_{\infty}$ denotes the $L^{\infty}$ norm and $\mathcal{N}$ is the Gaussian function defined in equation~\eqref{eq:Guassian distribution}. Here $\langle x c \rangle$ and $\mathrm{Var}(t)$ are the mean and variance of the scalar field, respectively. A smaller $E_{n,1}(t)$ indicates closer agreement between $\bar{c}$ and the corresponding Gaussian.

To quantify the approach to transverse uniformity, we define
\begin{equation}
E_{t,1} (t) =\frac{\lVert c (x,y,t) - f(x,t) \rVert_{\infty} }{ \lVert c\rVert_{\infty} },
\end{equation}
where the same Gaussian $f(x,t)$ is used as a reference. This measures the maximum pointwise deviation from a longitudinally varying but transversely uniform Gaussian state.

The Gaussian profile $f(x,t)$ corresponds to the first term in the asymptotic expansion of the scalar field \eqref{eq:ground state long time expansion}.
It is also useful to examine the approximation error when the first two terms in \eqref{eq:ground state long time expansion} are included:
\begin{equation}
\begin{aligned}
&E_{n,2} (t) =\frac{\lVert \bar{c}-  \left( f+ \partial_{x}f \left( c_{I}^{(1)}-\overline{\theta_{0}^{(1,1)}} c_{I}^{(0)} \right) \right) \rVert_{\infty} }{ \lVert \bar{c}\rVert_{\infty} }, \\
&E_{t,2} (t) =\frac{\lVert c- \left( f+ \partial_{x}f \left( c_{I}^{(1)}-\theta_{0}^{(1,1)} c_{I}^{(0)} \right) \right) \rVert_{\infty} }{ \lVert c\rVert_{\infty} }. \\ 
\end{aligned}
\end{equation}
If $E_{n,1} \gg E_{n,2}$ and $E_{t,1} \gg E_{t,2}$, the magnitude of the second term provides a practical estimate for the approximation error of the first term, and hence for the timescales of longitudinal normality and transverse uniformity. In this case, $c_{I}^{(0)} = 1$. Both $\theta_{0}^{(1,1)}$ and $c_{I}^{(1)}$ are computed numerically. The numerically obtained value of $c_{I}^{(1)}$ is small, on the order of $10^{-5}$.

For a channel with flat walls, deviation of $\bar{c}$ from the Gaussian profile typically appears as nonzero skewness. In periodically modulated channels, in addition to skewness, $\bar{c}$ may exhibit periodic fluctuations due to the domain geometry. As shown in Figure~\ref{fig:CrossSectionalAverage}(a), $\bar{c}$ is nearly symmetric but still contains periodic fluctuations, which are captured by the second term in \eqref{eq:ground state long time expansion}. This is confirmed by Figure~\ref{fig:CrossSectionalAverage}(b), which shows that $E_{n,1} > E_{n,2}$.

\begin{figure}
  \centering
  \subfigure[]{
    \includegraphics[width=0.46\linewidth]{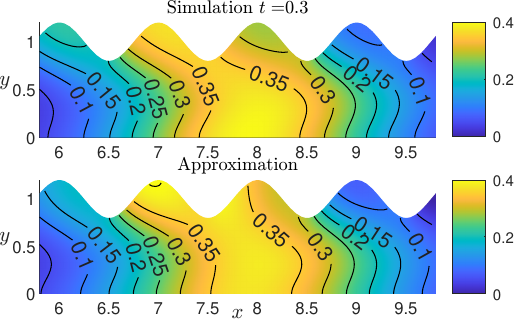}
  }
  \subfigure[]{
    \includegraphics[width=0.46\linewidth]{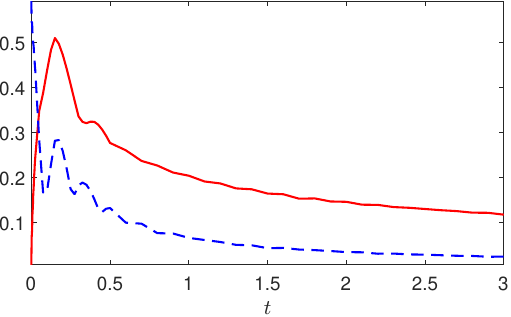}
  }
  \hfill
  \caption[]
 {(a) Top: solution of the advection–diffusion equation~\eqref{eq:AdvectionDiffusionEquationNon1} at $t = 0.2$.
Bottom: two-term approximation of the scalar field from~\eqref{eq:ground state long time expansion}.
(b) Red solid curve: $E_{t,1}(t)$; blue dashed curve: $E_{t,2}(t)$. }
  \label{fig:scalarTransverse}
\end{figure}

Figure~\ref{fig:scalarTransverse}(a) demonstrates that including the second term in \eqref{eq:ground state long time expansion} significantly improves the capture of transverse variations of the entire scalar field. Figure~\ref{fig:scalarTransverse}(b) demonstrates that $E_{t,1} > E_{t,2}$, supporting the use of the second term’s magnitude to estimate the timescales of both longitudinal normality and transverse uniformity.

\section{Characteristic time for converging to the slow manifold}
\label{sec:Characteristic time for converging to the slow manifold}
As seen in the previous section, reaching the time scale \(t_{s}=1/(\min_{k} \Re\lambda_{1}(k))\) is a necessary condition for dispersion, longitudinal normality, and transverse uniformity. In this section, we investigate how $t_{s}$   depends on physical parameters, flow, and domain geometry.

\subsection{Shear flow in the channel with flat boundaries}

We first examine our theory using a fundamental case: the diffusion of a passive scalar transported by a shear flow given by $u = u(y), v = 0$ in a channel with flat boundaries, represented as $\Omega = \{ (x, y) | x \in \mathbb{R}, 0 \leq y \leq 1 \}$.

In this scenario, the fundamental period of the system is zero, and the unit cell reduces to $\Omega(0) = \{ (0, y) | 0 \leq y \leq 1 \}$. Therefore, $\hat{\phi}$ is independent of $x$, and equation \eqref{eq:eigenvalue problem wavenumber} simplifies to
\begin{equation}\label{eq:shear flow eigenvalue problem}
\begin{aligned}
  &\lambda \hat{\phi} = \hat{\mathcal{L}}(k) \hat{\phi} = \mathrm{Pe} \mathrm{i} k u \hat{\phi} - \left( -k^{2} \hat{\phi} + \Delta_{\mathbf{y}} \hat{\phi} \right).
\end{aligned}
\end{equation}

\begin{figure}
  \centering
  \subfigure[$u=\sqrt{2}\cos\pi y$]{
    \includegraphics[width=0.46\linewidth]{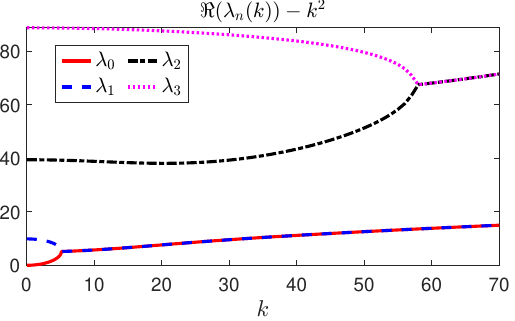}
  }
  \subfigure[$u=\sqrt{2}\cos\pi y$]{
    \includegraphics[width=0.46\linewidth]{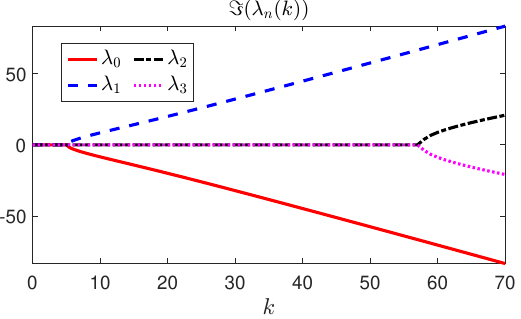}
  }
   \subfigure[$u=\sqrt{2}\cos2 \pi y$]{
    \includegraphics[width=0.46\linewidth]{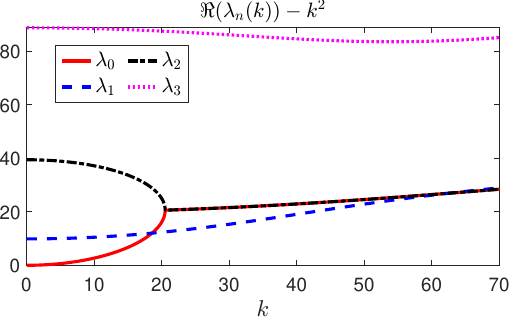}
  }
  \subfigure[$u=\sqrt{2}\cos2\pi y$]{
    \includegraphics[width=0.46\linewidth]{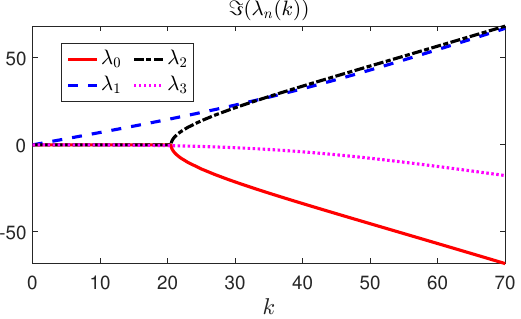}
 } 
  \hfill
  \caption[]
 {The first four eigenvalues for different shear flows with $\mathrm{Pe}=1$. }
  \label{fig:FlatChannelShearCos}
\end{figure}

\begin{figure}
  \centering
   \subfigure[$u=\sqrt{2}\cos\pi y$]{
     \includegraphics[width=0.46\linewidth]{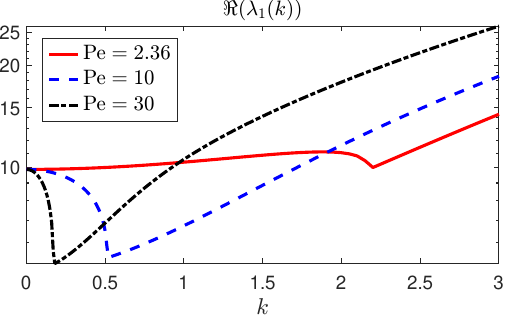}
   }
    \subfigure[$u=\sqrt{2}\cos 2\pi y$]{
      \includegraphics[width=0.46\linewidth]{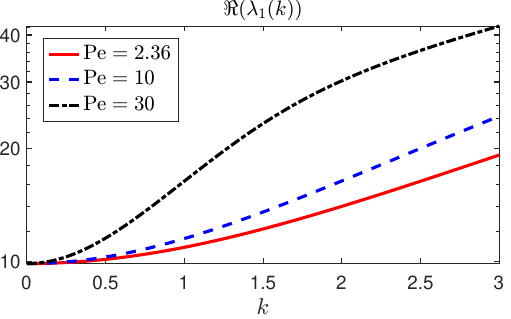}
      }
  \hfill
  \caption[]
{ Plot of $\Re (\lambda_{1} (k))$ as a function of the wavenumber $k$ for different $\mathrm{Pe}$ and different flows. The $y$-axis is in log scale. Results are shown for (a) $u=\sqrt{2}\cos \pi y$ and (b) $u=\sqrt{2}\cos 2\pi y$
  }
  \label{fig:FlatChannelShearCos1RePe}
\end{figure}

Next, we numerically compute the eigenvalues for several different shear flows. An interesting family of shear flow is $u = \sqrt{2} \cos(n \pi y)$, where $n \in \mathbb{N}^{+}$. These functions serve as the eigenfunctions of $\hat{\mathcal{L}}(0)$. Many analyses regarding Taylor dispersion focus on slowly varying solutions, where the energy is concentrated in the vicinity of zero wavenumber. Therefore, the eigenvalues of $\hat{\mathcal{L}}(0)$ can provide an estimate of the decay rate \cite{mercer1990centre, rosencrans1997taylor}. However, it is important to note that $\Re \lambda_{1}$ does not always reach its minimum value at $k = 0$, which is a crucial consideration when the spectrum of the solution is not concentrated near zero wavenumber.  For instance, $u = \sqrt{2} \cos(\pi y)$ exemplifies this behavior.  Figure \ref{fig:FlatChannelShearCos} (a, b) displays the real and imaginary parts of the first four eigenvalues for this shear flow with $\mathrm{Pe} = 1$. The $k^{2}$ term arises from the contribution of diffusion in the longitudinal direction of the channel, and we subtract this term from the eigenvalue to highlight the contribution from the flow. Here  $\Re (\lambda_{1}) - k^{2}$ decreases for small wavenumbers, with its global minimum occurring at $k = 5.1252$, yielding a value of 5.1537. The calculations in Appendix \ref{sec:Eigenvalue expansion shear flow} show the asymptotic expansion of $\lambda_{1}$ for small wavenumbers:
\begin{equation}
\lambda_{1} = \pi^{2} + \left( 1 - \frac{5\mathrm{Pe}^{2}}{6\pi^{2}} \right) k^{2} + \mathcal{O}(k^{3}).
\end{equation}
This implies that if $\mathrm{Pe} > \sqrt{\frac{6}{5}} \pi \approx 3.4414$, then $k = 0$ may not be a local minimum of $\lambda_{1}(k)$. In fact, when $\mathrm{Pe} > 2.36$, the global minimum is not at $k = 0$. Furthermore, as $\mathrm{Pe}$ increases, the location of the minimum approaches $k = 0$, as shown in Figure \ref{fig:FlatChannelShearCos1RePe} (a).

Interestingly, when the shear flow is given by $u = \sqrt{2} \cos(n \pi y)$ with $n \geq 2$, $k = 0$ is a local minimum of $\lambda_{1}(k)$, as supported by the asymptotic analysis for small wavenumbers presented in Appendix \ref{sec:Eigenvalue expansion shear flow}. Figure \ref{fig:FlatChannelShearCos} (c, d) displays the real and imaginary parts of the first four eigenvalues for the shear flow $u = \sqrt{2} \cos(2 \pi y)$,  further supporting this observation. Additionally, in conjunction with Figure \ref{fig:FlatChannelShearCos1RePe} (b), we observe that $\lambda_{1}(0) = \pi^{2}$ serves as the global minimum across different $\mathrm{Pe}$ values.

Therefore, for $u = \sqrt{2} \cos(n \pi y)$ with $n \geq 2$, we have \( \min_k \Re(\lambda_1(k) )= \pi^2 \) and \( t_s = 1/\min_k \Re\lambda_1(k) = 1/\pi^{2} \), whereas for $u = \sqrt{2} \cos(\pi y)$, it follows that $t_s \geq 1/\pi^{2}$. These results indicate that  if the flow strength is sufficiently large, the scalar diffusing under the shear flow $u = \sqrt{2} \cos(\pi y)$ could converge more slowly to the slow manifold than when advected by $u = \sqrt{2} \cos(n \pi y)$ with $n \geq 2$.

\begin{figure}
  \centering
  \subfigure[$u=2\sqrt{2} (y- \frac{1}{2})$]{
    \includegraphics[width=0.46\linewidth]{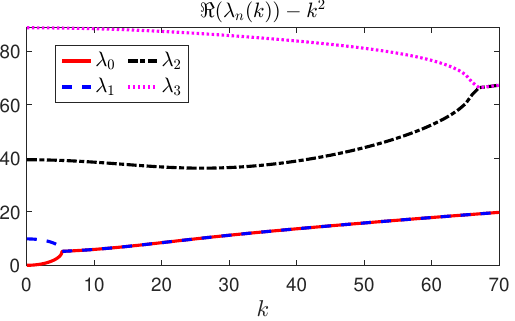}
  }
  \subfigure[$u=2\sqrt{2} (y- \frac{1}{2})$]{
    \includegraphics[width=0.46\linewidth]{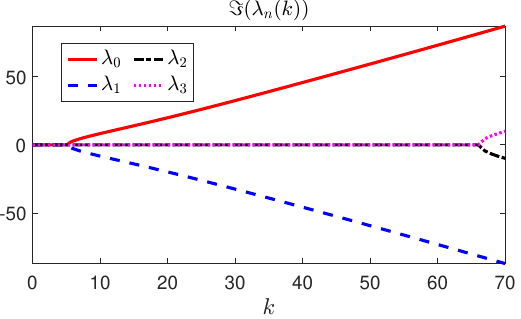}
  }
   \subfigure[$u=6\sqrt{5} (y (1-y)-\frac{1}{6})$]{
    \includegraphics[width=0.46\linewidth]{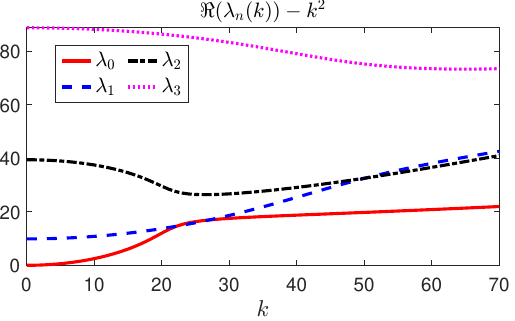}
  }
  \subfigure[$u=6\sqrt{5} (y (1-y)-\frac{1}{6})$]{
    \includegraphics[width=0.46\linewidth]{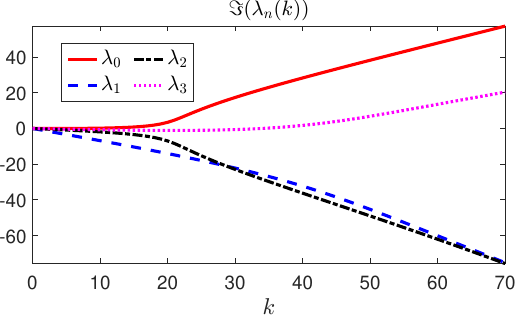}
 } 
  \hfill
  \caption[]
 {The first four eigenvalues for different shear flows with $\mathrm{Pe}=1$. }
  \label{fig:FlatChannelShearLinear}
\end{figure}
Next, we consider two practical flows: Couette flow given by $u = 2\sqrt{2} \left(y - \frac{1}{2}\right)$ and plane Poiseuille flow given by $u = 6\sqrt{5} \left(y (1-y) - \frac{1}{6}\right)$. The coefficients in these expressions are chosen to ensure that $\left\langle u, u \right\rangle = 1$, allowing for a fair comparison with the cosine shear flows. Couette flow describes the motion of a viscous fluid in the space between two surfaces, one of which moves tangentially relative to the other. Its velocity profile is similar to $u = \sqrt{2} \cos(\pi y)$. In contrast, plane Poiseuille flow represents the steady, laminar flow of a viscous fluid between two stationary, parallel plates, driven by a pressure gradient, structurally resembling $u = \sqrt{2} \cos(2\pi y)$.

Consequently, we can anticipate that the eigenvalues associated with these flows will be similar to those of the cosine flows. The results for $\mathrm{Pe} = 1$ are presented in Figure \ref{fig:FlatChannelShearLinear}. For Couette flow, $\Re \lambda_{1} - k^{2}$ reaches a global minimum value of 5.2149 at $k = 5.2149$. In contrast, for plane Poiseuille flow, $\Re \lambda_{1} - k^{2}$ attains its global minimum value of $\pi^{2}$ at $k = 0$. These results imply that the diffusing scalar advected by the plane Poiseuille flow converges more slowly to the slow manifold compared with the scalar advected by the Couette flow.

  Why may the global minimum of $\Re(\lambda_1(k))$ occur at a nonzero wavenumber? While a complete explanation remains open, we offer two conjectures.   From a mathematical point of view, the asymptotic expansion of the eigenvalue for a small wavenumber reveals that interactions among different eigenfunctions contribute to the derivative of $\lambda_1(k)$ and thereby determine whether $\Re(\lambda_1(k))$ attains a local minimum at $k = 0$. However, these interactions are generally complex and difficult to characterize analytically. For the flat parallel-plate channel, the eigenfunctions at zero wavenumber take the form $\phi_{n}= \sqrt{2} \cos(n \pi y)$. When the velocity field is expanded in this basis, calculations show that only the component corresponding to $\phi_1 = \sqrt{2} \cos(\pi y)$ causes $\Re(\lambda_1(k))$ to fail to attain a local minimum at $k = 0$ for certain values of Pe. Since $\phi_1$ is also the eigenfunction associated with $\lambda_1$, we conjecture that, in more general settings, the projection of the velocity field onto $\phi_1$ may similarly affect whether the global minimum of $\Re(\lambda_1(k))$ occurs at a nonzero wavenumber.

From a physical perspective, flows such as $u = \sqrt{2} \cos(\pi y)$ or $u = \frac{1}{2} - y$ shear the scalar field in opposite directions across the channel: fluid near $y = 0$ is advected downstream, while fluid near $y = 1$ is advected upstream. This antisymmetric shearing creates a configuration where scalar filaments are stretched in opposing directions, potentially suppressing transverse dispersion and slowing convergence to the asymptotic state. In contrast, flows like $u = \sqrt{2} \cos(n\pi y)$ for $n \geq 2$ contain more than two extrema across the channel height. These more oscillatory profiles lead to multiple shearing layers, which may enhance stretching and increase transverse concentration gradients, thereby accelerating the decay to the slow manifold.

These interpretations remain conjectural, but they suggest interesting connections between the spectral structure of the advection-diffusion operator and the geometry of the underlying flow. A more detailed investigation of this phenomenon is left for future work.

\subsection{Cellular flow in the channel with flat boundaries}
Shear flow is a unidirectional flow and has a transverse velocity component of zero. To examine the effect of the transverse velocity component in a channel, we consider scalar transport in a channel domain with flat boundaries but subject to a cellular flow given by  
\begin{equation}\label{eq:cellular flow}
\begin{aligned}
&u = \frac{2 \sin(2 \pi x) \cos(\pi y)}{\sqrt{5}}, \quad v = -\frac{4 \cos(2 \pi x) \sin(\pi y)}{\sqrt{5}}.
\end{aligned}
\end{equation}

\begin{figure}
  \centering
    \subfigure[]{
      \includegraphics[width=0.46\linewidth]{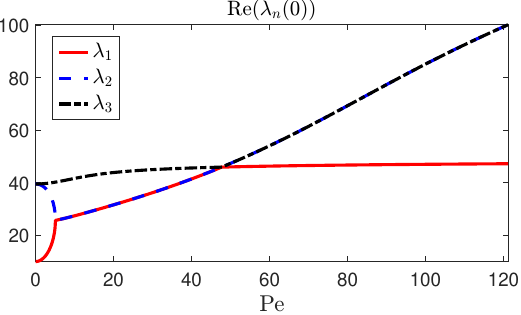}
    }
    \subfigure[]{
      \includegraphics[width=0.46\linewidth]{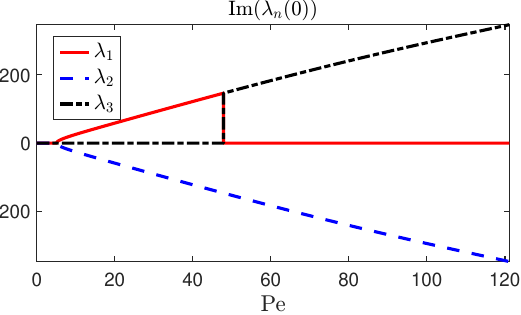}
    }
    \hfill
  \caption[]
  { Eigenvalues $\lambda_{n} (0), n=1,2,3$ as a function $\mathrm{Pe}$ for the cellular flow in the channel domain with flat walls.
    % Due to the labeling of the eigenvalue, $\lambda_{1} (0)$ may not be continuous as a function of $\mathrm{Pe}$ since there may be a jump in the imaginary part of $\lambda_{1} (0)$.
  }
  \label{fig:FlatChannelCellularK0PeRe}
\end{figure}

In this case, the fundamental period of the system is 1, and the unit cell is given by 
$\Omega (1) = \left\{(x,y) \, | \, -\frac{1}{2} \leq x \leq \frac{1}{2}, \, 0 \leq y \leq 1 \right\}.$
In the case of shear flow, since the eigenfunction is independent of $x$, $\hat{\mathcal{L}}(k)$ in equation \eqref{eq:eigenvalue problem} reduces to the Laplace operator when $k = 0$, yielding $\lambda_{1}(0) = \pi^{2}$ for all $\mathrm{Pe}$. In contrast, for cellular flow, the advection term persists at $k = 0$, causing $\lambda_{1}(0)$ to depend on $\mathrm{Pe}$. Figure \ref{fig:FlatChannelCellularK0PeRe} illustrates $\lambda_{1}(0)$ as a function of $\mathrm{Pe}$. As $\mathrm{Pe}$ increases, $\lambda_{1}(0)$ initially grows rapidly, but the rate of increase slows down after $\mathrm{Pe} \approx 46$. For example, at $\mathrm{Pe} = 200$, $\lambda_{1}(0) = 47.6839$, which is much larger than that in the shear flow case. This suggests that the scalar field under the advection of the cellular flow converges much faster to the slow manifold compared with the shear flow, a result attributed to the velocity component in the transverse direction. 

We note that for large $\mathrm{Pe}$, the effective longitudinal diffusivity $\kappa_{\mathrm{eff}}$ induced by shear flow exceeds that induced by cellular flow. Specifically, $\kappa_{\mathrm{eff}} \sim \mathrm{Pe}^{2}$ for steady shear flow \cite{camassa2010exact}, whereas our numerics show $\kappa_{\mathrm{eff}} \sim \mathrm{Pe}^{1/2}$ for the cellular flow defined in \eqref{eq:cellular flow}. The same scaling was rigorously proved under periodic $y$-boundary conditions in \cite{mclaughlin1994turbulent}. It is a reasonable result since the effective longitudinal diffusivity measures the ability of the flow to spread the scalar in the longitudinal direction, but it does not fully capture the mixing ability of the flow. The eigenvalue, however, serves as a better measure of the flow's mixing capability.

\begin{figure}
  \centering
    \subfigure[$\mathrm{Pe}=1$]{
    \includegraphics[width=0.46\linewidth]{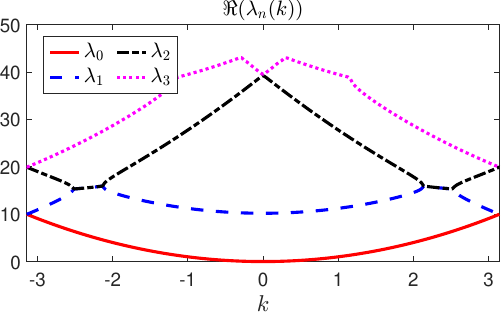}
  }
  \subfigure[$\mathrm{Pe}=50$]{
    \includegraphics[width=0.46\linewidth]{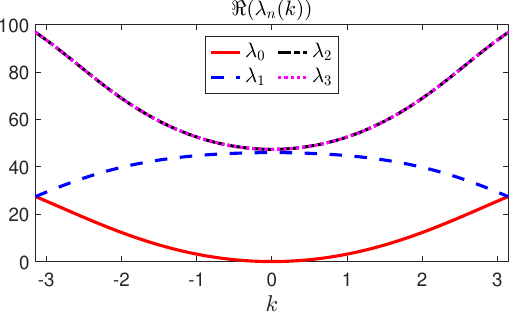}
  }
  \hfill
  \caption[]
  {The first four eigenvalues as functions of the wavenumber. $\lambda_{n} (k)$ is periodic in $k$ with period $2\pi$.  In panel (a), $\mathrm{Pe} = 1$, and in panel (b), $\mathrm{Pe} = 50$. 
  }
  \label{fig:FlatChannelCellular}
\end{figure}

Similar to the shear flow case $u = \sqrt{2} \cos \pi y$, the global minimum of the eigenvalue is not always located at $k = 0$. Figure \ref{fig:FlatChannelCellular} shows $\lambda_{1}(k)$ as a function of the wavenumber for different values of $\mathrm{Pe}$. For the $\mathrm{Pe}$ values considered in Table~\ref{tab:FlatChannelCellularK0PeRe}, the global minimum of $\lambda_{1}(k)$ occurs at $k = \pm \pi$. For large $\mathrm{Pe}$, as $\mathrm{Pe}$ increases, $\lambda_{1}(0)$ grows slowly, while the ratio $\min_{k} \lambda_{1}(k) / \lambda_{1}(0)$ decreases.

 \begin{table}
  \begin{center}
    \begin{tabular}{l|cccc}
          \hline
    $\mathrm{Pe}$   &$\mathrm{arg}\min\limits_{k}\lambda_{1} (k)$  &$\min\limits_{k}\lambda_{1} (k)$& $\lambda_{1} (0)$ & $\lambda_{1} (0)/\min\limits_{k}\lambda_{1} (k)$ \\
      \hline
     0& $\pm \pi$ &$\pi^{2}$&$\pi^{2}$ &1     \\
     1& $\pm \pi$ &9.980&10.20 &1.022     \\
    20& $\pm \pi$ &21.62&43.86 &2.029\\
    50& $\pm \pi$ &27.44&46.00 &1.677 \\
   100& $\pm \pi$ &31.67&47.00 &1.484\\
   200& $\pm \pi$ &35.61&47.68 &1.339\\
    \hline
  \end{tabular}
  \caption{Eigenvalues for the cellular flow case.}
  \label{tab:FlatChannelCellularK0PeRe}
  \end{center}
\end{table}

\subsection{Sinusoidal channel with pressure driven flow}
\begin{figure}
  \centering
    \includegraphics[width=0.46\linewidth]{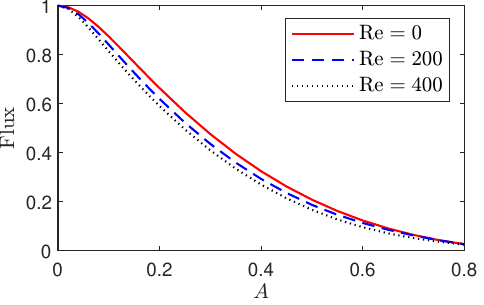}
  \hfill
  \caption[]
  {The averaged fluid flux at the inlet, $\frac{1}{|\Omega_{\text{inlet}}|} \int_{\text{inlet}} u (x,\mathbf{y}) \mathrm{d}\mathbf{y}$, as a function of $A$ for different Reynolds numbers, where $|\Omega_{\text{inlet}}|$ is the area of the inlet. }
  \label{fig:FluxAmplitude}
\end{figure}

Next, we consider a more practical flow: the pressure-driven flow in the sinusoidal channel governed by equation \eqref{eq:AdvectionDiffusionEquation}. The flow is solved for different Reynolds number in the cell domain given by
\begin{equation}\label{eq:sinusoidal channel cell domain}
\begin{aligned}
\Omega (L_{p}) =\left\{  (x, y) |  - \frac{L_p}{2}\leq x \leq \frac{L_p}{2},    0\leq y\leq  h(x) \right\},\quad  h(x) = 1 + A \sin\left(\frac{2\pi x}{L_{p}}\right).
\end{aligned}
\end{equation}
It is important to note that, for a fixed pressure gradient, the fluid flux at the inlet decreases as the amplitude of the boundary variation increases. This indicates that the flow strength becomes weaker with larger boundary undulations. Figure \ref{fig:FluxAmplitude} shows the relationship between the amplitude $A$ and the inlet flux for a fixed pressure gradient of $f_x = -12$. When $A = 0$, the flux is 1. In contrast, when $A = 0.8$, the flux drops to approximately $0.025 \sim 0.026$, which is about 40 times smaller. Therefore, to ensure that the characteristic velocity remains comparable across different values of $A$, we adjust the pressure gradient so that the inlet flux is fixed to be 1 for all $A$.

The constant-flux and constant-pressure-gradient cases are related by a nonlinear rescaling of $\mathrm{Re}$ and $\mathrm{Pe}$ via the pressure–flux relation shown in Figure~\ref{fig:FluxAmplitude}: at larger amplitudes, a fixed pressure drop yields a much smaller mean flux, which would correspond to smaller effective $\mathrm{Re}$ and $\mathrm{Pe}$ in the fixed-flux formulation. In other words, results for a constant pressure gradient can be interpreted as lying along the fixed-flux curves at appropriately reduced $\mathrm{Re}$ and $\mathrm{Pe}$.

From the previous section, we see that $\lambda_1(k)$ may not reach the global minimum at $k = 0$. However, $\lambda_1(0)$ does not differ significantly from $\min_k \lambda(k)$ and can serve as an estimate for the time scale $t_s$. Therefore, to reduce computational cost, we focus primarily on $\lambda_1(0)$ in this section for simplicity.

\begin{figure}
  \centering
      \subfigure[$\mathrm{Re}=0$]{
        \includegraphics[width=0.46\linewidth]{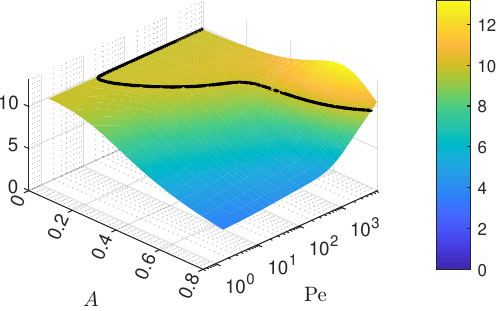}
      }
            \subfigure[$\mathrm{Re}=10$]{
        \includegraphics[width=0.46\linewidth]{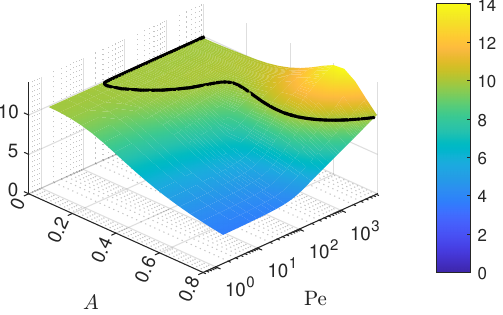}
      }
            \subfigure[$\mathrm{Re}=100$]{
        \includegraphics[width=0.46\linewidth]{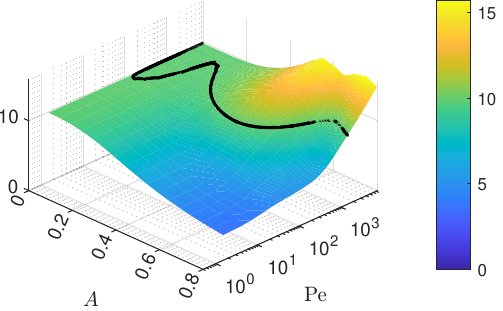}
      }
            \subfigure[$\mathrm{Re}=200$]{
        \includegraphics[width=0.46\linewidth]{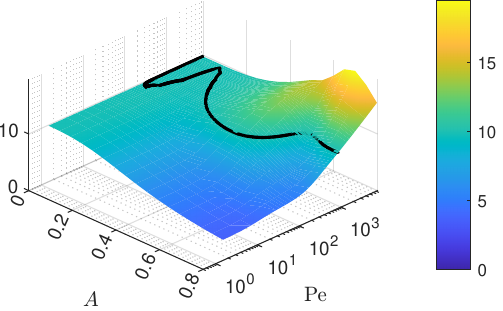}
      }
  \hfill
  \caption[]
 { The real part of $\lambda_{1} (0)$ as a function of $\mathrm{Pe}$ and $A$. The axis for $\mathrm{Pe}$ is in log scale. The black line indicate the contour line for $\lambda_{1} (0)=\pi^{2}$. }
  \label{fig:CurveChannelEigNSPeARe}
\end{figure}

The real part of $\lambda_1(0)$ for different Reynolds numbers, $A$, and $\mathrm{Pe}$ is shown in Figure \ref{fig:CurveChannelEigNSPeARe}. For $A = 0$, where the channel boundary is flat, the shear flow is the solution to the Navier-Stokes equation for any $\mathrm{Re}$. In this case, $\lambda_1(0) = \pi^2$ for all $\mathrm{Pe}$. The black curve in Figure \ref{fig:CurveChannelEigNSPeARe} indicates the contour for $\lambda_1(0) = \pi^2$, which separates the parameter regimes where the scalar field converges to the slow manifold faster than in the flat channel and where it converges more slowly.

For small $\mathrm{Pe}$, increasing the amplitude $A$ causes $\Re \lambda_{1}(0)$ to decrease below $\pi^2$. This suggests that, even though the non-flat boundary induces a non-zero transverse velocity component, the scalar may not converge to the slow manifold faster than in the case of pure shear flow. In contrast, for all Reynolds numbers considered here, if $\mathrm{Pe}$ is sufficiently large (above $10^{2}$), increasing $A$ leads to an increase in $\Re \lambda_{1}(0)$, making it larger than $\pi^2$.

This behavior is reasonable, as two competing effects are at play. On one hand, the non-flat boundary restricts diffusion compared with the flat case, which reduces the effective diffusivity and tends to slow down convergence. On the other hand, the induced transverse velocity enhances convergence. The relative importance of these effects depends on $\mathrm{Pe}$. When $\mathrm{Pe}$ is small, diffusion dominates, and the scalar converges more slowly. When $\mathrm{Pe}$ is large, advection becomes dominant, accelerating the convergence to the slow manifold.

If we view the contour as a curve $\mathrm{Pe} = f(A)$ in the $A-\mathrm{Pe}$ plane, its shape depends on the Reynolds number. When $\mathrm{Re} = 0$, the curve is convex, but it becomes non-convex for larger $\mathrm{Re}$. For nonzero Reynolds numbers ($\mathrm{Re} > 0$), there exists a range of $\mathrm{Pe}$ where the non-flat boundary enhances convergence to the slow manifold for either small or sufficiently large values of $A$.

\begin{figure}
  \centering
  \subfigure[$\lambda_{1,2}$]{
    \includegraphics[width=0.46\linewidth]{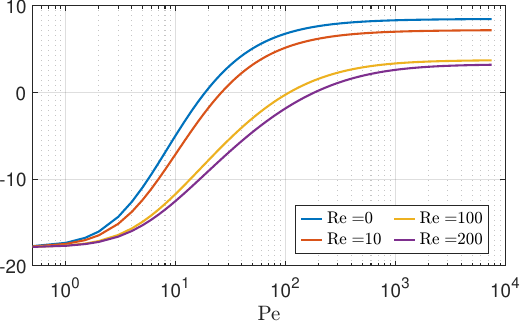}
  }
  \subfigure[$\kappa_{2}$]{
    \includegraphics[width=0.46\linewidth]{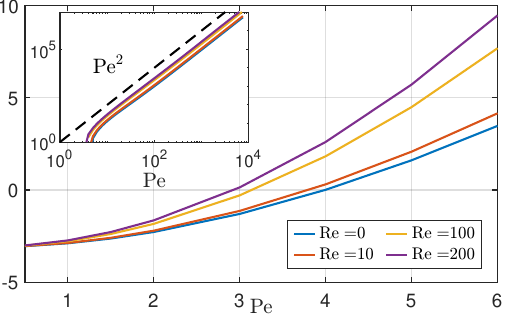}
  }
  \caption[]
  {Coefficients in the asymptotic expansion of the effective diffusivity (panel a) and the eigenvalue (panel b) in \eqref{eq:kappa eigenvalue small amplitude}. In panel b, the inset shows a log–log plot of $\lambda_{1,2}$ over a wider range of $\mathrm{Pe}$. The black dashed line represents $\mathrm{Pe}^{2}$. The largest relative fitting error across all investigated parameters is $3\%$.
}
  \label{fig:A2coef}
\end{figure}

We next investigate the asymptotic behavior in the limit of small $A$. Owing to symmetry, the expansions for both the effective diffusivity and the eigenvalue contain only even powers of $A$:

\begin{equation}\label{eq:kappa eigenvalue small amplitude}
\begin{aligned}
&\kappa_{\mathrm{eff}} = 1 + \frac{\mathrm{Pe}^{2}}{210} + \kappa_{2}A^{2} + \kappa_{4}A^{4} + \mathcal{O}(A^{6}), \;
\lambda_{1}(0) = \pi^{2} + \lambda_{1,2}A^{2} + \lambda_{1,4}A^{4} + \mathcal{O}(A^{6}).
\end{aligned}
\end{equation}
We fit the data presented in figure~\ref{fig:CurveChannelEigNSPeARe} to determine the coefficients in \eqref{eq:kappa eigenvalue small amplitude}. Figure~\ref{fig:A2coef} shows $\lambda_{1,2}$ and $\kappa_{2}$ as functions of $\mathrm{Pe}$.

Several trends can be observed for $\lambda_{1,2}$. First, $\lambda_{1,2}$ approaches a limiting value as $\mathrm{Pe}$ becomes large or small, with distinct limits in each regime. Second, for all $\mathrm{Re}$ considered, $\lambda_{1,2}$ is negative at small $\mathrm{Pe}$ and positive at large $\mathrm{Pe}$. This implies that, for small amplitudes, $\lambda_{1}(0) < \pi^{2}$ when $\mathrm{Pe}$ is small, and $\lambda_{1}(0) > \pi^{2}$ when $\mathrm{Pe}$ is large, consistent with figure~\ref{fig:CurveChannelEigNSPeARe}. The $\mathrm{Pe}$ values at which $\lambda_{1,2}$ changes sign are $18.5$, $26.1$, $108$, and $183$ for $\mathrm{Re} =$0, 10, 100, 200, respectively. For the $\mathrm{Re}$ values studied here, $\lambda_{1,2}$ decreases as $\mathrm{Re}$ increases, suggesting that higher $\mathrm{Re}$ slows convergence toward the slower manifold when the amplitude is small.

A similar trend holds for $\kappa_{2}$: it is negative at small $\mathrm{Pe}$ and positive at large $\mathrm{Pe}$. The sign-change $\mathrm{Pe}$ values are $3.99$, $3.81$, $3.15$, and $2.93$ for $\mathrm{Re} =$0, 10, 100, 200, respectively. For the range of $\mathrm{Re}$ considered, $\kappa_{2}$ increases with $\mathrm{Re}$. For large $\mathrm{Pe}$, $\lambda_{1,2}$ scales as $\mathrm{Pe}^{2}$, which is consistent with the fact that $\kappa_{\mathrm{eff}}$ scales as $\mathrm{Pe}^{2}$ when $A=0$. The coefficient in the small-amplitude expansion is therefore expected to exhibit the same scaling behavior.

The large-amplitude limit is challenging to simulate because maintaining a constant fluid flux requires a very large pressure drop, which in turn necessitates a highly refined mesh for accuracy. Studying this regime would require specialized analytical techniques, such as those in \cite{alexandre2025effective}.

 To investigate the asymptotic behavior of the effective diffusivity $\kappa_{\mathrm{eff}}$ and the eigenvalue $\lambda_{1}(0)$ in the long-wavelength limit $L_{p} \to \infty$, we compute their values numerically and present the results in table~\ref{tab:long wavelength limit}. Several observations can be made.
First, neither $\kappa_{\mathrm{eff}}$ nor $\lambda_{1}(0)$ converges to the flat‐channel values as $L_p$ increases, which is expected since boundary variations persist even in the long‐wavelength limit. Notably, the approach to the limiting values can be non-monotonic in some cases.
Second, for small $L_p$, inertia can increase $\kappa_{\mathrm{eff}}$ substantially. However, the limiting values of $\kappa_{\mathrm{eff}}$ depend only on the geometry and are independent of the Reynolds number. This follows from the fact that $\partial_{x} u$ and $v$ scale as $L_{p}^{-1}$, so inertial effects become negligible as $L_p \to \infty$.
Third, large boundary amplitudes slow the convergence to the limiting values dramatically.

\begin{table}
  \centering
  \begin{tabular}{lccccc |c ccccc}
    \hline
    & \multicolumn{5}{c}{$\kappa_{\mathrm{eff}}$} & & \multicolumn{5}{c}{$\lambda_{1} (0)$} \\
    \hline
    $\mathrm{Re}$ & 1& 2 & 5 & 10&20  & & 1 &2& 5 & 10&20  \\
    \hline
    0   &62.74 &24.33 &14.39 & 14.32 &14.52& &9.244&10.52& 10.41& 10.30 &10.27 \\
    50 & 94.61 &69.23 &20.78 & 15.73 &14.90&  &8.266&8.475& 10.07& 10.22&10.24\\
    \hline
    0   & 426.2 &160.4 &32.18 &23.90 &24.37  & &4.254&6.599& 15.54& 15.15&14.91 \\
    50  & 448.7 &321.8 &302.3 &90.98 &28.99  & &5.033&9.792& 9.529& 6.500&14.38 \\
    \hline  
  \end{tabular}

    \caption{Effective diffusivity $\kappa_{\mathrm{eff}}$ and eigenvalue $\lambda_{1}(0)$ for the domain defined in \eqref{eq:sinusoidal channel cell domain} are shown for $\mathrm{Pe} = 50$ with $A = 0.3$ (first two rows) and $A = 0.8$ (last two rows). Columns correspond to $L_p = 1, 2, 5, 10, 20$. The left block lists $\kappa_{\mathrm{eff}}$, and the right block lists $\lambda_{1}(0)$. For reference, a flat-walled channel yields $\kappa_{\mathrm{eff}} = 1 + \mathrm{Pe}^2/210 \approx 12.905$ and $\lambda_{1}(0) = \pi^2 \approx 9.870$.}
  \label{tab:long wavelength limit}
\end{table}

  The dependence of $\kappa_{\mathrm{eff}}$ on flow and the sinusoidal geometry in this type of domain has been extensively studied in previous work \cite{richmond2013flow,bouquain2012impact}. For conciseness, we do not reproduce these results here, but they can be computed from the present framework in a straightforward manner.

  \section{Extension to Porous Media}
  \label{sec:Extension to Porous Media}
The bi-orthogonal eigenfunction expansion framework presented in
Sections~\ref{eq:Governing Equation and asymptotic analysis} for
periodically modulated channels extends naturally to periodic porous
media. Let $Y \subset \mathbb{R}^d$ denote a unit cell, where $d$ is the
spatial dimension, containing the connected fluid subdomain $\Omega$
(pores and throats), repeated periodically in space. Let $\mathbf{u}$ be
the periodic steady flow in $\Omega$, subject to the no-slip condition
on $\partial\Omega$.

 Analogous to the cell eigenfunctions \eqref{eq:eigenfunction wavenumber} for the channel-flow problem, we consider eigenfunctions of the advection–diffusion equation of the form
 \begin{equation}
\begin{aligned}
\phi_{n}(\mathbf{x},\mathbf{k}) = e^{i \mathbf{k} \cdot \mathbf{x}} \hat{\phi}_{n}(\mathbf{x},\mathbf{k}), \quad \mathbf{k}\in \mathbb{R}^{d},
\end{aligned}
\end{equation}
 where $\hat{\phi}_{n}$ is periodic on $Y$. The corresponding cell
eigenvalue problems \eqref{eq:eigenvalue problem wavenumber} and
\eqref{eq:adjoint eigenvalue problem wavenumber} are defined in the same
manner as before.

Following the same reasoning used for
equation~\eqref{eq:eigenfunctionExpansion0th}, the mode associated with
$\lambda_0(\mathbf{k})$ dominates at long times and defines the slow
manifold of the system. The long-time asymptotics of the scalar field are
obtained by applying a perturbation analysis to the integral representation
of the slow manifold near zero wavenumber. This yields the homogenized
transport equation, providing the leading-order approximation to the scalar
field:
\begin{equation}
\begin{aligned}
\partial_t c + \left\langle \mathbf{u} \right\rangle \cdot \nabla c 
= \nabla \cdot \big( K_{\mathrm{eff}} \nabla \bar{c} \big),  
\quad (K_{\mathrm{eff}})_{ij} = \tfrac12 \, \partial_{k_i} \partial_{k_j} \lambda_0(\mathbf{0}),
\end{aligned}
\end{equation}
where $K_{\mathrm{eff}}$ is the \emph{tensorial} effective diffusivity,
which is symmetric and positive-definite for physically relevant flows.
The characteristic convergence time remains
$t_s = 1 / \min_{\mathbf{k}} \mathrm{Re}\,\lambda_1(\mathbf{k})$.

This formulation provides a rigorous route to Taylor--Aris-type dispersion
tensors in periodic porous microstructures, offering a direct spectral
characterization of $K_{\mathrm{eff}}$. The exploration of specific
three-dimensional geometries, and the influence of pore-scale topology on
anisotropic dispersion, is left for future work. Potential applications
include periodic microstructures encountered in porous media \cite{municchi2020macroscopic,auton2025homogenized}, structured
battery electrodes \cite{dussi2022less}, and patterned microelectrode arrays
in channel flows \cite{zhang1996mass}.

\section{Conclusion}
\label{sec:conclusion}
In this work, we investigate the long-time asymptotic behavior of a diffusing scalar passively advected in a channel with a periodically varying cross-section. By formulating the eigenvalue problem of the advection-diffusion operator over a unit cell and employing a biorthogonal eigenfunction expansion, we generalize the Fourier integral used in the flat channel case and obtain an integral representation \eqref{eq:eigenfunction expansion} of the scalar field. This integral representation allows us to identify the slow manifold \eqref{eq:eigenfunctionExpansion0th} of the system, which decays algebraically, while the difference between the scalar field and the slow manifold decays exponentially. We find that the long-time asymptotic expansion of the scalar field is related to the small-wavenumber expansion of the eigenfunctions and eigenvalues. Using the steepest descent method, we derive the asymptotic expansion of the slow manifold \eqref{eq:ground state long time expansion}. Since the correction term is algebraic, this expansion also serves as the long-time asymptotic expansion of the scalar field.

While the leading-order term in this expansion has previously been derived via multiple-scale analysis, the advantage of the present framework is that  it offers a clear understanding of the relationship between different time scales in the Taylor dispersion problem. The time scale for convergence to the slow manifold is determined by the smallest real part of the eigenvalue, given by  $t_s = 1/ \min_k \Re \lambda_1(k)$, which depends solely on the domain geometry and velocity field. This is also the time scale over which the variance begins to increase linearly once the exponential decay is complete, which is verified by numerical simulating the advection-diffusion equation in the entire domain. After this time scale is reached, the asymptotic expansion \eqref{eq:ground state long time expansion} becomes valid. The first term in the expansion is a Gaussian function with no variation in the transverse direction, while the remaining terms describe deviations from Gaussianity and transverse variations. Since these correction terms generally have small coefficients, they become negligible unless \( \mathrm{Pe} \) is large or the initial variance is small. In such cases, after time \( t_s \), the cross-sectionally averaged scalar field is well approximated by a Gaussian function. The second term provides the leading-order nontrivial description of the transverse variation of the scalar field. As time increases, when the second term becomes negligible compared with the first, the transverse uniformity time scale is reached.

We investigate the timescale $t_s$ across different flow and domain geometries. While $\Re \lambda_1(k)$ does not always attain its minimum at $k = 0$, we find that $\Re \lambda_1(0)$ still serves as a good estimate for the timescale. The presence of a transverse velocity component generally increases $\Re \lambda_1(0)$, promoting faster convergence, whereas a non-flat channel boundary restricts diffusion and tends to decrease $\Re \lambda_1(0)$.

We primarily test the theory using a sinusoidal channel with a smooth boundary; however, the framework applies equally to domains with non‐smooth periodic boundaries, such as the square‐wave cross‐section examined in \cite{haugerud2022solute}, or to non‐simply connected domains, as illustrated in figure~\ref{fig:CurveChannelSchematic}.

Future work could extend this framework in several directions. First, it could be applied to more complex flow conditions, including time-dependent flows. Second, investigating the impact of alternative boundary conditions or incorporating reactive transport could provide further insights into scalar mixing in environmental and industrial applications. Third, the example studied in this work exhibits symmetry in the flow direction; truly asymmetric periodic geometries, where no spatial translation can recover symmetry, may yield qualitatively different dispersion behavior and could be exploited to generate anisotropic scalar transport.

\section{ Declaration of Interests}
The author report no conflict of interest.

\appendix

\section{Eigenvalue}
\label{sec:Eigenvalue}
\subsection{Nonnegative real part}
\label{sec:Nonnegative real part}
First, we aim to show the real part of the eigenvalue is nonnegative.  Multiplying $\phi^{*}$ on both side of equation \eqref{eq:eigenvalue problem} and taking integration leads to
\begin{equation}
\begin{aligned}
&-\lambda \int\limits_{\Omega}^{}\phi \phi^{*}\mathrm{d}x\mathrm{d}\mathbf{y}=\int\limits_{\Omega}^{}\left(- \mathrm{Pe} \left(u \partial_{x}\phi+ \mathbf{v}  \cdot \nabla_{\mathbf{y}} \phi \right)+  \partial_{x}^{2}\phi+\Delta_{\mathbf{y}}\phi \right) \phi^{*}\mathrm{d}x\mathrm{d}\mathbf{y}. \\
\end{aligned}
\end{equation}
Then we take the real part of both side of the equation and use the fact that $\Re (\partial_{x} (\phi \phi^{*}))=2\Re ( \phi^{*}\partial_{x} \phi) $
\begin{equation}
\begin{aligned}
\Re(\lambda) \int\limits_{\Omega}^{}\phi \phi^{*}\mathrm{d}x\mathrm{d}\mathbf{y}&=\int\limits_{\Omega}^{} \frac{\mathrm{Pe} }{2} \Re \left( u \partial_{x}(\phi \phi^{*})+ \mathbf{v}  \cdot \nabla_{\mathbf{y}} (\phi \phi^{*})\right) -  \left( \partial_{x}^{2}\phi+ \Delta_{\mathbf{y}}\phi \right)  \phi^{*}\mathrm{d}x\mathrm{d}\mathbf{y}  \\
&=\int\limits_{\Omega}^{}  \partial_{x}\phi \partial_{x}\phi^{*}+\nabla_{\mathbf{y}}\phi \cdot \nabla_{\mathbf{y}}\phi^{*}   \mathrm{d}x\mathrm{d}\mathbf{y}. \\
\end{aligned}
\end{equation}
The second step follows the incompressibility of the flow, the no-flux boundary conditions of the flow and eigenfunction, and the periodicity of the eigenfunction and velocity field. Since the right hand side of the above equation is nonnegative, the real part of the eigenvalue is nonnegative. 

\subsection{Completeness of biorthogonal system}
\label{sec:complete orthonormal set}
We aim to show $\left\{ e^{\mathrm{i} k x}\hat{\phi}_{n}(x,\mathbf{y},k) | k\in \mathbb{R}, n \in \mathbb{N}\right\}$ and $\left\{ e^{\mathrm{i} k x}\hat{\varphi}_{n}(x,\mathbf{y},k) | k\in \mathbb{R}, n \in \mathbb{N}\right\}$ form a complete biorthogonal system, where $\mathbb{N}$ is the set including all nonnegative integers. Here $\hat{\phi}_{n}(x,\mathbf{y},k)$ and $\hat{\varphi}_{n}(x,\mathbf{y},k)$ solve equations \eqref{eq:eigenvalue problem wavenumber} and \eqref{eq:adjoint eigenvalue problem wavenumber} respectively.

Denoting $L^{2} (\Omega (L))$ as the set of all square integrable function defined on $\Omega (L)$.  Since $\hat{\phi}_{n}(x,\mathbf{y},k), \hat{\varphi}_{n}(x,\mathbf{y},k)$ are the eigenfunctions defined on $\Omega (L_{p})$ for each $k$, they form a complete biorthogonal system for $L^{2} (\Omega (L_{p}))$. 

Next, let's consider a function $f\in L^{2} (\Omega (2L_{p}))$.  Denoting $I_{[a,b]} (x) =
\begin{cases}
  1& x \in [a,b]\\
  0& x \not \in [a,b]]
\end{cases}
$ as the indicator function. We can approximate the indicator function using the Fourier series
\begin{equation}
\begin{aligned}
& I_{[-L_{p},0]}=\sum\limits_{k}^{}a_{1,k}e^{ikx}, \quad I_{[0,L_{p}]}=\sum\limits_{k}^{}a_{2,k}e^{ikx}. \\
\end{aligned}
\end{equation}
Then we have
\begin{equation}
\begin{aligned}
  f=&f  I_{[-L_{p},0]}+f I_{[0,L_{p}]} =f_{[-L_{p},0]}  I_{[-L_{p},0]}+f_{[0,L_{p}]} I_{[0,L_{p}]} \\
  =&f_{[-L_{p},0]} \sum\limits_{k}^{}a_{1,k}e^{ikx} +  f_{[0,L_{p}]}\sum\limits_{k}^{}a_{2,k}e^{ikx} \\
 = &\sum\limits_{k}^{}a_{1,k}e^{ikx} \sum\limits_{n}^{}\left\langle f_{[-L_{p},0]} , \hat{\varphi}_{n}(x,\mathbf{y},k) \right\rangle \hat{\phi}_{n}(x,\mathbf{y},k)\\
  &+\sum\limits_{k}^{}a_{2,k}e^{ikx} \sum\limits_{n}^{}\left\langle f_{[0,L_{p}]} , \hat{\varphi}_{n}(x,\mathbf{y},k) \right\rangle \hat{\phi}_{n}(x,\mathbf{y},k),
\end{aligned}
\end{equation}
where $f_{[0,L_p]}$ denotes the periodic extension of the restriction of the function on the subinterval $[0,L_{p}]$. Since $f_{[0,L_p]}$ has the fundamental period $L_p$, it can be approximated by the biorthogonal system $\left\{ \hat{\phi}_{n}(x,\mathbf{y},k) | n \in \mathbb{N}\right\}$ and $\left\{ \hat{\varphi}_{n}(x,\mathbf{y},k) |  n \in \mathbb{N}\right\}$ for each $k$. Following the same construction, we can extend the conclusion to the function defined on the unbounded interval $f\in L^{2} (\Omega (\infty))$.

\subsection{Eigenfunction expansion for small wavenumbers}
\label{sec:Eigenfunction expansion for small wavenumbers}
Assuming power series expansion of the eigenvalue and eigenfunctions
\begin{equation}\label{eq:Eigenfunction power series expansion for small wavenumbers}
\begin{aligned}
&\lambda_{n}= \sum\limits_{i=0}^{\infty} \lambda_{n}^{(i)}k^{i}, \quad \hat{\phi}_{n} (x,\mathbf{y} ) = \sum\limits_{i=0}^{\infty} \hat{\phi}_{n}^{(i)} (x,\mathbf{y}) k^{i},\quad  \hat{\varphi}_{n} (x,\mathbf{y} ) = \sum\limits_{i=0}^{\infty} \hat{\varphi}_{n}^{(i)} (x,\mathbf{y}) k^{i}.
\end{aligned}
\end{equation}
We consider normalized eigenfunctions satisfying
\begin{equation}
\begin{aligned}
\left\langle \hat{\phi}_{n}, \hat{\varphi}_{n} \right\rangle=\left\langle  \sum\limits_{i=0}^{\infty} \hat{\phi}_{n}^{(i)} (y) k^{i}, \sum\limits_{i=0}^{\infty} \hat{\varphi}_{n}^{(i)} (y) k^{i}, \right\rangle=1, \quad i \in \mathbb{N}^{+}.
\end{aligned}
\end{equation}
Since this constraint holds for arbitrary $k$, we have the following equations
\begin{equation}\label{eq:biorthogonal system constraint}
\begin{aligned}
&\left\langle \hat{\phi}_{n}^{(0)},\hat{\varphi}_{n}^{(0)}  \right\rangle =1, \quad  \sum\limits_{j=0}^{i} \left\langle \hat{\phi}_{n}^{(j)},\hat{\varphi}_{n}^{(i-j)}  \right\rangle=0,  \quad i \in \mathbb{N}^{+}.   \\
\end{aligned}
\end{equation}

For convenience, we use the following operators
\begin{equation}
\begin{aligned}
  &\hat{\mathcal{L}}_{0} (0) = - \mathrm{Pe}u (x,\mathbf{y})\partial_{x}- \mathrm{Pe} \mathbf{v}(x,\mathbf{y}) \cdot \nabla_{\mathbf{y}} + \partial_{x}^{2}+\Delta_{\mathbf{y}},  \\
  &\hat{\mathcal{L}}_{0}^{*} (0) =  \mathrm{Pe}u (x,\mathbf{y})\partial_{x}+ \mathrm{Pe} \mathbf{v}(x,\mathbf{y}) \cdot \nabla_{\mathbf{y}} + \partial_{x}^{2}+\Delta_{\mathbf{y}}.  \\
\end{aligned}
\end{equation}
Substituting the expansion \eqref{eq:Eigenfunction power series expansion for small wavenumbers} into the eigenvalue problem \eqref{eq:eigenvalue problem wavenumber}, we obtain a hierarchy of equations:
\begin{equation}
\begin{aligned}
  &\hat{\mathcal{L}}_{0} (0)\hat{\phi}_{n}^{(i)}=- \sum\limits_{j=0}^{i}\lambda_{n}^{(j)}\hat{\phi}_{n}^{(i-j)}+\mathrm{i} \mathrm{Pe} u (x,\mathbf{y})\hat{\phi}_{n}^{(i-1)}-2 \mathrm{i} \partial_{x}\hat{\phi}_{n}^{(i-1)}+\hat{\phi}_{n}^{(i-2)}, \\
 & \left. n_{x}\mathrm{i}\hat{\phi}_{n}^{(i-1)}+n_{x}\partial_{x}\hat{\phi}_{n}^{(i)}+ \mathbf{n}_{\mathbf{y}}\cdot \nabla_{\mathbf{y}}\hat{\phi}_{n}^{(i)} \right|_{\mathrm{wall}}=0,  \\
 &\hat{\mathcal{L}}_{0}^{*} (0) \hat{\varphi}_{n}^{(i)}=- \sum\limits_{j=0}^{i}\left( \lambda_{n}^{(j)} \right)^{*}\hat{\varphi}_{n}^{(i-j)}- \mathrm{i} \mathrm{Pe} u (x,\mathbf{y})\hat{\varphi}_{n}^{(i-1)}-2\mathrm{i} \partial_{x}\hat{\varphi}_{n}^{(i-1)}+\hat{\varphi}_{n}^{(i-2)},\\
 &\left. n_{x}\mathrm{i}\hat{\varphi}_{n}^{(i-1)}+n_{x}\partial_{x}\hat{\varphi}_{n}^{(i)}+ \mathbf{n}_{\mathbf{y}}\cdot \nabla_{\mathbf{y}}\hat{\varphi}_{n}^{(i)} \right|_{\mathrm{wall}}=0. \\
\end{aligned}
\end{equation}
for each $i\in \mathbb{N}$, and $\hat{\phi}_{n}^{(i)}=\hat{\varphi}_{n}^{(i)}=0$ if $i<0$.

The equation for $\lambda_{n}^{(0)}$ and $ \hat{\phi}_{n}^{(0)}$ is
\begin{equation}
\begin{aligned}
&\hat{\mathcal{L}}_{0} (0)\hat{\phi}_{n}^{(0)}=-\lambda_{n}^{(0)} \hat{\phi}_{n}^{(0)},\; \left. n_{x}\partial_{x}\hat{\phi}_{n}^{(0)}+ \mathbf{n}_{\mathbf{y}}\cdot \nabla_{\mathbf{y}}\hat{\phi}_{n}^{(0)} \right|_{\mathrm{wall}}=0.  \\
\end{aligned}
\end{equation}
When $n=0$, we have $\lambda_{0}^{(0)}=0$ and $\hat{\phi}_{0}^{(0)}=\hat{\varphi}_{0}^{(0)}=1$. When $n\in \mathbb{N}^{+}$, for general  flows and  boundary geometries, the closed-form expression of the solution is not available. For the shear flow in the  channel with flat boundaries, $\lambda_{n}^{(0)}= (n\pi)^{2}$ and $\hat{\phi}_{n}^{(0)}=\hat{\varphi}_{n}^{(0)}=\sqrt{2}\cos n \pi y$.

The equation for $\lambda_{n}^{(1)}$ and $\phi_{n}^{(1)}$ is
\begin{equation}\label{eq:eigenfunction small wavenumber order 1}
\begin{aligned}
&\left(\hat{\mathcal{L}}_{0} (0)+\lambda_{n}^{(0)} \right) \hat{\phi}_{n}^{(1)}=\mathrm{i} \mathrm{Pe} u \hat{\phi}_{n}^{(0)}-2 \mathrm{i} \partial_{x}\hat{\phi}_{n}^{(0)}-\lambda_{n}^{(1)} \hat{\phi}_{n}^{(0)}, \\
 & \left. n_{x}\mathrm{i}\hat{\phi}_{n}^{(0)}+n_{x}\partial_{x}\hat{\phi}_{n}^{(1)}+ \mathbf{n}_{\mathbf{y}}\cdot \nabla_{\mathbf{y}}\hat{\phi}_{n}^{(1)} \right|_{\mathrm{wall}}=0.\\
\end{aligned}
\end{equation}
Taking the inner product with $\hat{\varphi}_{n}^{(0)}$ gives
\begin{equation}
\begin{aligned}
  & \left\langle \left( \hat{\mathcal{L}}_{0} (0)+\lambda_{n}^{(0)} \right) \hat{\phi}_{n}^{(1)},\hat{\varphi}_{n}^{(0)}  \right\rangle= \left\langle \mathrm{i} \mathrm{Pe} u \hat{\phi}_{n}^{(0)}-2 \mathrm{i} \partial_{x}\hat{\phi}_{n}^{(0)}-\lambda_{n}^{(1)} \hat{\phi}_{n}^{(0)},\hat{\varphi}_{n}^{(0)} \right\rangle. \\
\end{aligned}
\end{equation}
Notice that $\left\langle \hat{\mathcal{L}}_{0} (0)\hat{\phi}_{n}^{(1)},\hat{\varphi}_{n}^{(0)}  \right\rangle=\left\langle \hat{\phi}_{n}^{(1)},\hat{\mathcal{L}}_{0}^{*} (0)\hat{\varphi}_{n}^{(0)}  \right\rangle-\mathrm{i} \left\langle \partial_{x} \hat{\phi}_{n}^{(0)},\hat{\varphi}_{n}^{(0)} \right\rangle-\mathrm{i} \left\langle  \hat{\phi}_{n}^{(0)},\partial_{x}\hat{\varphi}_{n}^{(0)} \right\rangle$. We have
\begin{equation}
\begin{aligned}
&\lambda_{n}^{(1)}= \mathrm{i} \left(  \left\langle\mathrm{Pe} u \hat{\phi}_{n}^{(0)} ,\hat{\varphi}_{n}^{(0)} \right\rangle- \left\langle \partial_{x} \hat{\phi}_{n}^{(0)},\hat{\varphi}_{n}^{(0)} \right\rangle+ \left\langle  \hat{\phi}_{n}^{(0)},\partial_{x}\hat{\varphi}_{n}^{(0)} \right\rangle \right).\\ 
\end{aligned}
\end{equation}
There are infinite possible solutions of equation \eqref{eq:eigenfunction small wavenumber order 1}. We are looking for the one constrained by equation \eqref{eq:biorthogonal system constraint}. More precisely, due to the linearity of the equation, we assume that $\hat{\phi}_{n}^{(1)}=\mathrm{i} \left( \theta_{n}^{(1,1)}+\hat{\phi}_{n}^{(0)}\theta_{n}^{(1,0)} \right)$ where $\theta_{n}^{(1,0)}$ is a constant and $\theta_{n}^{(1,1)}$ solves
\begin{equation}
\begin{aligned}
 &\left( \hat{\mathcal{L}}_{0} (0)+\lambda_{n}^{(0)} \right) \theta_{n}^{(1,1)}= \mathrm{Pe} u \hat{\phi}_{n}^{(0)}-2  \partial_{x}\hat{\phi}_{n}^{(0)}+\mathrm{i}\lambda_{n}^{(1)} \hat{\phi}_{n}^{(0)},  \\
 & \left. n_{x}\hat{\phi}_{n}^{(0)}+n_{x}\partial_{x}\theta_{n}^{(1,1)}+ \mathbf{n}_{\mathbf{y}}\cdot \nabla_{\mathbf{y}}\theta_{n}^{(1,1)} \right|_{\mathrm{wall}}=0.\\
\end{aligned}
\end{equation}
with $\left\langle \theta_{n}^{(1,1)},\hat{\varphi}_{n}^{(0)} \right\rangle=0$. Similarly, we assume that $\hat{\varphi}_{n}^{(1)}=\mathrm{i} \left( q_{n}^{(1,1)}+\hat{\varphi}_{n}^{(0)}q_{n}^{(1,0)} \right)$ where $q_{n}^{(1,0)}$ is a constant and $q_{n}^{(1,1)}$ solves
\begin{equation}
\begin{aligned}
 &\left( \hat{\mathcal{L}}_{0}^{*} (0)+ \left(\lambda_{n}^{(0)}  \right)^{*} \right) q_{n}^{(1,1)}=- \mathrm{Pe} u \hat{\varphi}_{n}^{(0)}-2  \partial_{x}\hat{\varphi}_{n}^{(0)}+\mathrm{i}\left( \lambda_{n}^{(1)} \right)^{*} \hat{\varphi}_{n}^{(0)},  \\
 & \left. n_{x}\hat{\varphi}_{n}^{(0)}+n_{x}\partial_{x}q_{n}^{(1,1)}+ \mathbf{n}_{\mathbf{y}}\cdot \nabla_{\mathbf{y}}q_{n}^{(1,1)} \right|_{\mathrm{wall}}=0.\\
\end{aligned}
\end{equation}
Then constraint \eqref{eq:biorthogonal system constraint} becomes
\begin{equation}
\begin{aligned}
0=&\left\langle \hat{\phi}_{n}^{(0)}, \hat{\varphi}_{n}^{(1)}\right\rangle+\left\langle \hat{\phi}_{n}^{(1)}, \hat{\varphi}_{n}^{(0)}\right\rangle =\left( \mathrm{i} q_{n}^{(1,0)} \right)^{*}+\mathrm{i} \theta_{n}^{(1,0)}. \\
\end{aligned}
\end{equation}
Therefore, we choose $\theta_{n}^{(1,0)}=q_{n}^{(1,0)}=0$. Then we have the property $\left\langle \hat{\phi}_{n}^{(1)},\hat{\varphi}_{n}^{(0)}  \right\rangle=0$.

The equation for $\lambda_{n}^{(2)}$ and $\hat{\phi}_{n}^{(2)}$ is
\begin{equation}
\begin{aligned}
  &\left( \hat{\mathcal{L}}_{0} (0)+\lambda_{n}^{(0)} \right) \hat{\phi}_{n}^{(2)}=-\lambda_{n}^{(2)} \hat{\phi}_{n}^{(0)}-\lambda_{n}^{(1)}\hat{\phi}_{n}^{(1)}+ \mathrm{i}\mathrm{Pe}  u \hat{\phi}_{n}^{(1)}-2 \mathrm{i} \partial_{x}\hat{\phi}_{n}^{(1)}+\hat{\phi}_{n}^{(0)},  \\
 & \left. n_{x}\mathrm{i}\hat{\phi}_{n}^{(1)}+n_{x}\partial_{x}\hat{\phi}_{n}^{(2)}+ \mathbf{n}_{\mathbf{y}}\cdot \nabla_{\mathbf{y}}\hat{\phi}_{n}^{(2)} \right|_{\mathrm{wall}}=0.\\  
\end{aligned}
\end{equation}
Taking the inner product with $\hat{\varphi}_{n}^{(0)}$ gives
\begin{equation}
\begin{aligned}
 \lambda_{n}^{(2)}&=1+\left\langle \left(  \mathrm{i}\mathrm{Pe}  u-\lambda_{n}^{(1)} \right) \hat{\phi}_{n}^{(1)}- \mathrm{i} \partial_{x}\hat{\phi}_{n}^{(1)},  \hat{\varphi}_{n}^{(0)}\right\rangle+\mathrm{i} \left\langle  \hat{\phi}_{n}^{(1)},\partial_{x}\hat{\varphi}_{n}^{(0)} \right\rangle.\\
\end{aligned}
\end{equation}
The equation for  $\hat{\phi}_{n}^{(2)}$ also has infinite solutions.  Due to the linearity of the equation, we consider the solution in the form $\hat{\phi}_{n}^{(2)}=- (\theta_{n}^{(2,2)}+\theta_{n}^{(2,0)}\hat{\phi}_{n}^{(0)})$ where $\theta_{n}^{(2,0)}$ is a constant. Here $\theta_{n}^{(2,2)}$ satisfies
$\left\langle \theta_{n}^{(2,2)},\hat{\varphi}_{n}^{(0)} \right\rangle=0$ and solves
\begin{equation}
\begin{aligned}
  &\left( \hat{\mathcal{L}}_{0} (0)+\lambda_{n}^{(0)} \right) \theta_{n}^{(2,2)}=\lambda_{n}^{(2)} \hat{\phi}_{n}^{(0)}+\lambda_{n}^{(1)}\hat{\phi}_{n}^{(1)}- \mathrm{i}\mathrm{Pe}  u \hat{\phi}_{n}^{(1)}+2 \mathrm{i} \partial_{x}\hat{\phi}_{n}^{(1)}-\hat{\phi}_{n}^{(0)},  \\
 & \left.- n_{x}\mathrm{i}\hat{\phi}_{n}^{(1)}+n_{x}\partial_{x}\theta_{n}^{(2,2)}+ \mathbf{n}_{\mathbf{y}}\cdot \nabla_{\mathbf{y}}\theta_{n}^{(2,2)} \right|_{\mathrm{wall}}=0.\\  
\end{aligned}
\end{equation}
For the adjoint problem, we assume that $\hat{\varphi}_{n}^{(2)}=- \left( q_{n}^{(2,2)}+\hat{\varphi}_{n}^{(0)}q_{n}^{(2,0)} \right)$ where $q_{n}^{(2,0)}$ is a constant and $q_{n}^{(2,2)}$ satisfies $\left\langle \hat{\phi}_{n}^{(0)}, q_{n}^{(2,2)}\right\rangle=0$ and solves
\begin{equation}
\begin{aligned}
 &\left( \hat{\mathcal{L}}_{0}^{*} (0)+\left( \lambda_{n}^{(0)} \right)^{*} \right) q_{n}^{(2,2)}=\left( \lambda_{n}^{(2)} \right)^{*} \hat{\varphi}_{n}^{(0)}+ \left( \lambda_{n}^{(1)} \right)^{*}\hat{\varphi}_{n}^{(1)}+ \mathrm{i}\mathrm{Pe}  u \hat{\varphi}_{n}^{(1)}+2 \mathrm{i} \partial_{x}\hat{\varphi}_{n}^{(1)}-\hat{\varphi}_{n}^{(0)},  \\
 & \left.- n_{x}\mathrm{i}\hat{\varphi}_{n}^{(1)}+n_{x}\partial_{x}q_{n}^{(2,2)}+ \mathbf{n}_{\mathbf{y}}\cdot \nabla_{\mathbf{y}}q_{n}^{(2,2)} \right|_{\mathrm{wall}}=0.\\  
\end{aligned}
\end{equation}

Then constraint \eqref{eq:biorthogonal system constraint} becomes
\begin{equation}
  \begin{aligned}
    0&= \left\langle \hat{\phi}_{n}^{(0)}, \hat{\varphi}_{n}^{(2)}\right\rangle+\left\langle \hat{\phi}_{n}^{(1)}, \hat{\varphi}_{n}^{(1)}\right\rangle+\left\langle \hat{\phi}_{n}^{(2)}, \hat{\varphi}_{n}^{(0)}\right\rangle=-\left( q_{n}^{(2,0)} \right)^{*}+\left\langle  \theta_{n}^{(1,1)}, q_{n}^{(1,1)} \right\rangle-\theta_{n}^{(2,0)}. 
\end{aligned}
\end{equation}
Therefore, we choose $\theta_{n}^{(2,0)}=\left( q_{n}^{(2,0)} \right)^{*}= \frac{1}{2}\left\langle  \theta_{n}^{(1,1)}, q_{n}^{(1,1)}\right\rangle$.

The equation for $\lambda_{n}^{(3)}$ and $\hat{\phi}_{n}^{(3)}$ is
\begin{equation}
\begin{aligned}
  &\left( \hat{\mathcal{L}}_{0} (0)+\lambda_{n}^{(0)} \right) \hat{\phi}_{n}^{(3)}=-\sum\limits_{i=1}^{3}\lambda_{n}^{(i)}\hat{\phi}_{n}^{(3-i)}+ \mathrm{i}  \mathrm{Pe} u \hat{\phi}_{n}^{(2)}-2 \mathrm{i} \partial_{x}\hat{\phi}_{n}^{(2)}+\hat{\phi}_{n}^{(1)}.  \\
\end{aligned}
\end{equation}
Taking the inner product with $\hat{\varphi}_{n}^{(0)}$ gives
\begin{equation}
\begin{aligned}
   \lambda_{n}^{(3)}&= \mathrm{i} \left\langle  \hat{\phi}_{n}^{(2)},\partial_{x}\hat{\varphi}_{n}^{(0)} \right\rangle+ \left\langle  \left( \mathrm{i}  \mathrm{Pe} u -\lambda_{n}^{(1)} \right)\hat{\phi}_{n}^{(2)}- \mathrm{i} \partial_{x}\hat{\phi}_{n}^{(2)},  \hat{\varphi}_{n}^{(0)} \right\rangle.
\end{aligned}
\end{equation}

\subsection{Eigenvalue expansion in the case with shear flows}
\label{sec:Eigenvalue expansion shear flow}
In the case of a flat channel with a shear flow,  we have $\lambda_{n}^{(0)}= (n\pi)^{2}$ and $\hat{\phi}_{n}^{(0)}=\hat{\varphi}_{n}^{(0)}=\sqrt{2}\cos n \pi y$. In addition, we have $\left( \hat{\varphi}_{n}^{(0)} \right)^{*}=\hat{\phi}_{n}^{(0)}$. Next, we compute the expansion of eigenvalue $\lambda_{1}$ for small $k$ when $u=\phi_{h}^{(0)}$ with the integer satisfying $h\geq 1$.  The following formulas will be used 
\begin{equation}
\begin{aligned}
&\hat{\phi}_{n}^{(0)}\hat{\phi}_{m}^{(0)}=2 \cos n \pi y  \cos m \pi y= \cos (n+m)\pi y+ \cos (n-m)\pi y= \frac{1}{\sqrt{2}} \left(  \hat{\phi}_{n+m}^{(0)}+\hat{\phi}_{n-m}^{(0)} \right), \\
&\hat{\phi}_{n}^{(0)}\hat{\phi}_{n}^{(0)}=2 \cos n \pi y  \cos n \pi y= \cos 2 n \pi y+ 1= \frac{1}{\sqrt{2}}  \hat{\phi}_{2n}^{(0)}+\hat{\phi}_{0}^{(0)}.  \\
\end{aligned}
\end{equation}
When $h=1$, $u=\hat{\phi}_{1}^{(0)}=\sqrt{2}\cos \pi y$, $u \hat{\phi}_{1}^{(0)}=\frac{1}{\sqrt{2}}  \hat{\phi}_{2}^{(0)}+1$ and $\lambda_{1}^{(1)}=\left\langle \mathrm{i} \mathrm{Pe}  u \hat{\phi}_{1}^{(0)} ,\hat{\varphi}_{1}^{(0)} \right\rangle  =0$. The equation for $\hat{\theta}_{1}^{(1,1)}$ becomes
\begin{equation}
\begin{aligned}
 & -\left( \partial_{y}^{2}+\lambda_{1}^{(0)} \right) \hat{\theta}_{1}^{(1,1)}=- \mathrm{Pe}\left( \frac{1}{\sqrt{2}}  \hat{\phi}_{2}^{(0)}+1 \right).  \\
\end{aligned}
\end{equation}
The solution is
\begin{equation}
\begin{aligned}
&\hat{\theta}_{1}^{(1,1)}=  \frac{\mathrm{Pe}}{\pi^{2}}  \left( \frac{-1}{3\sqrt{2}}\hat{\phi}_{2}^{(0)}+1 \right).
\end{aligned}
\end{equation}
Then we have
\begin{equation}
\begin{aligned}
& \lambda_{1}^{(2)}=1-\mathrm{Pe} \left\langle  \theta_{1}^{(1,1)} u\hat{\phi}_{1}^{(0)} \right\rangle =1-\mathrm{Pe}^{2} \left\langle \frac{1}{\pi^{2}}  \left( \frac{-1}{3\sqrt{2}}\hat{\phi}_{2}^{(0)}+1 \right) \left( \frac{1}{\sqrt{2}}  \hat{\phi}_{2}^{(0)}+1 \right) \right\rangle\\
&=1-\frac{\mathrm{Pe}^{2} }{ \pi^{2}} \left\langle \frac{-1}{6\sqrt{2}}\hat{\phi}_{2}^{(0)}+  \frac{-1}{6}+  \frac{1}{\sqrt{2}}  \hat{\phi}_{2}^{(0)} +\frac{-1}{3\sqrt{2}}\hat{\phi}_{2}^{(0)}+1\right\rangle=1-\frac{5\mathrm{Pe}^{2} }{6\pi^{2} }.
\end{aligned}
\end{equation}
Therefore, the approximation of the eigenvalue is
\begin{equation}
\begin{aligned}
&\lambda_{1}=\pi^{2}+ \left( 1-\frac{5\mathrm{Pe}^{2} }{6\pi^{2} }\right)k^{2}+ \mathcal{O} (k^{3}). 
\end{aligned}
\end{equation}
When $h= 2$,  $u \hat{\phi}_{1}^{(0)}=\frac{1}{\sqrt{2}}  \left( \hat{\phi}_{3}^{(0)}+ \hat{\phi}_{1}^{(0)} \right)$ and $\lambda_{1}^{(1)}=\left\langle \mathrm{i} \mathrm{Pe}  u \hat{\phi}_{1}^{(0)} ,\hat{\varphi}_{1}^{(0)} \right\rangle  =\frac{\mathrm{i} \mathrm{Pe}}{\sqrt{2}}$. The equation for $\hat{\theta}_{1}^{(1,1)}$ becomes
\begin{equation}
\begin{aligned}
 & -\left( \partial_{y}^{2}+\lambda_{1}^{(0)} \right) \hat{\theta}_{1}^{(1,1)}=- \frac{\mathrm{Pe}}{\sqrt{2}} \hat{\phi}_{3}^{(0)}.  \\
\end{aligned}
\end{equation}
The solution is
\begin{equation}
\begin{aligned}
&\hat{\theta}_{1}^{(1,1)}=  - \frac{\mathrm{Pe}}{8 \sqrt{2} \pi^{2}} \hat{\phi}_{3}^{(0)}.
\end{aligned}
\end{equation}
Then we have
\begin{equation}
\begin{aligned}
& \lambda_{1}^{(2)}=1-\mathrm{Pe}^{2} \left\langle  - \frac{1}{8 \sqrt{2} \pi^{2}} \hat{\phi}_{3}^{(0)}\frac{1}{\sqrt{2}}  \left( \hat{\phi}_{3}^{(0)}+ \hat{\phi}_{1}^{(0)} \right) \right\rangle\\
&=1+\frac{\mathrm{Pe}^{2} }{16 \pi^{2}} \left\langle   \frac{1}{\sqrt{2}}  \hat{\phi}_{6}^{(0)} +1 +\frac{1}{\sqrt{2}} \hat{\phi}_{4}^{(0)}+\frac{1}{\sqrt{2}} \hat{\phi}_{2}^{(0)}  \right\rangle=1+\frac{\mathrm{Pe}^{2} }{16 \pi^{2}}.
\end{aligned}
\end{equation}
Therefore, the approximation of the eigenvalue is
\begin{equation}
\begin{aligned}
&\lambda_{1}=\pi^{2}+ \frac{\mathrm{i} \mathrm{Pe}k}{\sqrt{2}}+  \left( 1+\frac{\mathrm{Pe}^{2} }{16 \pi^{2}} \right)k^{2}+ \mathcal{O} (k^{3}). 
\end{aligned}
\end{equation}

When $h\geq 3$,  $u \hat{\phi}_{1}^{(0)}=\frac{1}{\sqrt{2}}  \left( \hat{\phi}_{h+1}^{(0)}+ \hat{\phi}_{h-1}^{(0)} \right)$ and $\lambda_{1}^{(1)}=\left\langle  \mathrm{i} \mathrm{Pe} u \hat{\phi}_{1}^{(0)} ,\hat{\varphi}_{1}^{(0)} \right\rangle  =0$. The equation for $\hat{\theta}_{1}^{(1,1)}$ becomes
\begin{equation}
\begin{aligned}
 & -\left( \partial_{y}^{2}+\lambda_{1}^{(0)} \right) \hat{\theta}_{1}^{(1,1)}=- \frac{\mathrm{Pe}}{\sqrt{2}}  \left( \hat{\phi}_{h+1}^{(0)}+ \hat{\phi}_{h-1}^{(0)} \right).  \\
\end{aligned}
\end{equation}
The solution is
\begin{equation}
\begin{aligned}
&\hat{\theta}_{1}^{(1,1)}=  \frac{-\mathrm{Pe}}{\sqrt{2}\pi^{2}}  \left( \frac{\hat{\phi}_{h+1}^{(0)}}{(h+1)^{2}-1}+ \frac{\hat{\phi}_{h-1}^{(0)}}{(h-1)^{2}-1} \right).
\end{aligned}
\end{equation}
Then we have
\begin{equation}
\begin{aligned}
& \lambda_{1}^{(2)}=1-\mathrm{Pe} \left\langle  \theta_{1}^{(1,1)} u\hat{\phi}_{1}^{(0)} \right\rangle \\
&=1-\mathrm{Pe}^{2} \left\langle  \frac{-1}{\sqrt{2}\pi^{2}}  \left( \frac{\hat{\phi}_{h+1}^{(0)}}{(h+1)^{2}-1}+ \frac{\hat{\phi}_{h-1}^{(0)}}{(h-1)^{2}-1} \right) \frac{1}{\sqrt{2}}  \left( \hat{\phi}_{h+1}^{(0)}+ \hat{\phi}_{h-1}^{(0)} \right) \right\rangle\\
&=1+\frac{\mathrm{Pe}^{2} }{2 \pi^{2}} \left\langle    \left( \frac{\hat{\phi}_{h+1}^{(0)}}{(h+1)^{2}-1}+ \frac{\hat{\phi}_{h-1}^{(0)}}{(h-1)^{2}-1} \right)   \left( \hat{\phi}_{h+1}^{(0)}+ \hat{\phi}_{h-1}^{(0)} \right) \right\rangle\\
&=1+\frac{\mathrm{Pe}^{2} }{2\pi^{2} } \left( \frac{1}{(h+1)^{2}-1}+ \frac{1}{(h-1)^{2}-1}   \right).
\end{aligned}
\end{equation}
Therefore, the approximation of the eigenvalue is
\begin{equation}
\begin{aligned}
&\lambda_{1}=\pi^{2}+ \left( 1+\frac{\mathrm{Pe}^{2} }{2\pi^{2} } \left( \frac{1}{(h+1)^{2}-1}+ \frac{1}{(h-1)^{2}-1}   \right) \right)k^{2}+ \mathcal{O} (k^{3}). 
\end{aligned}
\end{equation}
\subsection{Asymptotic expansion in the limit of large channel length}
\label{sec:Asymptotic expansion}

In this section, we first consider the eigenvalue problem on a channel domain of finite length $L$, and then compute the asymptotic expansion of the eigenvalue and eigenfunction in the limit as $L$ approaches infinity. This is actually my original method for tackling this problem, and the results inspire the form of the eigenfunction in equation \eqref{eq:eigenvalue problem wavenumber}.

We denote $\lambda(L)$ and $\phi(x, \mathbf{y}, L)$ as the eigenvalues and eigenfunctions of the eigenvalue problem \eqref{eq:eigenvalue problem} on the domain $\Omega(L)$. We also denote $\lambda^{*}(L)$ and $\varphi(x, \mathbf{y}, L)$ as the eigenvalues and eigenfunctions of the adjoint operator in the adjoint eigenvalue problem \eqref{eq:adjoint eigenvalue problem} on the same domain. Here, we choose $\phi_n$ and $\varphi_n$ such that the sets $\{\phi_n\}_{n=0}^{\infty}$ and $\{\varphi_n\}_{n=0}^{\infty}$ form a biorthogonal system, satisfying $\langle \phi_n, \varphi_m \rangle = \delta_{n,m}$, where $\delta_{n,m}$ is the Kronecker delta. With this definition of the inner product and the properties of the biorthogonal system, we have $c_n = \langle c_I(x, \mathbf{y}), \varphi_n(x, \mathbf{y}) \rangle$.

For the advection-diffusion equation $\partial_t c = \mathcal{L} c$ on the domain $\Omega(L)$, the solution admits an eigenfunction expansion:
\begin{equation}
c(x, \mathbf{y}, t) = \sum_{n=0}^{\infty} \left( c_n \phi_n(x, \mathbf{y}) e^{-\lambda_n t} + c_n^{*} \phi_n^{*}(x, \mathbf{y}) e^{-\lambda_n^{*} t} \right) = \sum_{n=0}^{\infty} 2 \Re \left( c_n \phi_n(x, \mathbf{y}) e^{-\lambda_n t} \right),
\end{equation}
where $0 = \lambda_0 < \Re \lambda_1 < \ldots < \Re \lambda_n < \ldots$ are the eigenvalues with nonnegative imaginary parts, and $\phi_n$ are the associated eigenfunctions.

To approximate the scalar field in an infinitely long channel domain, we need to understand the structure of the eigenvalues in the limit as $L \rightarrow \infty$. Let's consider a simple but explicit example. When $a=0$, $u=v=0$, and $\kappa=1$, we have a pure diffusion equation. It is straightforward to verify that $\lambda_{n,m} = \left( \frac{2\pi m}{L} \right)^{2} + (n\pi)^{2}$ and $\phi_{n,m} = \exp \left( \frac{\mathrm{i} 2\pi m x}{L} \right) \cos(\pi n y)$. We observe that in the limit of infinite channel length, $\lambda_{n,0} = (n\pi)^{2}$ are cluster points; namely, for each $\lambda_{n,0}$, there exists a sequence $\lambda_{n,m} = \left( \frac{2\pi m}{L} \right)^{2} + (n\pi)^{2}$ that converges to $\lambda_{n,0}$. In general, $\lambda_{n,0}$ are the eigenvalues of the operator on the unit domain, $\Omega(L_{p}) = \{(x,\mathbf{y}) \mid -\frac{L_{p}}{2} \leq x \leq \frac{L_{p}}{2}, \mathbf{y} \in \Omega_{c}(x) \}$. The eigenvalues of the operator on $\Omega(L_{p})$ are also the eigenvalues of the operator on a larger domain $\Omega(L)$ if $L$ is a multiple of $L_{p}$. As the channel length increases, more and more eigenvalues fill the gaps between $\lambda_{n,0}$. In the case of an infinite-length channel, the spectrum of the operator $\mathcal{L}$ becomes continuous.

Therefore, we use two indices to label the eigenvalues and take the limit as the channel length approaches infinity:

\begin{equation}\label{eq:eigenfunction expansion L}
\begin{aligned}
&c(x, \mathbf{y}, t) = \lim_{L \rightarrow \infty} \sum_{n,m}^{\infty} 2 \Re \left( \left\langle c_{I}(x, \mathbf{y}), \varphi_{n,m}(x, \mathbf{y}, L) \right\rangle \phi_{n,m}(x, \mathbf{y}) e^{-\lambda_{n,m}(L) t} \right) \\
&= \lim_{L \rightarrow \infty} \sum_{n,m}^{\infty} 2 \Re \left( \phi_{n,m}(x, \mathbf{y}, L) e^{-\lambda_{n,m}(L) t} \frac{1}{|\Omega (L_{p})|} \int_{\Omega} c_{I}(x, \mathbf{y}) \varphi_{n,m}^{*}(x, \mathbf{y}, L) \mathrm{d}x \mathrm{d}\mathbf{y} \right) \\
&= \frac{1}{|\Omega(L_{p})|} \sum_{n=0}^{\infty} \int_{0}^{\infty} 2 \Re \left( \phi_{n}(x, \mathbf{y}, k) e^{-\lambda_{n}(k) t} \left( \int_{\Omega} c_{I}(x, \mathbf{y}) \varphi_{n}^{*}(x, \mathbf{y}, k) \mathrm{d}x \mathrm{d}\mathbf{y} \right) \right) \mathrm{d}k,
\end{aligned}
\end{equation}
where $k = \frac{2 \pi m}{L}$, $\phi_{n}(x, \mathbf{y}, k) = \lim_{L \rightarrow \infty} \phi_{n,m}(x, \mathbf{y}, L)$, and $\lambda_{n}(k) = \lim_{L \rightarrow \infty} \lambda_{n,m}(L)$. The variable $k$ resembles the wavenumber in the Fourier transform. Since $\lambda_{n}(k)^{*} = \lambda_{n}(-k)$, equation \eqref{eq:eigenfunction expansion L} is equivalent to equation \eqref{eq:eigenfunction expansion}.

The asymptotic expansion of equation \eqref{eq:eigenfunctionExpansion0th} involves the derivatives of the eigenvalue and eigenfunction with respect to $k$ at $k=0$. We can compute $\lambda_{n}'(0)$ using the definition of the derivative:
\begin{equation}
\begin{aligned}
&\lambda_{n}'(0)= \lim\limits_{k\rightarrow 0}\frac{\lambda_{n} (k)-\lambda_{n} (0)}{k} =\lim\limits_{L\rightarrow \infty} \frac{\lambda_{n,m}(,L)- \lambda_{n,0} (L)}{\frac{2 \pi m}{L}}=\lim\limits_{L\rightarrow \infty} \frac{\lambda_{n,1}(,L)- \lambda_{n,0} (L)}{\frac{2 \pi }{L}}. \\
\end{aligned}
\end{equation}
The second step uses the definition $k = \frac{2 \pi m}{L}$. Therefore, in order to compute the derivative of $\lambda_{0}(k)$ at $k=0$, we need to determine the asymptotic expansion of the smallest non-zero eigenvalue $\lambda_{0,1}(L)$ as $L \rightarrow \infty$.

To find the asymptotic expansion of the smallest nonzero eigenvalue and corresponding eigenfunction when the interval length approaches infinity, one may consider the a regular power series expansion
\begin{equation}
\begin{aligned}
&\lambda_{n,m}=\sum\limits_{i=0}^{\infty}L^{-i} \lambda_{n,m}^{(i)}, \quad \phi_{n,m} (x,\mathbf{y}) =\sum\limits_{i=0}^{\infty}L^{-i}\phi_{n,m}^{(i)} (x,\mathbf{y}). \\
\end{aligned}
\end{equation}
The expansion for the eigenvalue is fine. However, the expansion of the eigenfunction may not provide a desirable approximation. For example, in the flat channel, the eigenvalue problem associated with the pure diffusion equation has the smallest nonzero eigenvalue  $ \left( \frac{2\pi}{L}  \right)^{2}$ and corresponding eigenfunction $e^{ \frac{\mathrm{i} 2 \pi x}{L} }$. The regular power series expansion of this eigenfunction is $e^{\frac{\mathrm{i} 2 \pi x}{L} }=1+  \frac{\mathrm{i} 2 \pi x}{L} - \frac{1}{2}(\frac{ 2 \pi x}{L})^{2}+ \mathcal{O} \left( L^{-3} \right)$. Obviously, this asymptotic expansion doesn't converge uniformly for all $x$. 

Inspired by \cite{rosencrans1997taylor},  we consider a multiscale expansion to capture the slowly varying part of the function and obtain a uniform asymptotic expansion
\begin{equation}\label{eq:multiscale expansion}
\begin{aligned}
  &\phi_{n,m} (x,\mathbf{y}) =\phi_{n,m}^{(0)} \left( x, \frac{x}{L}, \mathbf{y} \right)+ L^{-1} \phi_{n,m}^{(0)}\left( x, \frac{x}{L}, \mathbf{y} \right)+L^{-2}\phi_{n,m}^{(0)} \left( x, \frac{x}{L}, \mathbf{y} \right)+ \hdots, \\
  &\varphi_{n,m} (x,\mathbf{y}) =\varphi_{n,m}^{(0)} \left( x, \frac{x}{L}, \mathbf{y} \right)+ L^{-1} \varphi_{n,m}^{(0)}\left( x, \frac{x}{L}, \mathbf{y} \right)+L^{-2}\varphi_{n,m}^{(0)} \left( x, \frac{x}{L}, \mathbf{y} \right)+ \hdots \\
\end{aligned}
\end{equation}
For convenience, we denote $\xi=\frac{x}{L}$. With the chain rule, the differentiation operator $\partial_{x}$  leads to the operator $\partial_{x} + \frac{1}{L}\partial_{\xi}$ when we substitute the expansion into the equation.  We also assume that $\phi_{n,m}^{(i)}$ has a fundamental period $1$ in $\xi$ and a fundamental period $L_{p}$ in $x$.  For convenience, we define the following operators
\begin{equation}
\begin{aligned}
  &\mathcal{L}_{0} =  -\mathrm{Pe}u (x,\mathbf{y})\partial_{x}+ \mathrm{Pe} \mathbf{v}(x,\mathbf{y}) \cdot \nabla_{\mathbf{y}} +\partial_{x}^{2}+\Delta_{\mathbf{y}},  \\
  &\mathcal{L}_{0}^{*} =  \mathrm{Pe}u (x,\mathbf{y})\partial_{x}+ \mathrm{Pe} \mathbf{v}(x,\mathbf{y}) \cdot \nabla_{\mathbf{y}}+ \partial_{x}^{2}+\Delta_{\mathbf{y}},  \\
  &\mathcal{L}_{1}=-\mathrm{Pe} u (x,\mathbf{y})\partial_{\xi}+2 \partial_{x}\partial_{\xi}, \quad \mathcal{L}_{1}^{*}=\mathrm{Pe} u (x,\mathbf{y})\partial_{\xi}+2 \partial_{x}\partial_{\xi}, \\
&\mathcal{L}_{2}= \partial_{\xi}^{2},\quad \mathcal{L}_{2}^{*}=\mathcal{L}_{2}.
\end{aligned}
\end{equation}
Substituting the above expansions into the eigenvalue problem yields a hierarchy of equations:
\begin{equation}
\begin{aligned}
&\mathcal{L}_{0}\phi_{n,m}^{(i)}=- \sum\limits_{j=0}^{i}\lambda_{n,m}^{(j)}\phi_{n,m}^{(i-j)}-\mathcal{L}_{1}\phi_{n,m}^{(i-1)}-\mathcal{L}_{2}\phi_{n,m}^{(i-2)}, \quad i=0,1,2,..., \\
&\mathcal{L}_{0}^{*}\varphi_{n,m}^{(i)}=- \sum\limits_{j=0}^{i}\left( \lambda_{n,m}^{(j)} \right)^{*}\varphi_{n,m}^{(i-j)}-\mathcal{L}_{1}^{*}\varphi_{n,m}^{(i-1)}-\mathcal{L}_{2}^{*}\varphi_{n,m}^{(i-2)}, \quad i=0,1,2,..., 
\end{aligned}
\end{equation}
where $\phi_{n,m}^{(i)}=\varphi_{n,m}^{(i)}=0$ if  $i< 0$. An additional constraint arises from the orthogonality condition $\left\langle \phi_{n,m}, \varphi_{h,l} \right\rangle= \delta_{n,h}\delta_{m,l}$. Since the expansions are valid  for arbitrary large $L$, we have
 \begin{equation}\label{eq:unit norm condition}
\begin{aligned}
  &\left\langle\phi_{n,m}^{(0)}, \varphi_{n,m}^{(0)}  \right\rangle=1, \quad \left\langle\phi_{n,m}^{(0)}, \varphi_{h,l}^{(0)}  \right\rangle=0,\quad  \sum\limits_{j=0}^{i}\left\langle \phi_{n,m}^{(j)}, \varphi_{h,l}^{(j)} \right\rangle=0, \quad i=1,2,3... 
\end{aligned}
\end{equation}
There remains a degree of freedom since multiplying an eigenfunction by a constant yields another valid eigenfunction. However, any such choice provides a satisfactory eigenfunction expansion of the scalar field. In the following calculations, we will select the one with the simplest expression.

The integrals in equation \eqref{eq:unit norm condition} depend on $L$. As we focus on the limit of large $L$, we can simplify the asymptotic calculations using the following proposition related to the Riemann–Lebesgue lemma.
\begin{proposition}
Assume that $f (x, \xi)$ is periodic with period $1$ in $x$ and period $1$ in $\xi$. If the $k$-th order partial derivatives of $f$ exists and $ \left. \partial_{\xi}^{k}f  \right|_{\xi=-\frac{L}{2} }= \left. \partial_{\xi}^{k}f  \right|_{\xi=\frac{L}{2} }$, for $k=0,1,... ,\alpha$, then as the integer $L\rightarrow \infty$, we have
  \begin{equation}
\begin{aligned}
&\frac{1}{L}\int\limits_{-\frac{L}{2}}^{\frac{L}{2}}f \left( x, \frac{x}{L} \right)\mathrm{d} x=\int\limits_{-\frac{1}{2}}^{\frac{1}{2}}f \left( Lx, x \right)\mathrm{d} x=\int\limits_{-\frac{1}{2}}^{\frac{1}{2}} \int\limits_{-\frac{1}{2}}^{\frac{1}{2}}f (x, \xi)\mathrm{d} x \mathrm{d} \xi+ \mathcal{O} (L^{-\alpha}). \\
\end{aligned}
\end{equation}

\end{proposition}
If $ \left. \partial_{\xi}^{k} f \right|_{\xi=-\frac{L}{2}} = \left. \partial_{\xi}^{k} f \right|_{\xi=\frac{L}{2}}$ for all nonnegative integers $k$, we can approximate the integral on the left-hand side of the above equation by the double integral on the right-hand side, with a correction term that is asymptotically smaller than any power of $L$. This leads us to introduce the following notation:
\begin{equation}
\begin{aligned}
& \left\langle f(x, \xi, \mathbf{y}), g(x, \xi, \mathbf{y}) \right\rangle_{x} = \frac{1}{L_{p} |\Omega_{c}|} \int_{\Omega(L_{p})} f(x, \xi, \mathbf{y}) g^{*}(x, \xi, \mathbf{y}) \mathrm{d} x \mathrm{d} \mathbf{y}, \\
&\left\langle f(x, \xi, \mathbf{y}) \right\rangle_{\xi} = \int_{-\frac{1}{2}}^{\frac{1}{2}} f(x, \xi, \mathbf{y}) \mathrm{d} \xi.
\end{aligned}
\end{equation}
We then approximate the inner products in expansion \eqref{eq:unit norm condition} as follows:
\begin{equation}
\begin{aligned}
& \left\langle f, g \right\rangle = \left\langle \left\langle f, g \right\rangle_{x} \right\rangle_{\xi}.
\end{aligned}
\end{equation}

To simplify the expression, we use the notation \( \left\langle f \right\rangle_{x} = \left\langle f, 1 \right\rangle_{x} \) to denote the average of the function with respect to \( x \) and \( \mathbf{y} \) over \( \Omega(L_{p}) \).

 The equation for $\lambda_{n,m}^{(0)}$ and $\phi_{n,m}^{(0)}$ is
\begin{equation}
\begin{aligned}
  & -\mathrm{Pe}\left( u \partial_{x}\phi_{n,m}^{(0)}+ \mathbf{v}_{\mathbf{y}}\cdot \nabla_{\mathbf{y}} \phi_{n,m}^{(0)} \right)+\partial_{x}^{2}\phi_{n,m}^{(0)}+\Delta_{\mathbf{y}}\phi_{n,m}^{(0)}=-\lambda_{n,m}^{(0)}\phi_{n,m}^{(0)}=0, \\
  &\left. n_{x}\partial_{x}\phi_{n,m}^{(0)}+ \mathbf{n}_{\mathbf{y}}\cdot \nabla_{\mathbf{y}}\phi_{n,m}^{(0)} \right|_{\mathrm{wall}}=0.
\end{aligned}
\end{equation}
For $n=0$, the solution is $\lambda_{0,m}^{(0)}=0$, $\phi_{0,m}^{(0)} (x,\xi,\mathbf{y})=\phi_{0,n}^{(0)} (\xi)$. The equation for $\lambda_{n,m}^{(1)}$ and $\phi_{n,m}^{(1)}$ is
\begin{equation}
\begin{aligned}
 &\mathcal{L}_{0}\phi_{n,m}^{(1)}=\mathrm{Pe}u \partial_{\xi}\phi_{n,m}^{(0)} -2\partial_{x}\partial_{\xi}\phi_{n,m}^{(0)}- \lambda_{n,m}^{(0)}\phi_{n,m}^{(1)}-\lambda_{n,m}^{(1)}\phi_{n,m}^{(0)},  \\
 &\left. n_{x}\partial_{\xi}\phi_{n,m}^{(0)}+n_{x}\partial_{x}\phi_{n,m}^{(1)}+ \mathbf{n}_{\mathbf{y}}\cdot \nabla_{\mathbf{y}}\phi_{n,m}^{(1)} \right|_{\mathrm{wall}}=0.  \\
\end{aligned}
\end{equation}
If $n>0$, the calculation is relatively complicated and is irreverent to the asymptotic expansion of equation \eqref{eq:eigenfunctionExpansion0th} at long times. Therefore, we first focus on the case $n=0$:
\begin{equation}\label{eq:order1 n0}
\begin{aligned}
 &-\mathrm{Pe}\left(u \partial_{x}\phi_{0,m}^{(1)}+\mathbf{v}_{\mathbf{y}}\cdot \nabla \phi_{0,m}^{(1)}  \right)+\partial_{x}^{2}\phi_{0,m}^{(1)}+\Delta_{\mathbf{y}} \phi_{0,m}^{(1)} =-\lambda_{0,m}^{(1)}\phi_{0,m}^{(0)}+ \mathrm{Pe}u \partial_{\xi}\phi_{0,m}^{(0)},  \\
    &\left.  n_{x}\partial_{\xi}\phi_{0,m}^{(0)}+n_{x}\partial_{x}\phi_{0,m}^{(1)}+ \mathbf{n}_{\mathbf{y}}\cdot \nabla_{\mathbf{y}}\phi_{0,m}^{(1)} \right|_{\mathrm{wall}}=0.
\end{aligned}
\end{equation}
Averaging with respect to  $x,y$, the left hand side becomes
\begin{equation}
\begin{aligned}
\left\langle \mathcal{L}_{0}\phi_{0,m}^{(1)}  \right\rangle&= \left\langle -\mathrm{Pe}\left(u \partial_{x}\phi_{0,m}^{(1)}+\mathbf{v}_{\mathbf{y}}\cdot \nabla \phi_{0,m}^{(1)}  \right)+\partial_{x}^{2}\phi_{0,m}^{(1)}+\Delta_{\mathbf{y}} \phi_{0,m}^{(1)}  \right\rangle \\
&=  \left\langle  -\mathrm{Pe} \left( \partial_{x} , \nabla_{\mathbf{y}} \right) \cdot \left( \phi_{0,m}^{(1)} u, \phi_{0,m}^{(1)} \mathbf{v} \right)  +   \left(  \partial_{\xi}^{2}+\Delta_{\mathbf{y}} \right)\phi_{0,m}^{(1)}  \right\rangle  \\
&=\frac{1 }{|\Omega_{c}|L_{p}} \left(   -\mathrm{Pe}   \int\limits_{\mathrm{wall}}^{} n_{x}\phi_{0,m}^{(1)} u+ n_{\mathbf{y}} \cdot (\phi_{0,m}^{(1)} \mathbf{v}) \mathrm{d} x \mathrm{d} \mathbf{y} - \int\limits_{\mathrm{wall}}^{} \partial_{\xi}\phi_{0,m}^{(0)} n_{x} \mathrm{d} x \mathrm{d} \mathbf{y} \right)  \\
&= -\frac{\partial_{\xi}\phi_{0,m}^{(0)} }{|\Omega_{c}|L_{p}}    \int\limits_{\mathrm{wall}}^{} (1,\mathbf{0})\cdot (n_{x}, n_{\mathbf{y}}) \mathrm{d} x\mathrm{d} \mathbf{y}\\
&=0.
\end{aligned}
\end{equation}
The second step follows from  the divergence theorem, the incompressibility of the flow,  and the no-flux boundary condition of $\phi_{0,m}^{(1)}$. In the third step, we use the vanishing boundary condition of the flow. In the last step, we use the divergence theorem again. Therefore, we have
\begin{equation}
\begin{aligned}
&0=-\lambda_{0,m}^{(1)}\phi_{0,m}^{(0)}+ \mathrm{Pe}\left\langle u \right\rangle \partial_{\xi}\phi_{0,m}^{(0)}.\\
\end{aligned}
\end{equation}
The solution is $\lambda_{0,m}^{(1)}=\mathrm{i} 2 \pi m \mathrm{Pe} \left\langle u \right\rangle $, $\phi_{0,m}^{(0)}=e^{\mathrm{i} 2\pi m\xi}$. Certainly, $\phi_{0,m}^{(0)}$ multiplied with a nonzero constant is still a solution. For simplicity, we choose the expression of $\phi_{0,m}^{(0)}$ defined here. With a similar calculation, for the adjoint problem, we have
\begin{equation}
\begin{aligned}
&0=-\left( \lambda_{0,m}^{(1)} \right)^{*}\phi_{0,m}^{(0)}- \mathrm{Pe}\left\langle u \right\rangle \partial_{\xi}\phi_{0,m}^{(0)}.
\end{aligned}
\end{equation}
The solution is  $\left( \lambda_{0,m}^{(1)} \right)^{*}=-\mathrm{i} 2 \pi m \mathrm{Pe} \left\langle u \right\rangle$, $\varphi_{0,m}^{(0)}=\phi_{0,m}^{(0)}=e^{\mathrm{i} 2\pi m\xi}$.

%\textbf{With the condition $\left\langle\phi_{n,m}^{(0)},\phi_{n,m}^{(0)}  \right\rangle=1$, Notice that their complex conjugate is another solution. However, we choose this expression for $\lambda_{0,m}^{(0)}$ since based on our definition, $\lambda_{n,m}$ is the one with positive imaginary part.}

Then equation \eqref{eq:order1 n0} becomes
\begin{equation}
\begin{aligned}
& -\mathrm{Pe} \left( u \partial_{x}\phi_{0,m}^{(1)}+\mathbf{v}_{\mathbf{y}}\cdot \nabla \phi_{0,m}^{(1)}\right)+\partial_{x}^{2}\phi_{0,m}^{(1)}+\Delta_{\mathbf{y}}\phi_{0,m}^{(1)}= \mathrm{Pe}\left( u- \left\langle u \right\rangle \right) \partial_{\xi}\phi_{0,m}^{(0)} . \\
\end{aligned}
\end{equation}
Due to the linearity of the equation, we can assume that $\phi_{0,m}^{(1)}= \theta_{0,m}^{(1,1)} (x,y)\partial_{\xi}\phi_{0,m}^{(0)}+\theta_{0,m}^{(1,0)} (\xi)$ where $\theta_{0,m}^{(1,1)}$ satisfies $\left\langle \theta_{0,m}^{(1,1)} \right\rangle=0$ and solves
\begin{equation}\label{eq:theta11}
\begin{aligned}
&-\mathrm{Pe}\left( u \partial_{x}\theta_{0,m}^{(1,1)}+  \mathbf{v}_{\mathbf{y}}\cdot \nabla \theta_{0,m}^{(1,1)} \right)+\partial_{x}^{2}\theta_{0,m}^{(1,1)}+\Delta_{\mathbf{y}}\theta_{0,m}^{(1,1)} =\mathrm{Pe}  \left( u- \left\langle u \right\rangle \right), \\
   &\left. n_{x}+n_{x}\partial_{x}\theta_{0,m}^{(1,1)}+ \mathbf{n}_{\mathbf{y}}\cdot \nabla_{\mathbf{y}}\theta_{0,m}^{(1,1)} \right|_{\mathrm{wall}}=0, \quad \left. \theta_{0,m}^{(1,1)} \right|_{x=-\frac{L_{p}}{2}} = \left. \theta_{0,m}^{(1,1)} \right|_{x=\frac{L_{p}}{2}}.  \\
\end{aligned}
\end{equation}
Similarly, we can assume that $\varphi_{0,m}^{(1)}= q_{0,m}^{(1,1)} (x,y)\partial_{\xi}\varphi_{0,m}^{(0)}+q_{0,m}^{(1,0)} (\xi)$ where $q_{0,m}^{(1,1)}$ satisfies $\left\langle q_{0,m}^{(1,1)} \right\rangle=0$ and solves
\begin{equation}\label{eq:q11}
\begin{aligned}
 &-\mathrm{Pe}\left( u \partial_{x}q_{0,m}^{(1,1)}+  \mathbf{v}_{\mathbf{y}}\cdot \nabla q_{0,m}^{(1,1)} \right)+\partial_{x}^{2}q_{0,m}^{(1,1)}+\Delta_{\mathbf{y}}q_{0,m}^{(1,1)} =-\mathrm{Pe}  \left( u- \left\langle u \right\rangle \right), \\
   &\left. n_{x}+n_{x}\partial_{x}q_{0,m}^{(1,1)}+ \mathbf{n}_{\mathbf{y}}\cdot \nabla_{\mathbf{y}}q_{0,m}^{(1,1)} \right|_{\mathrm{wall}}=0, \quad \left. q_{0,m}^{(1,1)} \right|_{x=-\frac{L_{p}}{2}} = \left. q_{0,m}^{(1,1)} \right|_{x=\frac{L_{p}}{2}}.  \\
\end{aligned}
\end{equation}
When $i=1$, equation \eqref{eq:unit norm condition} becomes 
\begin{equation}
  \begin{aligned}
  &0=\left\langle \phi_{0,m}^{(1)}, \varphi_{0,l}^{(0)} \right\rangle+\left\langle \phi_{0,m}^{(0)}, \varphi_{0,l}^{(1)} \right\rangle =\theta_{0,m}^{(1,0)} (\xi) e^{-\mathrm{i} 2\pi l\xi} + q_{0,l}^{(1,0)} (\xi) e^{\mathrm{i} 2\pi m\xi}
 .
\end{aligned}
\end{equation}
If $m \neq l$, additional averaging with respect to $\xi$ leads to zero. To make this expression to be zero when  $m=l$,  we have to choose  $\theta_{0,m}^{(1,0)} =q_{0,m}^{(1,0)} =0$.

The equation for $\lambda_{n,m}^{(2)}$ and $\phi_{n,m}^{(2)}$ is
\begin{equation}\label{eq:order 2}
\begin{aligned}
 &\mathcal{L}_{0}\phi_{n,m}^{(2)}=-\lambda_{n,m}^{(0)}\phi_{n,m}^{(2)}-\lambda_{n,m}^{(1)}\phi_{n,m}^{(1)}-\lambda_{n,m}^{(2)}\phi_{n,m}^{(0)}+\mathrm{Pe}u \partial_{\xi}\phi_{n,m}^{(1)} - 2\partial_{x}\partial_{\xi}\phi_{n,m}^{(1)}-\partial_{\xi}^{2}\phi_{n,m}^{(0)},  \\
&\left. n_{x}\partial_{\xi}\phi_{n,m}^{(1)}+n_{x}\partial_{x}\phi_{n,m}^{(2)}+ \mathbf{n}_{\mathbf{y}}\cdot \nabla_{\mathbf{y}}\phi_{n,m}^{(2)} \right|_{\mathrm{wall}}=0, \quad \left. \phi_{n,m}^{(2)} \right|_{x=-\frac{L_{p}}{2}} = \left. \phi_{n,m}^{(2)} \right|_{x=\frac{L_{p}}{2}}.  \\
\end{aligned}
\end{equation}
A worth noting observation is that
\begin{equation}
\begin{aligned}
  &\lambda_{0,m}^{(1)}\phi_{0,m}^{(1)}= \lambda_{0,m}^{(1)}\theta_{0,m}^{(1,1)} \partial_{\xi}\phi_{0,m}^{(0)}+\lambda_{0,m}^{(1)}\theta_{0,m}^{(1,0)} (\xi)\\
  &= \mathrm{Pe}\left\langle u \right\rangle \left( \theta_{0,m}^{(1,1)}  \partial_{\xi}^{2}\phi_{0,m}^{(0)} +  \theta_{0,m}^{(1,0)} \partial_{\xi}\phi_{0,m}^{(0)} \right)= \mathrm{Pe}\left\langle u \right\rangle  \partial_{\xi}\phi_{0,m}^{(1)}.
\end{aligned}
\end{equation}
Therefore, when $n=0$, equation \eqref{eq:order 2} simplifies to
\begin{equation}
\begin{aligned}
&\mathcal{L}_{0}\phi_{0,m}^{(2)}= -\lambda_{0,m}^{(2)}\phi_{0,m}^{(0)}+\mathrm{Pe} \left( u- \left\langle u \right\rangle \right) \partial_{\xi}\phi_{0,m}^{(1)} -2\partial_{x}\partial_{\xi}\phi_{0,m}^{(1)}-\partial_{\xi}^{2}\phi_{0,m}^{(0)}.  \\
\end{aligned}
\end{equation}
Averaging the above equation with respect to  $x,y$,  the left hand side becomes
\begin{equation}\label{eq:order 2 LHS}
\begin{aligned}
\left\langle \mathcal{L}_{0}\phi_{0,m}^{(2)}\right\rangle&= \left\langle  -\mathrm{Pe}\left( u \partial_{x}\phi_{0,m}^{2}+\mathbf{v}_{\mathbf{y}}\cdot \nabla \phi_{0,m}^{2}\right)+\partial_{x}^{2}\phi_{0,m}^{2}+\Delta_{\mathbf{y}}\phi_{0,m}^{2} \right\rangle\\
 &= \left\langle (\partial_{x}, \nabla_{\mathbf{y}})\cdot (-\mathrm{Pe}u\phi_{0,m}^{2}+\partial_{x}\phi_{0,m}^{2}, -\mathrm{Pe}\mathbf{v}\phi_{0,m}^{2}+\nabla_{\mathbf{y}}\phi_{0,m}^{2} ) \right\rangle \\
&= \frac{-1}{|\Omega_{c}|L_{p}}\int\limits_{\mathrm{wall}}^{} n_{x} \partial_{\xi}\phi_{0,m}^{(1)} \mathrm{d} x \mathrm{d} \mathbf{y}=-\left\langle \partial_{\xi} \partial_{x}\phi_{0,m}^{(1)} \right\rangle
=- \partial_{\xi}^{2}\phi_{0,m}^{(0)}  \left\langle \partial_{x} \theta_{0,m}^{(1,1)} \right\rangle,\\
\end{aligned}
\end{equation}
where the first step follows from the incompressibility of the flow. The second step uses the divergence theorem and the flux boundary of $\phi_{0,m}^{(2)}$. Therefore, we have
\begin{equation}
\begin{aligned}
  \lambda_{0,m}^{(2)}\phi_{0,m}^{(0)}&= \left\langle \partial_{x} \theta_{0,m}^{(1,1)} \right\rangle\partial_{\xi}^{2}\phi_{0,m}^{(0)}+\mathrm{Pe}\left\langle \left( u- \left\langle u \right\rangle \right) \theta_{0,m}^{(1,1)} \right\rangle \partial_{\xi}^{2}\phi_{0,m}^{(0)}-\partial_{\xi}^{2}\phi_{0,m}^{(0)}  -2 \left\langle \partial_{x} \theta_{0,m}^{(1,1)} \right\rangle\partial_{\xi}^{2}\phi_{0,m}^{(0)}\\
&=- \left( 1- \mathrm{Pe}\left\langle \left( u- \left\langle u \right\rangle \right) \theta_{0,m}^{(1,1)} \right\rangle+\left\langle \partial_{x} \theta_{0,m}^{(1,1)} \right\rangle\ \right)\partial_{\xi}^{2}\phi_{0,m}^{(0)}.
\end{aligned}
\end{equation}
Substituting the expression of $\phi_{0,m}^{(0)}$ leads to 
\begin{equation}
\begin{aligned}
  \lambda_{0,m}^{(2)}&=4 \pi^{2}m^{2}\left\langle1-  \mathrm{Pe} \left( u- \left\langle u \right\rangle \right) \theta_{0,m}^{(1,1)} + \partial_{x} \theta_{0,m}^{(1,1)} \right\rangle\\
  &=4 \pi^{2}m^{2}\left\langle \nabla_{\mathbf{y}} \theta_{0,m}^{(1,1)}\cdot \nabla_{\mathbf{y}} \theta_{0,m}^{(1,1)}+ (1+ \partial_{x}\theta_{0,m}^{(1,1)})^{2} \right\rangle\geq 0, \\
\end{aligned}
\end{equation}
where the second step follows from equation \eqref{eq:theta11} and integration by parts.

Then the equation for $\lambda_{0,m}^{(2)}$ and $\phi_{0,m}^{(2)}$ becomes
\begin{equation}
\begin{aligned}
  &\mathcal{L}_{0}\phi_{0,m}^{(2)}=\left( -\mathrm{Pe} \left\langle u \theta_{0,m}^{(1,1)} \right\rangle+\mathrm{Pe}\left( u-\left\langle u \right\rangle \right)\theta_{0,m}^{(1,1)} +  \left\langle \partial_{x}\theta_{0,m}^{(1,1)} \right\rangle -2\partial_{x}\theta_{0,m}^{(1,1)} \right)  \partial_{\xi}^{2}\phi_{0,m}^{(0)},\\
 & \left. \left( n_{x}\partial_{x}+ \partial_{\mathbf{n}_{y}} \right)\phi_{0,m}^{2}= -n_{x}\theta_{0,m}^{(1,1)}\partial_{\xi}^{2}\phi_{0,m}^{(0)} \right|_{wall }.  \\  
\end{aligned}
\end{equation}
The linearity of the equation inspires us to  assume that $\phi_{0,m}^{(2)}=\theta_{0,m}^{(2,2)} (x,y) \partial_{\xi}^{2}\phi_{0,m}^{(0)}+ \theta_{0,m}^{(2,0)} (\xi)$, where $\theta_{0,m}^{(2,2)}$ satisfies $\left\langle \theta_{0,m}^{(2,2)} \right\rangle=0$ and solves
\begin{equation}
\begin{aligned}
  &\mathcal{L}_{0}\theta_{0,m}^{(2,2)}= -\mathrm{Pe} \left\langle u \theta_{0,m}^{(1,1)} \right\rangle+\mathrm{Pe}\left( u-\left\langle u \right\rangle \right)\theta_{0,m}^{(1,1)} + \left\langle \partial_{x}\theta_{0,m}^{(1,1)} \right\rangle -2\partial_{x}\theta_{0,m}^{(1,1)}, \\
  & \left. \left( n_{x}\partial_{x}+ \partial_{\mathbf{n}_{y}} \right)\theta_{0,m}^{(2,2)}= -n_{x}\theta_{0,m}^{(1,1)}\right|_{wall }.  \\
\end{aligned}
\end{equation}
With a similar calculation, we have $\varphi_{0,m}^{(2)}=q_{0,m}^{(2,2)} (x,y) \partial_{\xi}^{2}\varphi_{0,m}^{(0)}+ q_{0,m}^{(2,0)} (\xi)$, where $q_{0,m}^{(2,2)}$ satisfies $\left\langle q_{0,m}^{(2,2)} \right\rangle=0$ and solves
\begin{equation}
\begin{aligned}
&\mathcal{L}_{0}^{*}q_{0,m}^{(2,2)}= \mathrm{Pe} \left\langle u q_{0,m}^{(1,1)} \right\rangle-\mathrm{Pe}\left( u-\left\langle u \right\rangle \right)q_{0,m}^{(1,1)} +  \left\langle \partial_{x}q_{0,m}^{(1,1)} \right\rangle -2\partial_{x}q_{0,m}^{(1,1)}, \\
  & \left. \left( n_{x}\partial_{x}+ \partial_{\mathbf{n}_{y}} \right)q_{0,m}^{(2,2)}= -n_{x}q_{0,m}^{(1,1)}\right|_{wall }.  \\
\end{aligned}
\end{equation}

When $i=2$, equation \eqref{eq:unit norm condition} becomes 
\begin{equation}
  \begin{aligned}
 &0=\left\langle \phi_{0,m}^{(0)}, \varphi_{0,l}^{(2)} \right\rangle+\left\langle \phi_{0,m}^{(1)}, \varphi_{0,l}^{(1)} \right\rangle+\left\langle \phi_{0,m}^{(2)}, \varphi_{0,l}^{(0)} \right\rangle=\\   
&+ e^{\mathrm{i} 2\pi m\xi}  \left( q_{0,l}^{(2,0)} (\xi) \right)^{*}   + e^{-\mathrm{i} 2\pi l\xi}  \theta_{0,m}^{(2,0)} (\xi) +(2\pi)^{2}lm e^{\mathrm{i} 2\pi (m-l)\xi}  \left\langle \theta_{0,m}^{(1,1)} ,q_{0,l}^{(1,1)}  \right\rangle.\\
\end{aligned}
\end{equation}
When $l \neq m$, additional averaging with respect to $\xi$ gives zero. To make the expression vanishing when $l =m$, we choose $\left\langle \theta_{0,m}^{(2,2)}\right\rangle=\left\langle q_{0,m}^{(2,2)}\right\rangle=0$, and
\begin{equation}
\begin{aligned}
  &\theta_{0,m}^{(2,0)}=q_{0,m}^{(2,0)}=\frac{-(2\pi m)^{2}} {2} e^{\mathrm{i} 2\pi m\xi} \left\langle \theta_{0,m}^{(1,1)} (x,y), q_{0,m}^{(1,1)} (x,y) \right\rangle. \\
\end{aligned}
\end{equation}

 For $\mathcal{O}(\epsilon^3)$,  when $n=0$, we have
\begin{equation}
\begin{aligned}
  \mathcal{L}_{0}\phi_{0,m}^{(3)}&=-\sum\limits_{i=0}^{3}\lambda_{0,m}^{(i)}\phi_{0,m}^{(3-i)}  -\partial_{\xi}^{2}\phi_{0,m}^{(1)}+\mathrm{Pe}u\partial_{\xi}\phi_{0,m}^{2}  -2\partial_{x}\partial_{\xi}\phi_{0,m}^{2}\\
  &=-\lambda_{0,m}^{(3)}\phi_{0,m}^{(0)}-\lambda_{0,m}^{(2)} \theta_{0,m}^{(1,1)}\partial_{\xi}\phi_{0,m}^{(0)}-\theta_{0,m}^{(1,1)}\partial_{\xi}^{3}\phi_{0,m}^{(0)}\\
  &+\mathrm{Pe}\left( u-\left\langle u \right\rangle \right) \left( \theta_{0,m}^{(2,2)}\partial_{\xi}^{3}\phi_{0,m}^{(0)}+ \partial_{\xi}\theta_{0,m}^{(2,0)}  \right) -2\partial_{x}\theta_{0,m}^{(2,2)}\partial_{\xi}^{3}\phi_{0,m}^{(0)},\\
 &\left. \left( n_{x}\partial_{x}+ \partial_{\mathbf{n}_{y}} \right)\phi_{0,m}^{(3)}= -n_{x}\partial_{\xi}\phi_{0,m}^{(2)} \right|_{\mathrm{wall}}.  \\
\end{aligned}
\end{equation}
Averaging the above equation with respect to  $x,y$ leads to
\begin{equation}
\begin{aligned}
  -\left\langle \partial_{\xi}\partial_{x}\phi_{0,m}^{(2)}  \right\rangle=&\left\langle-\lambda_{3}\phi_{0,m}^{(0)} +\mathrm{Pe}\left( u-\left\langle u \right\rangle \right)  \theta_{0,m}^{(2,2)}\partial_{\xi}^{3}\phi_{0,m}^{(0)} -2\partial_{x}\theta_{0,m}^{(2,2)}\partial_{\xi}^{3}\phi_{0,m}^{(0)}\right\rangle,\\
\end{aligned}
\end{equation}
which simples to 
\begin{equation}
\begin{aligned}
&\lambda_{3}\phi_{0,m}^{(0)}= \left\langle-\partial_{x}\theta_{0,m}^{(2,2)} +\mathrm{Pe} \left( u-\left\langle u \right\rangle \right) \theta_{0,m}^{(2,2)} \right\rangle\partial_{\xi}^{3}\phi_{0,m}^{(0)}.
\end{aligned}
\end{equation}

Now we have the expansion for the smallest eigenvalue and its associated eigenfunction as $L\rightarrow \infty$
\begin{equation}
\begin{aligned}
  \lambda_{0,m}=& \mathrm{i} 2 \pi m  \mathrm{Pe} \left\langle u \right\rangle L^{-1} +(2\pi m)^{2}\left\langle1-  \mathrm{Pe} \left( u- \left\langle u \right\rangle \right) \theta_{0,m}^{(1,1)} + \partial_{x} \theta_{0,m}^{(1,1)} \right\rangle L^{-2}\\
  &+ (\mathrm{i} 2\pi m)^{3}\left\langle-\partial_{x}\theta_{0,m}^{(2,2)} +\mathrm{Pe} \left( u-\left\langle u \right\rangle \right) \theta_{0,m}^{(2,2)} \right\rangle L^{-3}+\mathcal{O} (L^{-4}), \\
  \phi_{0,m}=&e^{ \mathrm{i} 2\pi m \xi}+L^{-1}\theta_{0,m}^{(1,1)}  \partial_{\xi}e^{ \mathrm{i} 2\pi m \xi} +L^{-2} \left( \theta_{0,m}^{(2,2)}  \partial_{\xi}^{2}e^{ \mathrm{i} 2\pi m \xi}+ \theta_{0,m}^{(2,0)} \right) +\mathcal{O} (L^{-3}), \\
 \varphi_{0,m}=&e^{\mathrm{i} 2\pi m \xi}+L^{-1}q_{0,m}^{(1,1)} \partial_{\xi}e^{ \mathrm{i} 2\pi m \xi} +L^{-2} \left( q_{0,m}^{(2,2)}  \partial_{\xi}^{2}e^{ \mathrm{i} 2\pi m \xi}+ q_{0,m}^{(2,0)} \right) +\mathcal{O} (L^{-3}). \\ 
\end{aligned}
\end{equation}
Recall that $k=\frac{2 \pi m}{L}$ and $\xi=\frac{x}{L}$, we have the expansion of $\lambda_{0} (k)$ and $\phi_{0} (k)$ around $k=0$
\begin{equation}\label{eq:eigenvalue eigenvector small k expansion 1}
\begin{aligned}
  &\lambda_{0}= \lambda_{0}' (0) k +\lambda_{0}'' (0) \frac{k^{2}}{2}+\lambda_{0}''' (0) \frac{k^{3}}{6}+\mathcal{O} (k^{4}), \\
  &\lambda_{0}' (0)=\mathrm{i} \mathrm{Pe} \left\langle u \right\rangle, \quad \lambda_{0}'' (0)=2\left\langle1-  \mathrm{Pe} \left( u- \left\langle u \right\rangle \right) \theta_{0,0}^{(1,1)} + \partial_{x} \theta_{0,0}^{(1,1)} \right\rangle,\\
 &\lambda_{0}''' (0)= -6\mathrm{i}\left\langle-\partial_{x}\theta_{0,0}^{(2,2)} +\mathrm{Pe} \left( u-\left\langle u \right \rangle \right) \theta_{0,0}^{(2,2)} \right\rangle, \\
 &\phi_{0}=e^{ \mathrm{i}  k x} \left( 1+\mathrm{i}  k \theta_{0,0}^{(1,1)} -k^{2} \left( \theta_{0,0}^{(2,2)} + \frac{1}{2}\left\langle \theta_{0,0}^{(1,1)}, q_{0,0}^{(1,1)} \right\rangle \right)+ \mathcal{O} (k^{3})  \right), \\
 &\varphi_{0}=e^{ \mathrm{i}  k x} \left( 1+\mathrm{i}  k q_{0,0}^{(1,1)} - k^{2} \left( q_{0,0}^{(2,2)} + \frac{1}{2}\left\langle \theta_{0,0}^{(1,1)}, q_{0,0}^{(1,1)} \right\rangle \right)+ \mathcal{O} (k^{3})  \right), \\
\end{aligned}
\end{equation}
which is exactly the same as the result we obtained by expanding the eigenfunction around the small wave number in appendix \ref{sec:Eigenfunction expansion for small wavenumbers}.

\section{Numerical method}
\label{sec:Numerical method}
This section describes the algorithm used to obtain the steady solution of the Navier-Stokes equation and advection-diffusion equation.

We employ Newton's method to compute the steady-state flow solution. The Newton's method has a faster convergence rate if a sufficiently accurate approximation is known. We  obtain such an initial guess by the Picard iteration, which is commonly used for solving the steady Navier-Stokes equation due to its stability and global convergence properties  \cite{pollock2019anderson}.
\begin{equation}\label{eq:NS Oseen}
\begin{aligned}
&\mathrm{Re} \mathbf{u}^{n} \cdot \nabla \mathbf{u}^{n+1}  =  \Delta \mathbf{u}^{n+1} -\nabla p^{n+1}+ \mathbf{f}^{n}, \quad \nabla \cdot \mathbf{u}^{n+1} = 0, \quad \left. \mathbf{u}^{n+1} \right|_{\mathrm{wall}}=0,
\end{aligned}
\end{equation}
where the superscript $n$ denotes the solution at the $n$-th iteration. To maintain a unit flux at the inlet during the iteration, we compute $F=\frac{1}{|\Omega_{\text{inlet}}|} \int_{\text{inlet}} u (x,\mathbf{y}) \mathrm{d}\mathbf{y}$ and normalize the velocity field by $1/F$ to ensure the total flux equals unity. The Picard iteration is terminated once the difference between successive solutions falls below a prescribed tolerance, and the resulting field is then used as the initial guess for Newton’s method.

The finite element method is used to discretize the system of equations.  A triangular mesh with quadratic elements (P2) is employed.  We implement the algorithm using the software FreeFEM++ \cite{hecht2012new}.  After rewriting  the steady Navier-Stokes equation in the weak form, the problem becomes finding the $\mathbf{u},p$ such that $F(\mathbf{u},p)=0$ for all test function $\mathbf{v}$ (zero on the boundary) and $q$  , where
\begin{equation}
\begin{aligned}
&F (\mathbf{u},p)= \int\limits_{\Omega}^{} \mathrm{Re}(\mathbf{u} \cdot \nabla \mathbf{u}) \cdot \mathbf{v}+ \nabla \mathbf{u} :: \nabla \mathbf{v}- p \nabla \cdot \mathbf{v}- q \nabla \cdot \mathbf{u} +\mathbf{f}\cdot \mathbf{v} \mathrm{d} \mathbf{x},
\end{aligned}
\end{equation}
where $\nabla \mathbf{u} :: \nabla \mathbf{v}= \sum\limits_{i,j}^{} \partial_{x_{i}}u_{j}\partial_{x_{i}}v_{j}$.
Linearizing $F(\mathbf{u},p)$ around $(\mathbf{u}^{n}, p^{n})$ gives $F (\mathbf{u}^{n}+\delta \mathbf{u}^{n},p^{n}+\delta p^{n}) \approx F (\mathbf{u}^{n},p^{n})+ DF (\mathbf{u}^{n}, p^{n}) (\delta u^{n}, \delta p^{n})$, where
\begin{equation}
\begin{aligned}
&DF (\mathbf{u}, p)(\delta u, \delta p)=\int\limits_{\Omega}^{} \mathrm{Re} \left( (\delta \mathbf{u} \cdot \nabla \mathbf{u}) \cdot \mathbf{v}+ ( \mathbf{u} \cdot \nabla \delta \mathbf{u}) \cdot \mathbf{v} \right)+ \nabla \delta \mathbf{u} :: \nabla \mathbf{v}-\delta p \nabla \cdot \mathbf{v}- q \nabla \cdot \delta\mathbf{u}\mathrm{d} \mathbf{x}.
\end{aligned}
\end{equation}
Assuming that $F (\mathbf{u}^{n}+\delta \mathbf{u}^{n},p^{n}+\delta p^{n})\approx 0$,  at each iteration, we solve the linear equation $0=F (\mathbf{u}^{n},p^{n})+ DF (\mathbf{u}^{n}, p^{n}) (\delta \mathbf{u}^{n}, \delta p^{n})$ for $\delta \mathbf{u}^{n}$ and $\delta p^{n}$ with a no slip boundary condition at the solid wall of the channel and a periodic boundary condition at $x=0,L$.  Next we update the solution with $\mathbf{u}^{(n+1)}= \mathbf{u}^{n}+ \delta \mathbf{u}^{n}$. As in the Picard iteration step above, the velocity field is normalized at each step using the inlet flux.  The iteration terminates if $\lVert \delta \mathbf{u}^{n}\rVert_{\infty}/ \left| \mathbf{u}^{n+1} \right|$ is smaller than the given tolerance.

The advection-diffusion equation \eqref{eq:AdvectionDiffusionEquationNon1} is discretized using P2 elements. The Crank-Nicolson method is applied for time integration, resulting in the following equation:
\begin{equation}
\begin{aligned}
&\frac{c^{n+1}-c^{n}}{\frac{1}{2}h}+ \mathrm{Pe} \mathbf{u} \cdot \nabla \left( c^{n+1}+c^{n} \right) =   \Delta \left( c^{n+1}+c^{n} \right) ,\quad  \left.  \mathbf{n}\cdot \nabla c^{n+1} \right|_{\mathrm{wall}}=0.
\end{aligned}
\end{equation}

\bibliographystyle{jfm}

\end{document}